\pdfoutput=1
\documentclass{article} 
\usepackage[font=large]{subfig}
\usepackage[export]{adjustbox}
\usepackage{graphicx} 
\usepackage{pdfpages} 
\usepackage{amssymb}
\usepackage{amsmath}
\usepackage{url}
\usepackage{appendix}
\usepackage{booktabs}
\usepackage{carlito}
\usepackage{comment}
\usepackage{float}
\usepackage{tikz}

\makeatletter
\def\@normalsize{\@setsize\normalsize{10pt}\xpt\@xpt
\abovedisplayskip 10pt plus2pt minus5pt\belowdisplayskip 
\abovedisplayskip \abovedisplayshortskip \z@ 
plus3pt\belowdisplayshortskip 6pt plus3pt 
minus3pt\let\@listi\@listI}

\def\subsize{\@setsize\subsize{12pt}\xipt\@xipt}
\def\section{\@startsection {section}{1}{\z@}{1.0ex plus
1ex minus .2ex}{.2ex plus .2ex}{\large\bf}}
\def\subsection{\@startsection 
   {subsection}{2}{\z@}{.2ex plus 1ex} {.2ex plus .2ex}{\subsize\bf}}
\makeatother
\begin{document}
\date{}

\title{\bf {\carlito Poisson\_CCD}: A dedicated simulator for modeling CCDs}

\author{Craig Lage, Andrew Bradshaw, J. Anthony Tyson, and the LSST Dark Energy Science Collaboration\\
  Department of Physics\\
  University of California - Davis\\
cslage@ucdavis.edu}

\maketitle

\subsection*{\centering Abstract}
{\em Keywords: 
LSST, modeling, camera, CCD, simulation, diffusion, image processing.  
}
\\
\\
A dedicated simulator, {\carlito Poisson\_CCD}, has been constructed which models astronomical CCDs by solving Poisson's equation numerically and simulating charge transport within the CCD.  The potentials and free carrier densities within the CCD are self-consistently solved for, giving realistic results for the charge distribution within the CCD storage wells.  The simulator has been used to model the CCDs which are being used to construct the LSST digital camera.  The simulator output has been validated by comparing its predictions with several different types of CCD measurements, including astrometric shifts, brighter-fatter induced pixel-pixel covariances, saturation effects, and diffusion spreading.  The code is open source and freely available.

\section{Introduction}

Charge Coupled Devices (CCDs) have been the workhorse devices for astronomical imaging for some time.  George Smith's Nobel lecture at \cite{Smith_Nobel} gives an excellent summary of the early history.  While other detectors are making inroads, CCDs are still the dominant imaging device in astronomical applications.  In recent years thick, fully depleted CCDs with their wide spectral response have been applied to spectroscopic applications as well as imaging.  Although these devices have high quantum efficiency, relatively good linearity, and acceptable dynamic range, they have a number of problematic effects that can impact the precision and accuracy of astronomical data.  It is important that these effects are well understood so that they can be accounted for during image processing.  To help understand these effects, we have built a dedicated simulator, {\carlito Poisson\_CCD}, which solves the electrostatics in the bulk silicon of the CCD, and propagates incoming charges down to the collecting wells where they are collected and stored.  The simulator has proven very useful for understanding a number of CCD effects, which will be described in this paper.

Of course, detailed semiconductor modeling codes already exist, are commercially available, and have been validated against silicon results.  What is the purpose of developing yet another simulator?  The answer is severalfold.  First, commercial semiconductor codes typically use proprietary source codes and are quite expensive, while the code described here is open source and freely available \cite{Poisson-CCD-code}.  Second, using a commercial semiconductor device simulator requires spending quite a bit of time learning to use the code and set up the initial conditions.  The code described here sets up the initial conditions for a typical CCD with a few simple configuration parameters.  A third reason is that some of the commercial tools have trouble handling grids physically large enough to simulate a meaningful number of pixels.  This code circumvents this by using a variable grid spacing, as described in Section \ref{Basics}.  Also, it is hoped that this code is simple enough that it can be mastered by people who are not semiconductor experts.  The target user group is people in the astronomy field who want to answer questions about CCDs without investing a great deal of time learning the details of semiconductor physics.

The code was developed as part of the development effort of the LSST.  The LSST, originally known as the Large Synoptic Survey Telescope, is now known as the Simonyi Survey Telescope at the Vera Rubin Observatory, and will be used to conduct a ten-year survey of the southern sky known as the Legacy Survey of Space and Time.  This instrument is an innovative, large, fast survey facility currently under construction at Cerro Pachon in Chile \cite{LSST_2019}.  The digital camera, also currently under construction, consists of approximately 3.2 gigapixels and is the largest digital camera ever constructed.  The camera uses fully-depleted silicon CCDs which are back illuminated and 100 microns thick in order to optimize quantum efficiency in the near infrared.  The imaging area consists of 189 CCDs, with each CCD containing 16 imaging regions laid out in an 8x2 array.  Each imaging region has a pixel array with approximately 500x2000 10 micron square pixels, giving 16 Megapixels total.  Each imaging region also has its own independent amplifier (\cite{oconnor2019uniformity}, \cite{oconnor_2016}).  The focal plane contains CCDs from two different vendors, the ITL STA3800C from the  University of Arizona Imaging Technology Laboratory \cite{ITL_website}, and the E2V CCD250 from Teledyne E2V \cite{E2V_website}.  However, although this code was developed and tested against these two CCDs from the LSST project, it has already found more general use on other CCDs (\cite{Villasenor_2017}), and the hope is that this will continue.

This paper is divided into several sections.  In Section 2, we give an overview of the simulator, describing the basic structure of the simulation volume, how we solve the semiconductor equations, and how we treat incoming photons.  We also show a number of examples of the outputs available from the simulator.  In Section 3, we review a number of the validation tests that were performed to validate the simulation results against measured data of different kinds, and finally we conclude.

\section{Overview}
\begin {figure}[H]
   \subfloat[b][Silicon volume]{\includegraphics[trim=0.0in 1.0in 0.4in 0.5in,clip,width=0.59\textwidth]{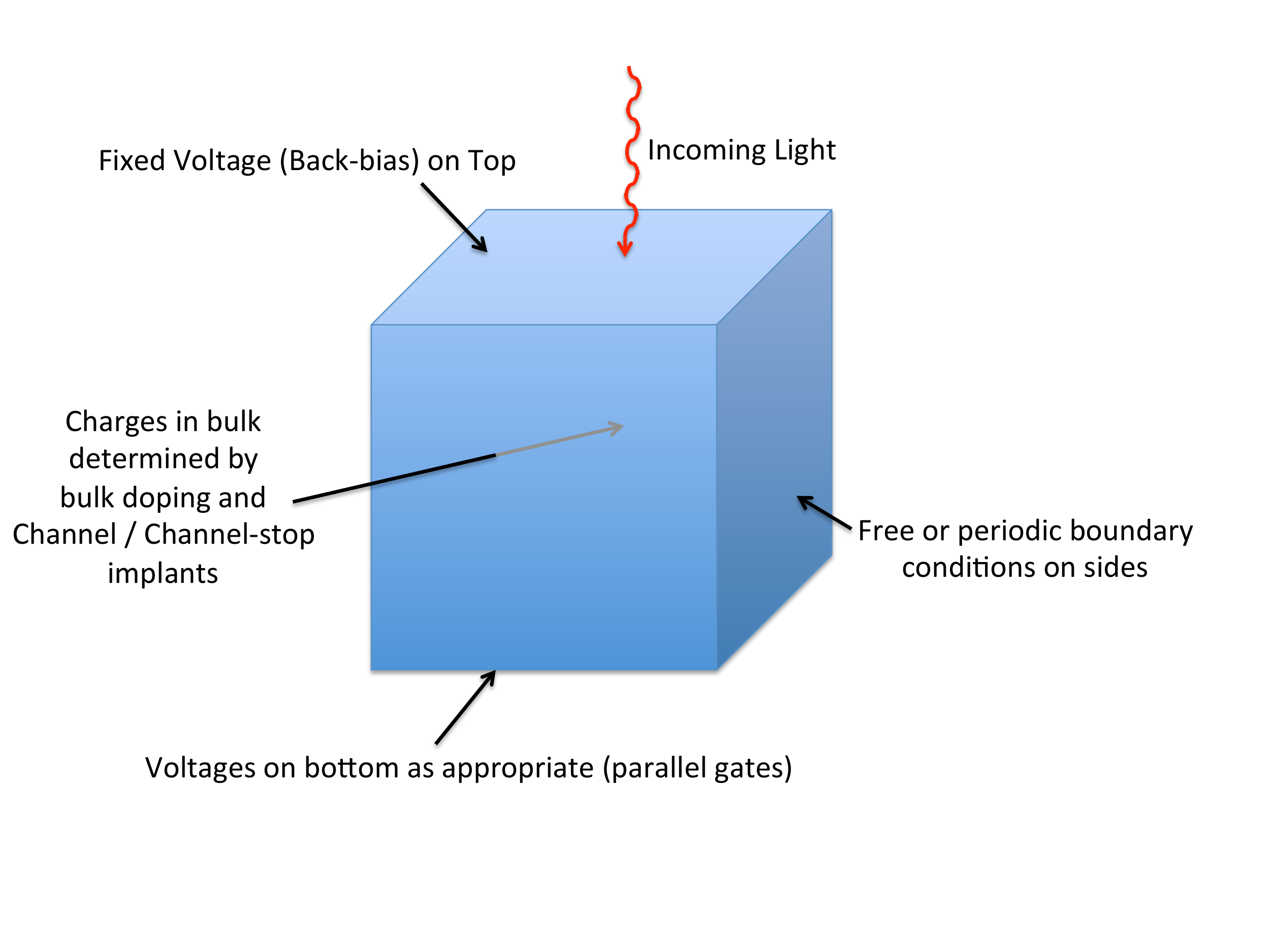}}
   \subfloat[b][Charge tracking]{\includegraphics[trim=2.0in 0.0in 1.0in 0.1in,clip,width=0.39\textwidth]{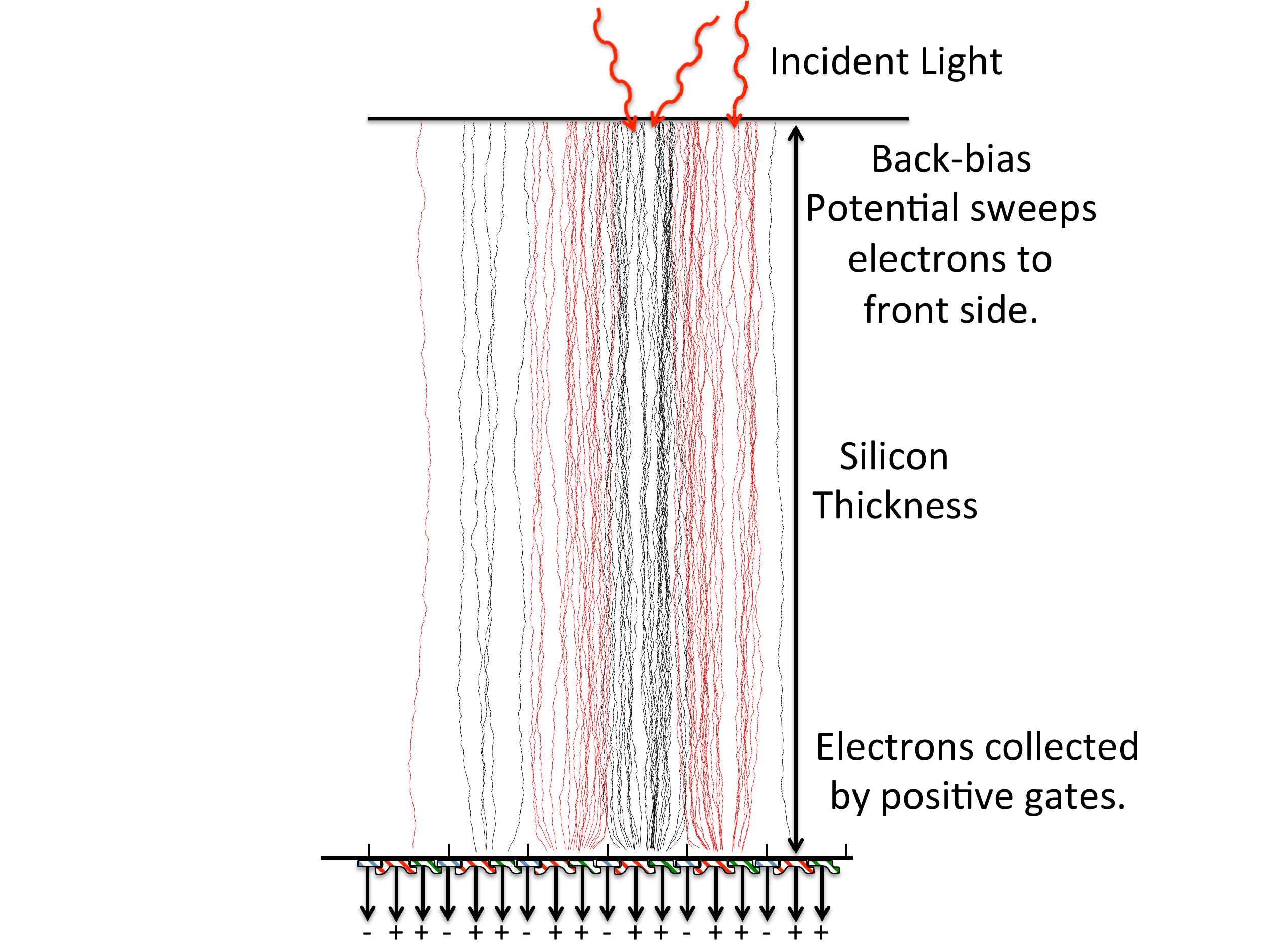}}
   \caption{Major tasks of the {\carlito Poisson\_CCD} simulator.  The left-hand figure shows the silicon volume.  Given charges in the silicon bulk and voltages applied to the silicon boundaries, one solves for the electric fields and free carrier densities within the silicon.  The right-hand figure shows incoming charges, created by incident photons, which are propagated down through the silicon bulk to determine the number of charges in each pixel.  The red and black colors are used to delineate the charges associated with individual pixels.}
  \label{Major_tasks}
  \end{figure}

The simulator performs two basic tasks, as shown in Figure \ref{Major_tasks}.  First, given the charges in the silicon bulk and the boundary conditions determined by potentials applied to the silicon surface, the simulator solves Poisson's equation numerically to determine the potentials and electric fields in the silicon bulk.  The solution to Poisson's equation is determined using the technique of successive over-relaxation (SOR - see for example \cite{HADJIDIMOS2000177}), and using multi-grid methods (see Section \ref{Multi-grid}) to speed convergence.  In regions where there are mobile carriers (holes and electrons) quasi-Fermi level methods are used to simultaneously solve for the potentials and free carrier densities in the device.  The electric fields are determined by numerically differentiating the electrostatic potential.  In general, the simulator only solves for the potentials and free-carrier densities in equilibrium, and is not intended to give transient solutions.  However, it is possible to repeatedly solve the equations with slight changes in initial conditions in order to give transient results.  This technique has been used to generate movies of the CCD charge transport, as discussed in Section \ref{Other_tests}.

After solving for the device potentials, the second major task of the simulator comes into play.  As incoming photons enter the CCD, they generate hole-electron pairs.  The electric field in the CCD separates these charges, and the electrons propagate down to the collecting wells where they are collected, stored, and later counted.  The simulator models this carrier transport in a physically realistic way, in order to determine in which pixel a generated carrier ends up.  This is very useful for modeling pixel distortions that result from electric fields in the device, either built-in electric fields, such as those due to ``tree rings''\cite{plazas2014}, or electric fields due to collected charges, such as those that lead to the brighter-fatter (BF) effect (\cite{antilogus2014}, \cite{gruen2015}, \cite{guyonnet2015}, \cite{Lage_2017}, \cite{Coulton_2018}).  Note that in an astronomical CCD, the time between incoming charges (on the order of msec) is typically much longer than the time required for a charge to propagate down to the bottom (on the order of nsec).  Also, a single charge has little impact on the existing potentials and fields.  So it is an excellent approximation to assume that a single charge propagates in a frozen electric field.  The simulator is designed so that if multiple charges are being added, the user can choose how often to re-solve Poisson's equation.  A comment here about the nature of the charges is in order.  The simulator is designed for N-channel CCDs (with P-type substrate), where the collected charges are electrons, and for the rest of the paper we will make that assumption.  There is no physical reason why it will not work for P-channel CCDs (with N-type substrate), but it is not currently designed to support them.  Most of the work in adapting it for P-channel CCDs would be nomenclature related, basically interchanging the names of holes and electrons, but adjustment of parameters like mobilities and effective masses would also be needed.  This could be done if there is demand for it.

\subsection{Basic structure of the simulations}
\label{Basics}
The simulator is written in C++, and is controlled by a text-based configuration file, which contains all of the information about the silicon volume, pixel sizes, number of pixels, any non-pixel regions, etc.  The configuration file also defines the problem being solved.   By convention the configuration file has a .cfg extension, but this is not necessary.  Appendix \ref{Parameter_Appendix} lists the configuration parameters.  As the simulation progresses, it writes out a number of files.  Large files containing information like potentials, charge densities, electric fields at each grid point are written as high-density HDF5 files, having file extension .hdf5.  Several smaller text files, with a .dat extension, are written which contain information on the grids or the number of electrons in each pixel.  After the simulation has completed, easily modifiable Python scripts are used to plot out results as desired.

\begin {figure}[H]
\begin{center}
   \includegraphics[trim=0.0in 0.0in 0.0in 0.0in,clip,width=0.49\textwidth]{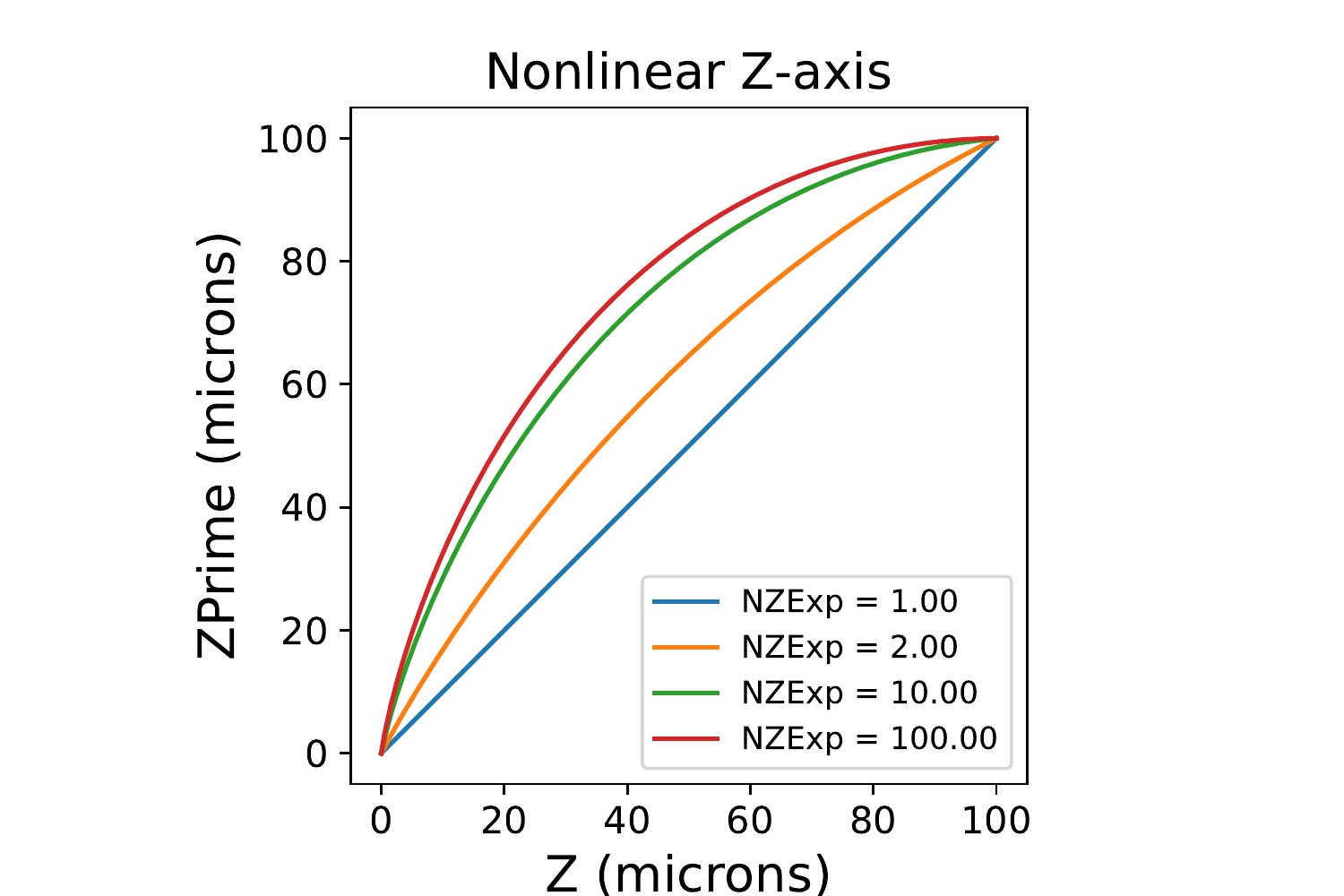}
   \caption{Non-linear Z-axis scheme. The parameter NZExp allows one to have smaller Z-axis grid cells near the bottom of the simuation volume, where the fields and charges are changing more rapidly.  A value NZExp=1 is a linear grid.  The recommended value is to Use NZExp=10.0, which increases the resolution at z=0 by a factor of 10, and decreases the resolution at the top of the CCD by a factor of 10.}
  \label{Nonlinear_Z}
\end{center}
\end{figure}

The simulation volume is set up on a fixed three-dimensional rectangular grid, which does not change once the simulation has started.    One begins by deciding the number of grid cells in each dimension.  Because multi-grid methods are used (see Section \ref{Multi-grid}), the number of grid cells in each dimension must be a multiple of 32.  Note that as a convention, we refer to the side of the CCD where the circuitry is patterned as the bottom, and the side where the incident light comes in as the top.  For the LSST CCDs, which are 100 microns thick and have pixels 10 microns square, a typical resolution is to have 32 simulation grid cells per pixel, so that each grid cell is 0.31 microns.  Initially, the grid was defined to be completely uniform in all three dimensions.   However, for thick CCDs like those used in the LSST, the potentials and fields change rapidly in the region near the bottom, and only very slowly near the top.  When the simulation had enough resolution to be accurate in the rapidly changing region at the bottom, most grid cells were wasted near the top.  Of course, adaptive grid methods solve this problem, but also make the code much more complex, which defeated the purpose of having a relatively simple simulator.  The solution chosen in this work was to use a non-linear grid in the Z-dimension only.   This has proven to give high resolution where needed, without adding significant complexity to the simulation code.  Figure \ref{Nonlinear_Z} shows the scheme.  This introduces some additional partial derivatives, as discussed more in section \ref{Solving}, but these values can be pre-calculated and this is much simpler than an adaptive grid scheme.

\subsection{Setting up the initial conditions}
Because the simulator is not a general purpose semiconductor solver, and is intended to model devices with a given structure, certain assumptions are made about the structure of the CCD which simplifies building the device structure.  This is one of the reasons why this code is so much simpler than a commercial device simulator.  It is assumed that the CCD is a slab of silicon with a given thickness given by the parameter ``SensorThickness''.  It is assumed that the top surface of the CCD is at a fixed voltage given by the parameter ``Vbb''.  The bottom surface of the CCD has voltages specified by the various gate potentials.  The doping deep in the silicon is asumed constant with a value given by ``BackgroundDoping'', although a periodic variation in this doping can be introduced using the ``TreeRing'' parameters.  The doping level is assumed to be modified by the introduction of implants from the bottom side.  There are several options for these doping profiles, including a square profile of a given depth or a sum of N Gaussian profiles.  Use of 1 or 2 Gaussian profiles has been found to accurately reproduce the measured doping profiles on commercial CCDs.  For more details on this, see \cite{CCD-Physical-Analysis}.
\subsubsection{Pixel arrays}
Setting up the initial conditions in the periodic pixel array is straightforward, and is specified by a relatively small number of parameters which describe the gate voltages and doping levels.  Rather than go through these in detail, the reader is referred to Appendix \ref{Parameter_Appendix} or the ``pixel-itl'' and ``pixel-e2v'' examples at \cite{Poisson-CCD-code}.

\subsubsection{Fixed regions}
Setting up the initial conditions in non-periodic regions outside the pixel array is straightforward, but more laborious than setting up the pixel arrays.  The extents, dopings, applied voltages, and quasi-Fermi levels need to be specified for each region.  At present only rectangular regions are supported.  Also, it is assumed that the same doping profiles which are used in the pixel array are used in the surrounding circuitry, so the only options for doping profiles are the channel doping, the channel stop doping, and no doping.  Examples of simulations setting up non-periodic regions are the ``edge.cfg'', ``trans.cfg'', and ``io.cfg'' files at \cite{Poisson-CCD-code}, and the results of these simulations are detailed in Section \ref{Validation}.
\subsection{Solving for the potentials, fields, and free carrier densities}
\label{Solving}
In this section we give a brief description of the methods that are used to solve Poisson's equation on the grid.  We are trying to solve the following equation, where $\varphi$ is the potential, $\rho$ is the charge density, and $\rm \epsilon_{Si}$ is the dielectric constant of silicon:
\begin{equation}\rm
  \nabla^2 \varphi = \frac{\rho}{\epsilon_{Si}}
\end{equation} 
Each of the partial derivatives can be discretized on a rectangular grid with grid spacing h as follows:
\begin{equation}\rm
  \frac{\partial^2 \varphi_{i,j,k}}{\partial x ^2} = \frac{\left(\varphi_{i+1,j,k} - \varphi_{i,j,k}\right) - \left(\varphi_{i,j,k} - \varphi_{i-1,j,k}\right)}{h^2}
\end{equation} 
Giving for the discretized Poisson's equation:
\begin{equation}\rm
  \varphi_{i+1,j,k} + \varphi_{i-1,j,k} + \varphi_{i,j+1,k} + \varphi_{i,j-1,k} + \varphi_{i,j,k+1} + \varphi_{i,j,k-1} - 6 * \varphi_{i,j,k} = \frac{h^2}{\epsilon_{Si}} * \rho_{i,j,k}
\end{equation} 
This can be turned into an iterative equation, and basically one just iterates until convergence:
\begin{equation}\rm
  \varphi^{\left(n+1\right)}_{i,j,k} = \frac{1}{6} * \left(\varphi^{\left(n\right)}_{i+1,j,k} + \varphi^{\left(n\right)}_{i-1,j,k} + \varphi^{\left(n\right)}_{i,j+1,k} + \varphi^{\left(n\right)}_{i,j-1,k} + \varphi^{\left(n\right)}_{i,j,k+1} + \varphi^{\left(n\right)}_{i,j,k-1} - \frac{h^2}{\epsilon_{Si}} * \rho_{i,j,k}\right)
\end{equation}
Which we write in shorthand as follows:
\begin{equation}\rm
  \varphi^{\left(n+1\right)} = \frac{1}{6} * \left(\varphi_{pm}^{\left(n\right)} - \frac{h^2}{\epsilon_{Si}} * \left(\rho_f + \rho_m\right)\right)
\end{equation}
where we define:
\begin{equation}\rm
  \varphi_{pm}^{\left(n\right)} = \varphi^{\left(n\right)}_{i+1,j,k} + \varphi^{\left(n\right)}_{i-1,j,k} + \varphi^{\left(n\right)}_{i,j+1,k} + \varphi^{\left(n\right)}_{i,j-1,k} + \varphi^{\left(n\right)}_{i,j,k+1} + \varphi^{\left(n\right)}_{i,j,k-1}
\end{equation}
and we have split $\rho$ into a fixed charge density $\rm \rho_f$ and a mobile charge density $\rm \rho_m$.  However, $\rm \rho_m$ is a highly nonlinear function of the potential $\varphi$, as described below.  In quasi-equilibrium, the drift current $\rm J_E$ and the diffusion current $\rm J_D$ are equal, giving a net current of zero, so we can write (see Sze \cite{sze1981physics}, for example):

  \begin{equation} \rm
    J_E =  q_e \mu_n n \frac{d \varphi}{d x} = -J_D = -q_e D_n \frac {d n}{d x} 
  \end{equation}
  Here $\rm q_e$ is the electron charge, $\rm \mu_n$ is the electron mobility, $\rm n$ is the electron density in electrons per unit volume, and $\rm D_n$ is the electron diffusion coefficient.  This gives: 
  \begin{equation} \rm
    \mu_n d \varphi = -D_n \frac {d n}{n} 
  \end{equation}
We also know from the Einstein relations that the mobility and diffusion coefficent are related by the following equation, where $\rm k$ is Boltzmann's constant and $\rm T$ is the absolute temperature::
\begin{equation} \rm
    \frac{\mu_n}{D_n} = \frac{q_e}{kT}
\end{equation}
so, integrating both sides:
\begin{equation} \rm
    \frac {q_e \varphi}{kT} = \log\left(n\right) + C
\end{equation}
We take the constant of integration into the exponential and define the quasi-Fermi level $\rm \varphi_F$ in terms of the intrinsic carrier density $\rm n_i$, giving:
\begin{equation} \rm
    n = n_i \exp \left( \frac {q_e \left( \varphi - \varphi_F \right)}{kT} \right)
  \end{equation}
So we need to solve the following equation for $\rm \varphi$, where $\rm \varphi_F$ is a constant:    
    \begin{equation} \rm
    \nabla^2 \varphi = \frac{1}{\epsilon_{Si}} \left(\rho_f + q_e n_i \exp\left(\frac {q_e \left(\varphi - \varphi_F\right)}{kT}\right)\right)
  \end{equation}
which, when discretized is:
\begin{equation}\rm
  \varphi^{\left(n+1\right)} = \frac{1}{6} * \left(\varphi_{pm}^{\left(n\right)} - \frac{h^2}{\epsilon_{Si}} * \rho_f - \frac{h^2 q_e n_i}{\epsilon_{Si}} \exp\left(\frac {q_e \left(\varphi^{\left(n+1\right)} - \varphi_F\right)}{kT}\right)\right)
\end{equation}

\begin {figure}[H]
  \begin{center}
    \includegraphics[trim=0.0in 0.0in 0.0in 0.0in,clip,width=0.99\textwidth]{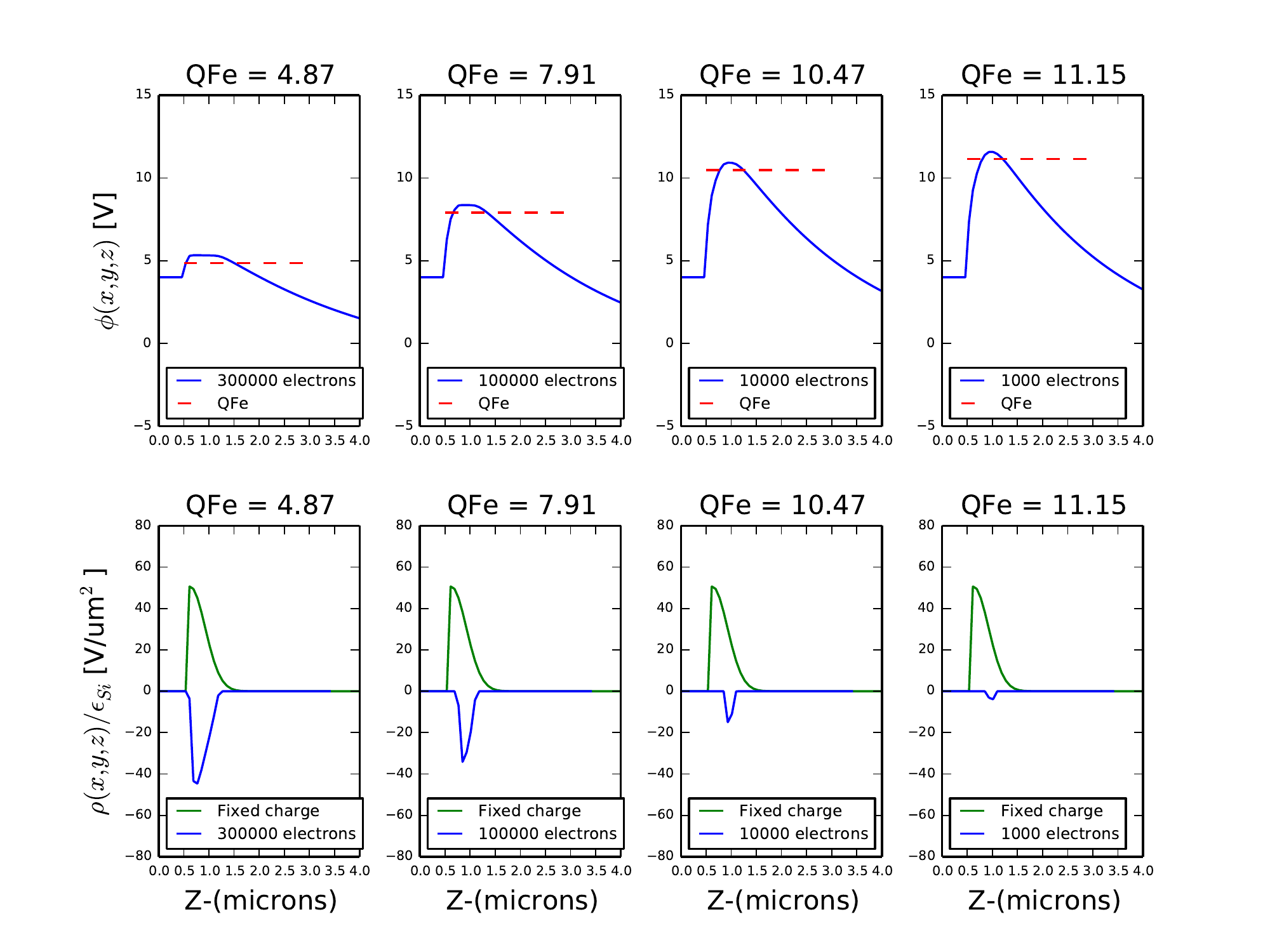}
  \end{center}
\caption{Impact of varying $\rm \varphi_F$ (called QFe in the code, with units of Volts) on the potential and electron density.}
\label{QFe}
\end{figure}

Because of the strong non-linearity, simply iterating is numerically unstable.  The method that works, described in detail by Rafferty, et.al. \cite{rafferty1985iterative} and buillding on the work of Gummel (\cite{Gummel}), is to take this  last equation as a non-linear equation for $\rm \varphi^{\left(n+1\right)}$ in terms of $\rm \varphi^n$ and run a Newton's method ``inner loop'' to find $\rm \varphi^{\left(n+1\right)}$ at each grid point.  Then we iterate to convergence as before.  This allows us to simultaneously solve for the potential and the carrier density.  Of course, we just described the electron density here, but there is a similar equation for holes, but with opposite signs.  Figure \ref{QFe} shows an example of varying $\rm \varphi_F$ on the solution.  Note that the quasi-Fermi level is constant in each region containing mobile carriers.  In the CCD, each collecting well contains a different number of mobile carriers, so $\rm \varphi_F$ is constant in each well, but is different from well to well.  However, when simulating the device, instead of knowing the value of $\rm \varphi_F$, we typically know the number of electrons in each well.  So how do we translate from the known number of electrons to the unknown value of $\rm \varphi_F$?  The code provides two methods, selected by the value of the parameter ``ElectronMethod''.  With this parameter set to 1, a test simulation is run where the parameter $\rm \varphi_F$ (called QFe in the code) is varied through a range.  For each value of QFe, we integrate over a pre-determined region to count the number of electrons in the well, building a look-up table of the number of electrons vs QFe.  The results for a few values of QFe are shown in Figure \ref{QFe}.   Then the code interpolates to determine the value of QFe which gives the appropriate number of electrons.  In practice one can get close to the desired number of electrons, but the non-linearity causes variations from the desired number, so a second method was developed.  When ``ElectronMethod'' has a value of 2, what is done is to place the correct number of electrons in the well, uniformly distributed in the center of the well.  The code then moves the electrons around until the value of QFe is constant in the well.  This allows one to get the correct number of electrons in the well without needing to know the value of the quasi-Fermi level.

There is one more complication.  As discussed in Section \ref{Basics}, a non-linear Z-axis is used to concentrate grid cells in the bottom region where the potentials and charge densities are changing most strongly. In principle, any smooth funcation can be used for the Z-axis mapping.  Here we chose an easily differentiable polynomial function of the following form, where z is the linear z coordinate, and zp is the non-linear coordinate, which is the actual value used in solving and plotting.  Here $\rm T_{Si}$ is the silicon thickness, and NZExp is a user-defined constant that controls the non-linearity, as described in Figure \ref{Nonlinear_Z}.

    \begin{equation}\rm
  zp = - T_{Si} * \left(NZExp - 1.0\right) * \left(z / T_{Si}\right)^{\left(NZExp + 1.0\right)/NZExp} + NZExp * z;
    \end{equation} 

The non-linear z-axis modifies Poisson's equation from:

    \begin{equation}\rm
      \nabla^2 \varphi = \frac{\partial^2 \varphi}{\partial x^2} + \frac{\partial^2 \varphi}{\partial y^2} + \frac{\partial^2 \varphi}{\partial z^2}
    \end{equation} 
to:
    \begin{equation}\rm
      \nabla^2 \varphi = \frac{\partial^2 \varphi}{\partial x^2} + \frac{\partial^2 \varphi}{\partial y^2} + \frac{\partial^2 \varphi}{\partial z^2} \left(\frac{\partial z'}{\partial z}\right)^2 + \frac{\partial \varphi}{\partial z'} \frac{\partial^2 z'}{\partial z^2}
    \end{equation}
In practice, the additional partial derivatives can be pre-computed, so this is simply accounted for in the discretized equations and has only a minor impact on the speed of iteration.

\subsection{Multi-grid methods}
\label{Multi-grid}
\begin {figure}[H]
\begin{center}
\includegraphics[trim=0.0in 0.0in 0.0in 0.5in,clip,width=0.80\textwidth]{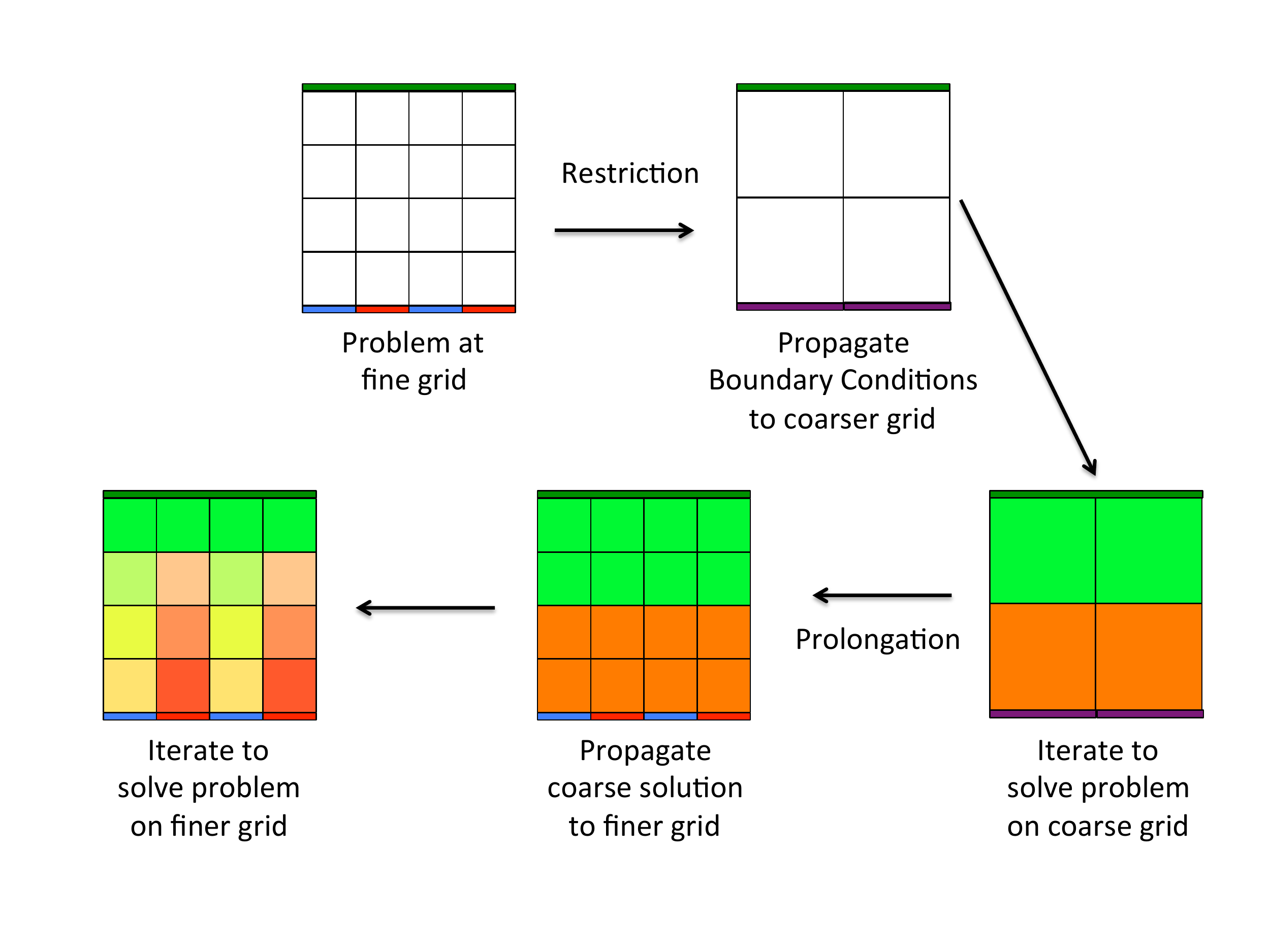}
\includegraphics[trim=0.0in 1.0in 0.0in 0.7in,clip,width=0.80\textwidth]{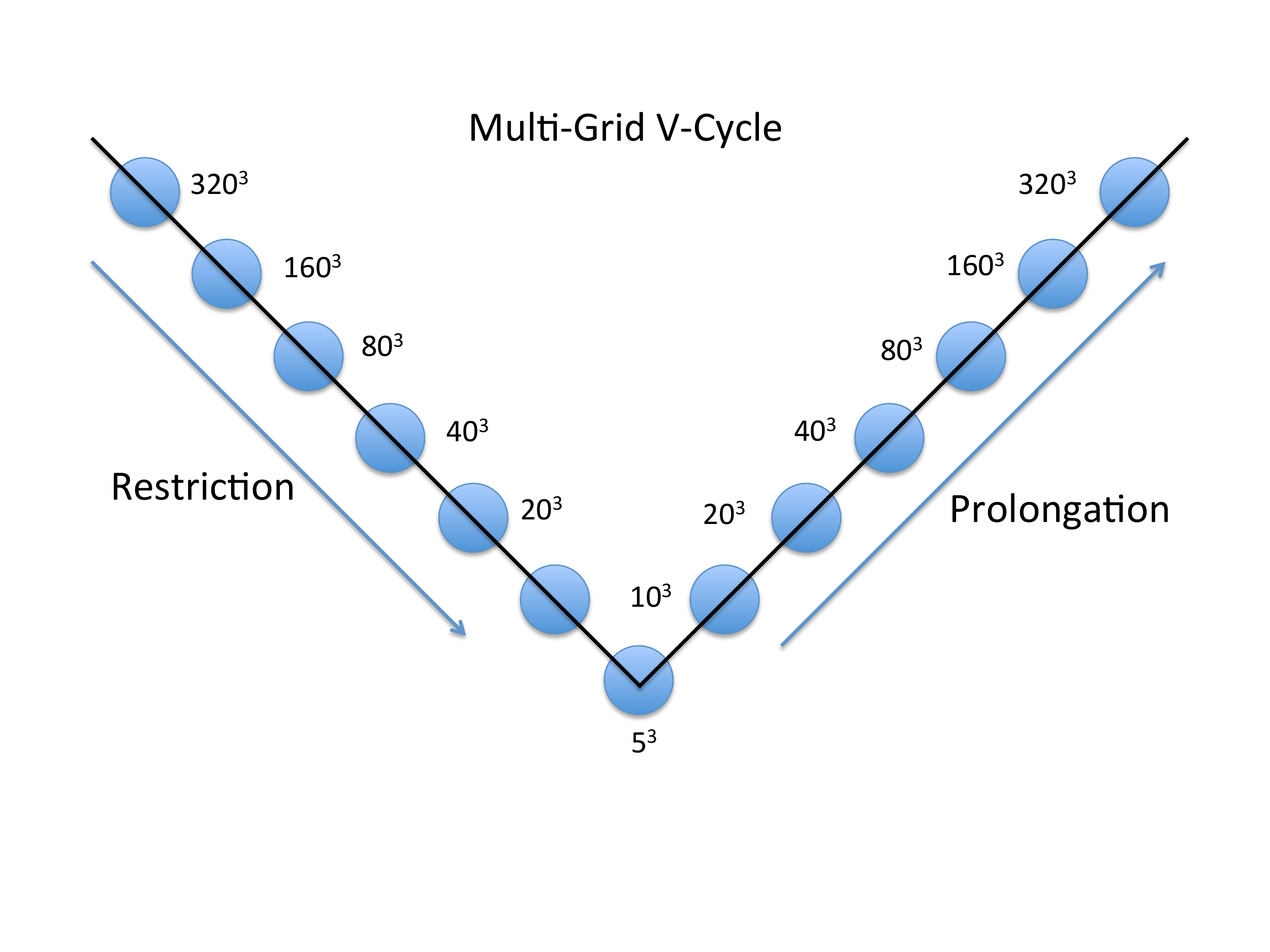}
\caption{Basic idea of multi-grid methods.  Long wavelength modes are solved on a coarser grid, which is then propagated to a finer grid (``Prolongation''), where more iterations are performed to find the fine details of the solution.}
\label{Multi}
\end{center}
\end{figure}

It is well known that multi-grid methods speed convergence of solutions to Poisson's equation by getting correct solutions to the long-wavelength modes at a coarser grid where convergence is much more rapid.  There is a wealth of literature on the subject, and Briggs \cite{briggs2000multigrid} or Press \cite{NumericalRecipes} give excellent summaries.  The basic idea is shown in Figure \ref{Multi}.  In practice in this code, we have adopted a simpler method.  Rather than use ``Restriction'' to propagate the boundary conditions down to the coarser grid, we simply set up the boundary conditions on all of the sub-grids at the outset of the problem.  In addition, we have found that there is little value in running coarser grids than $\rm 40^3$, because at this resolution the problem converges very rapidly.  So for a typical problem which has perhaps $\rm 320^3$ grid cells, we define the finest grid and three subgrids, with the coarsest grid having $\rm 40^3$ grid cells.  We then set up the boundary conditions on all of the subgrids, iterate the coarsest grid, ``Prolongate'' the solution up to the next grid, and continue until we have reached the finest grid.

\begin {figure}[H]
  \centering
  \includegraphics[trim=0.0in 0.0in 0.0in 0.27in,clip,width=0.85\textwidth]{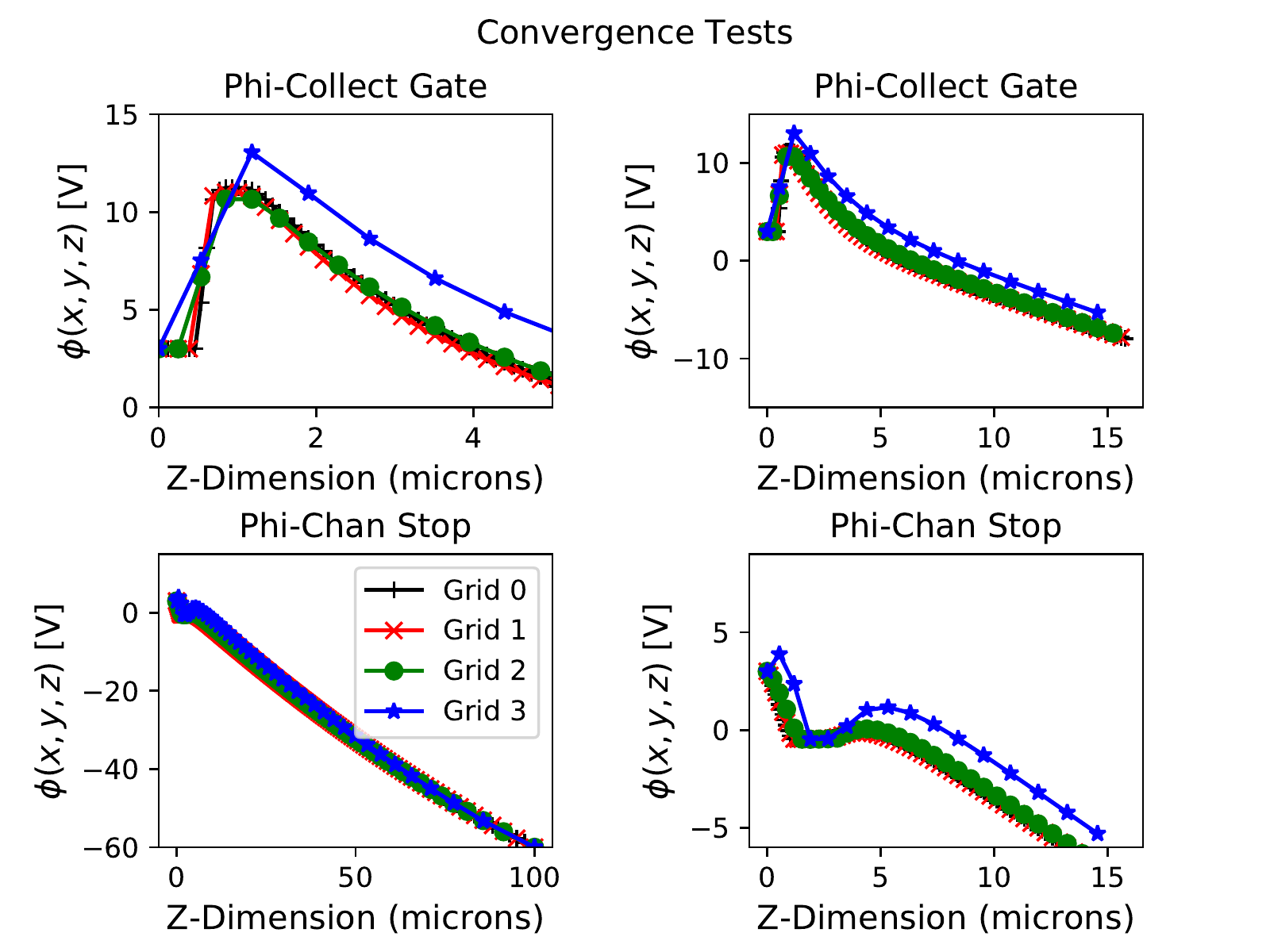}\\
  \caption{Convergence of the multi-grid subgrids.  The parameter ``ncycle'' has a value of 64 in these plots.  Each subgrid is a factor of two coarser than the preceding grid.}
  \label{Convergence_1}
\end{figure}

\begin {figure}[H]
  \centering

\includegraphics[trim = 0.0in 0.0in 0.0in 0.0in, clip, width=0.85\textwidth]{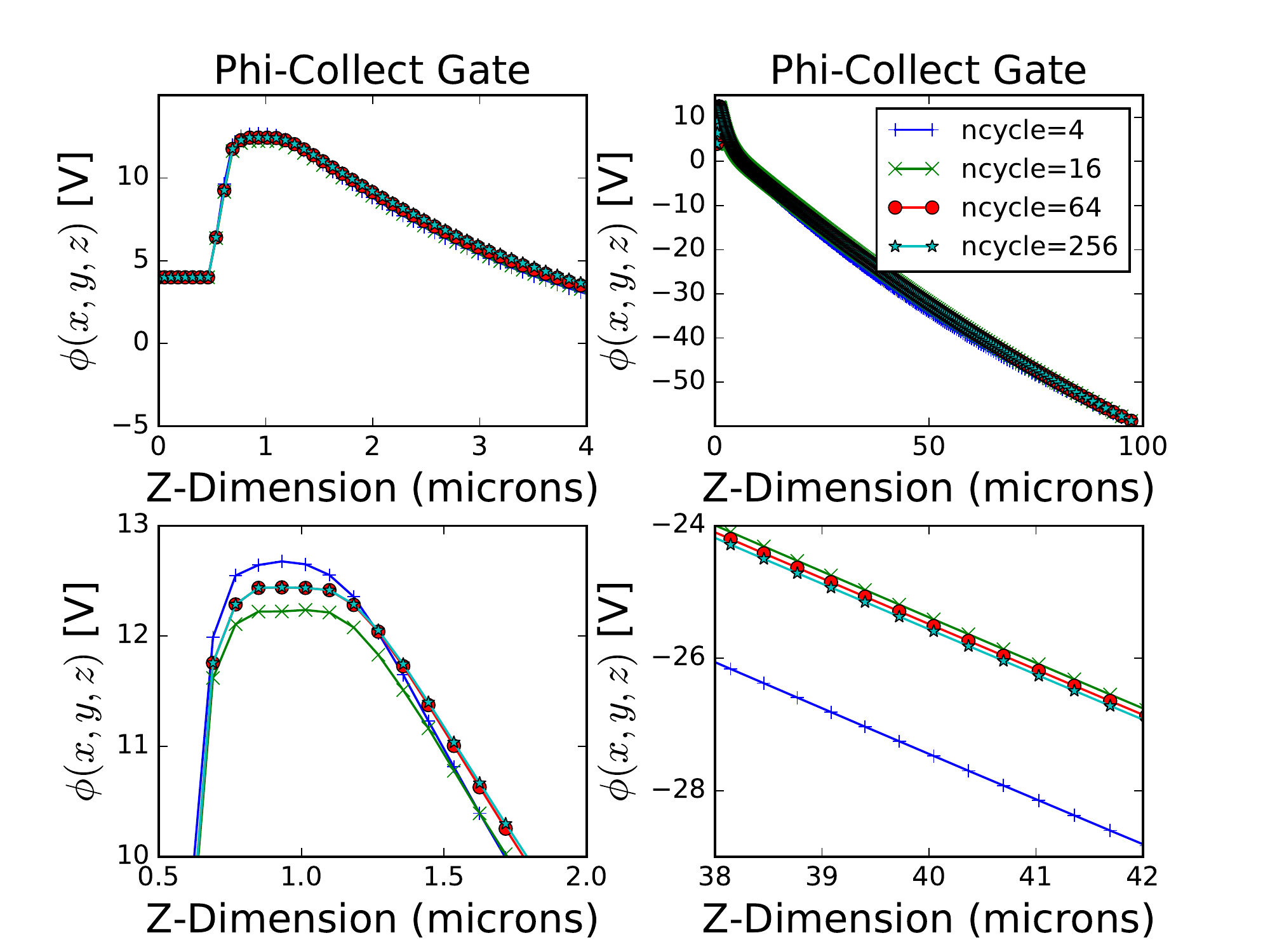}
  \caption{Convergence of the highest resolution grid as a function of the parameter ``ncycle''.  For most uses a value of 64 is adequate.}
  \label{Convergence_2}
\end{figure}

At this point it is appropriate to discuss the problem of convergence.  Convergence of the SOR algorithm is notoriously slow.  Multi-grid methods help a great deal, however, care must still be taken to ensure that the solution has converged adequately for your problem.  The parameter ``ncycle'' controls the number of iterations taken at the finest grid.  Each coarser grid increases the number of iterations by a factor of 4.  So for example, a typical problem like one of the ``pixel'' examples, which has ncycle=128, the coarsest grid has 1/8 the resolution, and will run $\rm 128 \times 4^3 = 8192$ SOR cycles at the coarsest grid.  Figures \ref{Convergence_1} and \ref{Convergence_2} show the convergence of a typical problem.  For most problems, a value of ncycle=64 is adequate.  The problems most dependent on accurate convergence have proven to be the pixel distortion simulations like those in Section \ref{Covariances}.  Since we are dealing with very small deviations in the pixel shapes due to the BF effect, it is important to make sure the results have converged. 

\subsection{Modeling carrier transport}

\begin {figure}[H]
\begin{minipage}{0.40\textwidth}
 \includegraphics[trim=2.0in 2.0in 5.2in 1.0in,clip,width=0.60\textwidth]{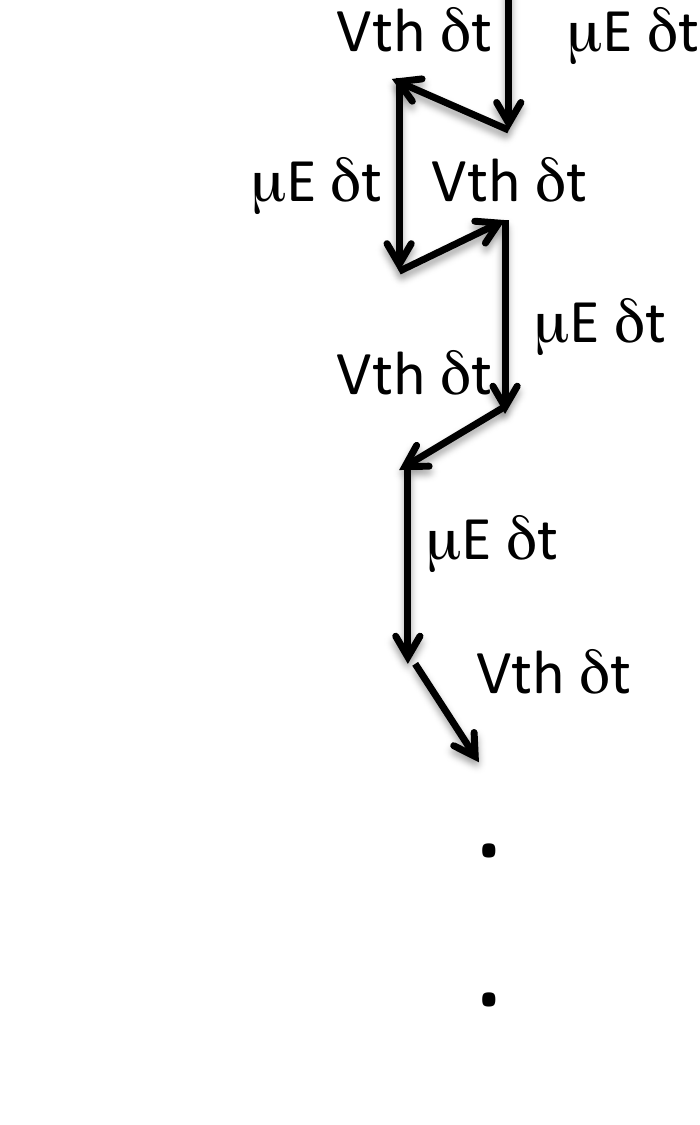}
 \end{minipage}
\begin{minipage}{0.58\textwidth}
\begin{tabular}{l} 
    Mobility: $\mu (E, T)$ calculated from Jacobini model \cite{jacobini1977}\\
    $\rm \; \; \; \; \mu \approx 1500 \frac{cm^2}{V-sec}$ at $\rm E = 6000 \frac{V}{cm}$\\
    Collision time:\\
    $\rm \; \; \; \; \tau = \frac{m_e^*}{q_e} \mu$\\
    $\rm \; \; \; \; \tau$ typically about 0.9 ps.\\
    $\rm \; \; \; \; \delta t$ drawn from exponential distribution with mean of $\tau$\\
    $\rm \; \; \; \; V_{th} = DiffMultiplier \, \sqrt{\frac{8 k T}{\pi m_e^*}}$\\
    $\rm \; \; \; \; V_{drift} = \mu E$\\
    Each thermal step in a random direction in 3 dimensions.\\
    Typically about 1000 steps to propagate to the collecting well.\\
\end{tabular}
\end{minipage}
   \caption{Diffusion model}
  \label{Diffusion}
\end{figure}

\begin {figure}[H]
\begin{minipage}{0.49\textwidth}
\begin{center}
  \includegraphics[trim=0.8in 0.0in 1.2in 0.5in,clip,width=0.75\textwidth]{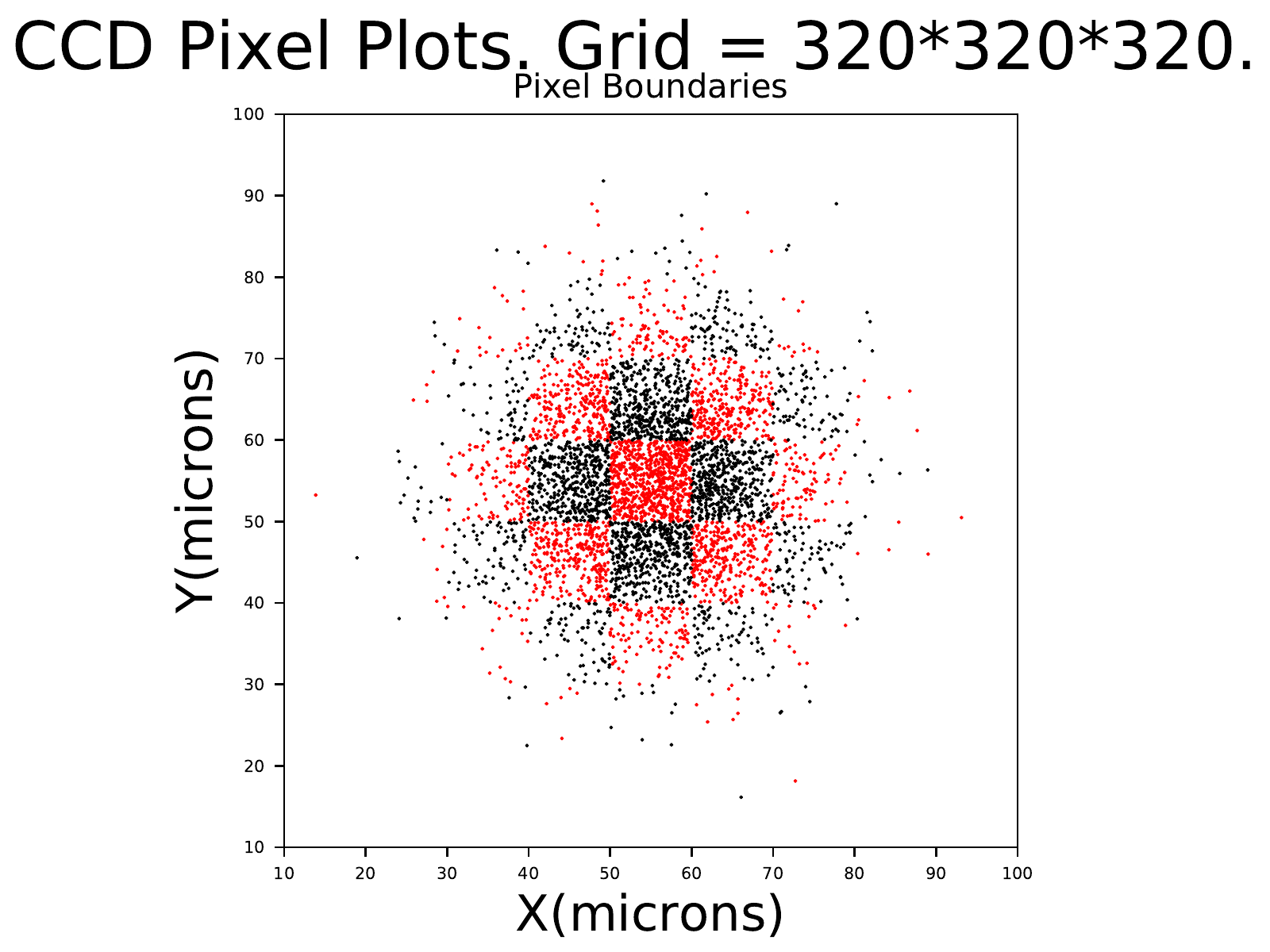}\\
  \includegraphics[trim=0.2in 0.5in 3.5in 1.0in,clip,width=0.55\textwidth]{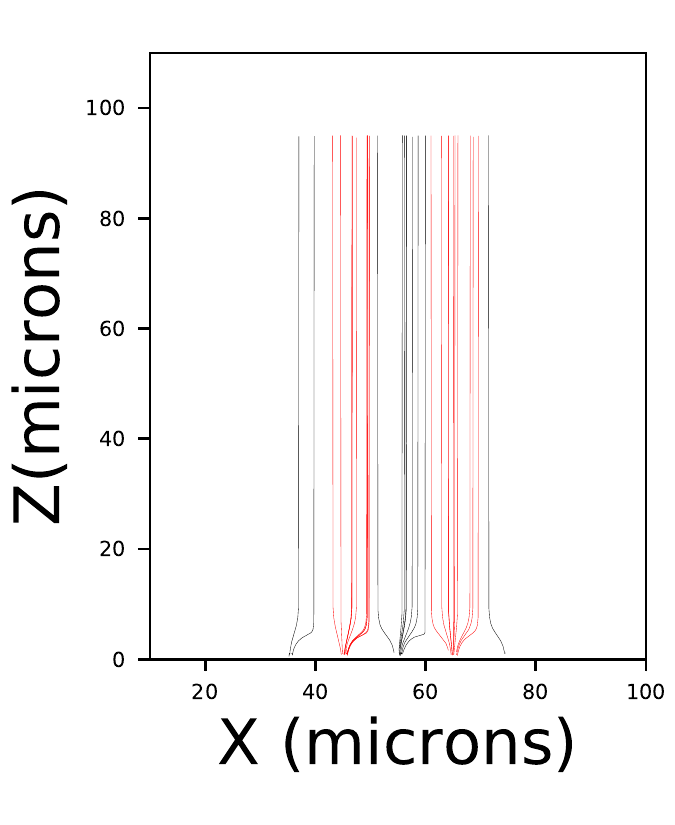}\\
  Diffusion turned off
\end{center}
\end{minipage}
\begin{minipage}{0.49\textwidth}
\begin{center}
  \includegraphics[trim=0.8in 0.0in 1.2in 0.5in,clip,width=0.75\textwidth]{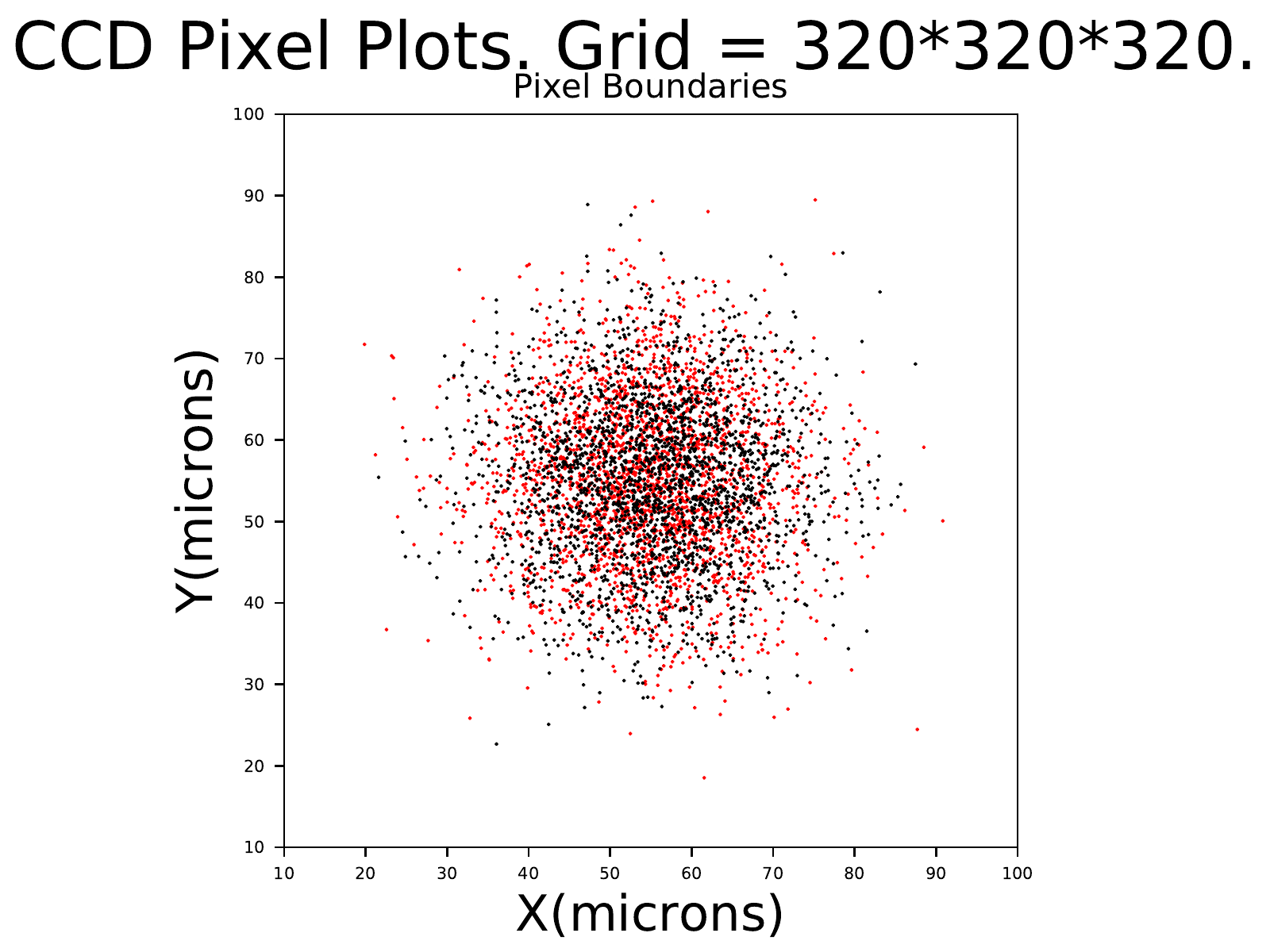}\\
  \includegraphics[trim=0.2in 0.5in 3.5in 1.0in,clip,width=0.55\textwidth]{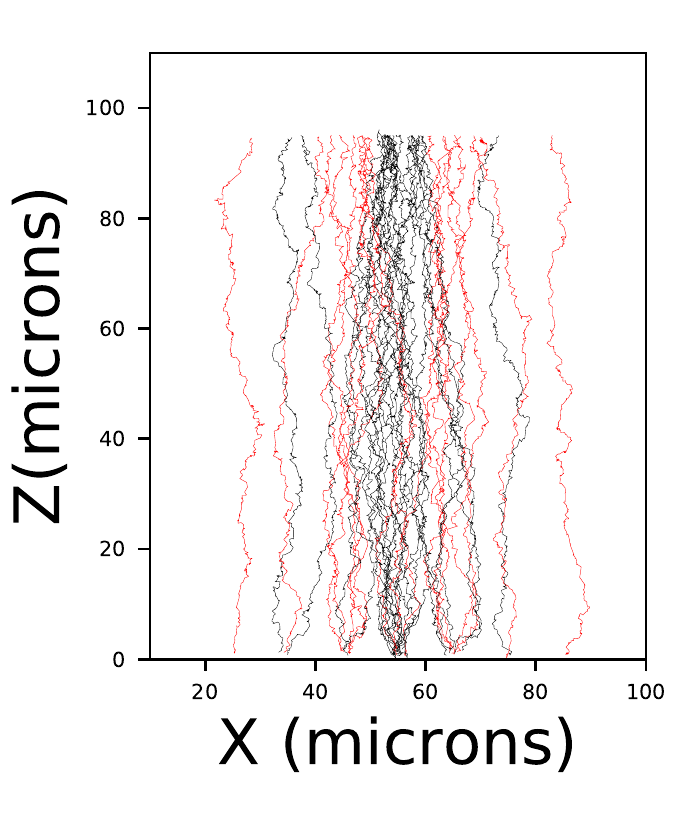}\\
  Realistic Diffusion
\end{center}
\end{minipage}
   \caption{Impact of diffusion on electron paths of a Gaussian spot with a sigma of 1 pixel. With diffusion turned off, the electrons simply propagate down and end up in the same pixel they started.  With realistic diffusion, the electrons can cross pixel boundaries.}
  \label{Diffusion_impact}
\end{figure}

The basic scheme for modeling carrier transport is shown in Figure \ref{Diffusion}.  Electrons are assumed to have lattice collisions on a time scale $\tau$, which is on the order of picoseconds.  At each collision, the electron is assumed to pick up a thermal velocity $\rm V_{th}$ which is in a random direction.  In addition to this thermal velocity, it also has a drift velocity given by $\rm V_{drift} = \mu E$, where the mobility $\rm \mu(E, T)$ is calculated as a function of electric field and temperature using the mobility model of Jacobini \cite{jacobini1977}.  These two velocities are added vectorially and it travels in this direction at this velocity for a time $\rm \delta t$ until the next collision, and this continues until the electron reaches the bottom.  The electron path is logged in the *\_Pts.dat file.  If the parameter ``LogPixelPaths'' is zero, only the initial and final positions are logged.  If this parameter is one, the entire path is logged.  The thermal velocity has a multiplier (``DiffMultiplier'') which can be used to tune the amount of diffusion.  If this is set to zero, diffusion is turned off, with the impact as shown in Figure \ref{Diffusion_impact}.  A value $\rm DiffMultiplier = 2.30$ has been found to accurately reproduce the amount of diffusion seen in $\rm Fe^{55}$ data (see Section \ref{Fe55}).  Since the value of $\rm m_e^*$ in the code is the bare electron mass, this value is equivalent to an electron effective mass of about 0.19.  This is somewhat low, as Green \cite{green_1990} finds a value of 0.27.  This value of DiffMultiplier is easily adjusted by the user, however.

The initial electron locations in X and Y can be determined in a number of ways, as determined by the ``PixelBoundaryTestType'' parameter.  These include an equally spaced grid, a Gaussian spot, a random location within a boundary, an $\rm Fe^{55}$ event, or reading in a list of locations.  The starting location in Z can either be specified, or calculated given a filter band.

\subsection{Example outputs}
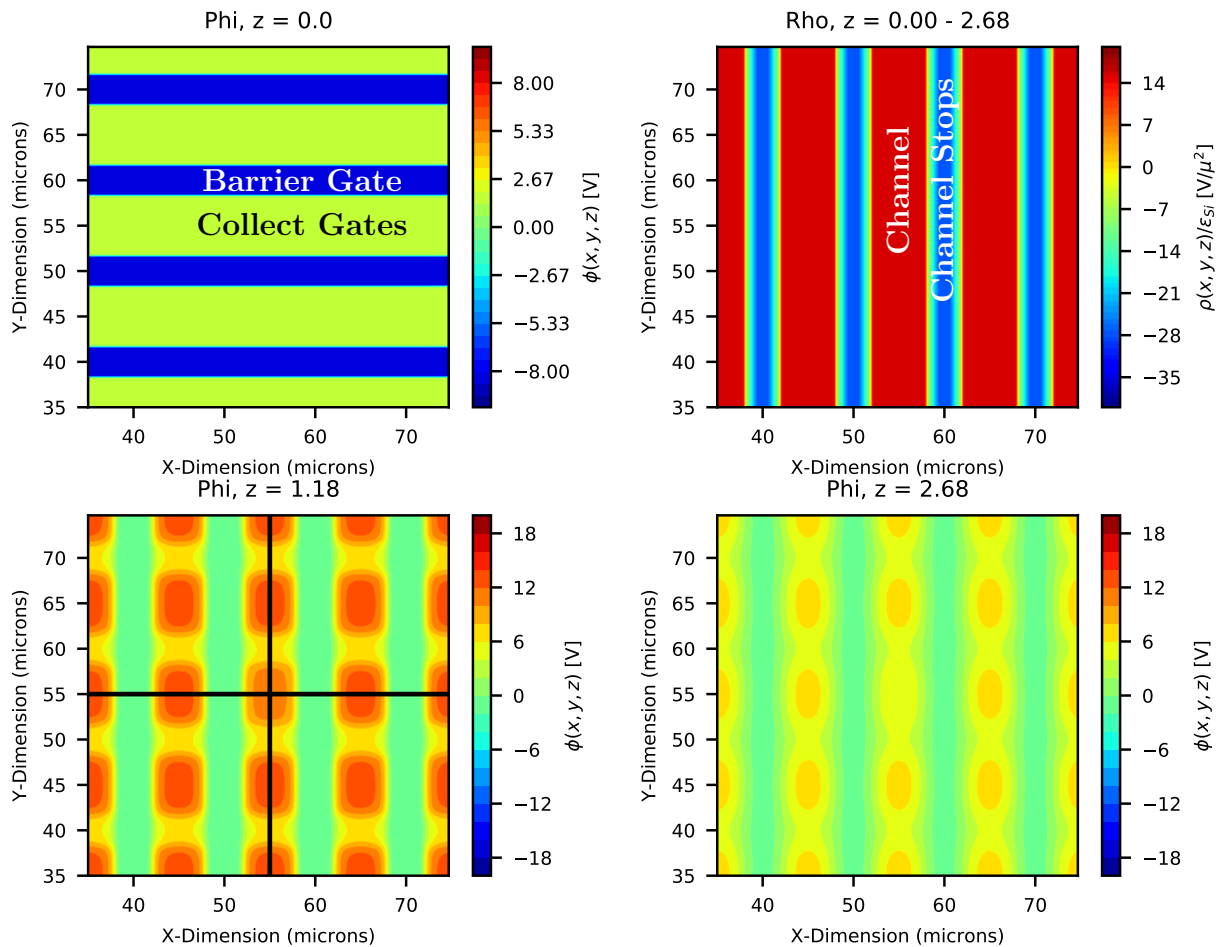
\begin{figure}[H]
  \begin{center}
    \begin{tikzpicture}
    \node[anchor=south west,inner sep=0] (image) at (0,0) {\includegraphics[trim=0.5in 0.0in 0.4in 0.36in,clip,width=0.95\textwidth]{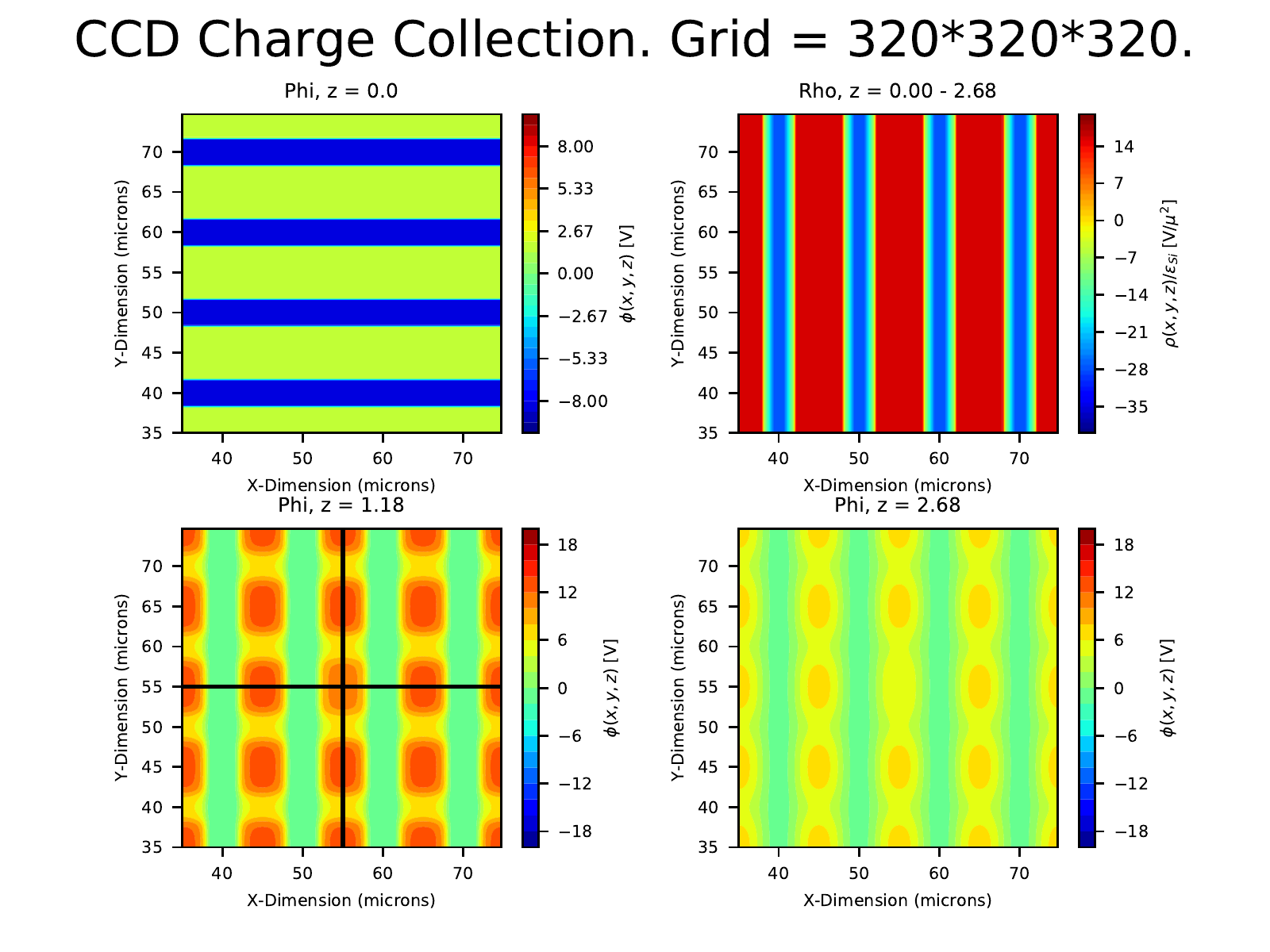}};
    \node[align=center,text=black,font={\large\bfseries}] at (4.1,10.25) {Collect Gates};
    \node[align=center,text=white,font={\large\bfseries}] at (4.1,10.84) {Barrier Gate};
    \node[align=center,text=white,font={\large\bfseries},rotate=90] at (12.65,10.7) {Channel Stops};
    \node[align=center,text=white,font={\large\bfseries}, rotate=90] at (12.02,10.7) {Channel};            
  \end{tikzpicture}
\end{center}
  \caption{A summary of the region near the bottom of the ITL STA3800C CCD.  The upper left shows the applied parallel gate voltages, and the upper right shows a 2D projection of the fixed charges.  The lower two plots show the potential at two different z-values above the bottom of the CCD.  Here the center pixel has 100,000 electrons and the surrounding pixels are empty.}
  \label{Summary_plot}
\end{figure}

\begin{figure}[H]
  \begin{center}
      \includegraphics[trim=0.2in 0.0in 0.5in 0.36in,clip,width=0.95\textwidth]{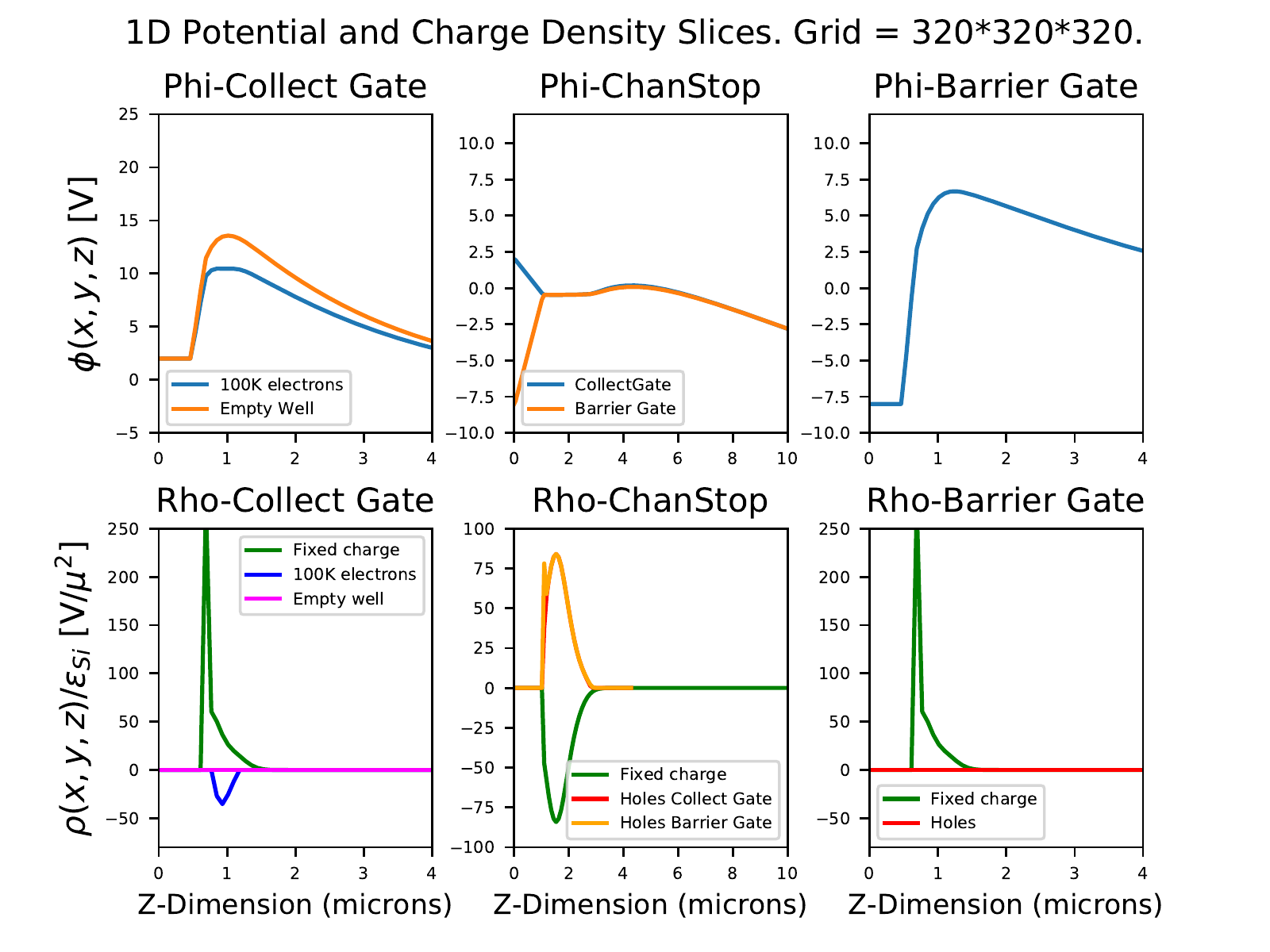}
     \end{center}
  \caption{A set of vertical 1D profiles of potential and charge density at various locations of the ITL STA3800C CCD.  Again, the center pixel has 100,000 electrons and the surrounding pixels are empty.  The sharp discontinuities near the left edge of the plots are at the Si/SiO2 interface where the device silicon begins.  This is at a z-coordinate of about 0.69 in the channel region and 1.09 in the channel stop region.}
  \label{1D_plot}
\end{figure}

\begin{figure}[H]
\begin{minipage}{0.49\textwidth}
    \begin{center}
      \includegraphics[trim=0.0in 0.0in 0.0in 0.0in,clip,width=1.10\textwidth]{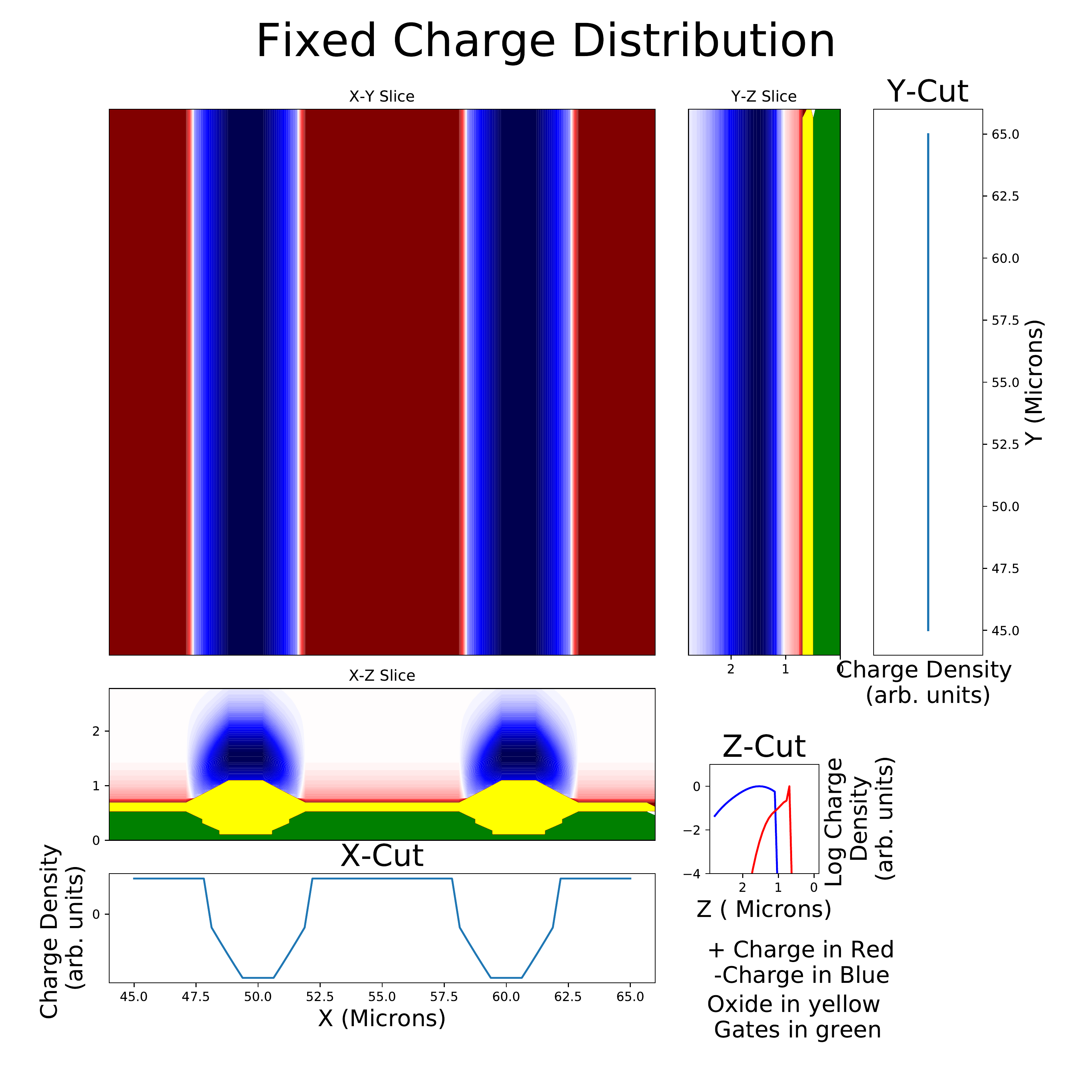}
    \end{center}
    \end{minipage}
    \begin{minipage}{0.49\textwidth}
    \begin{center}
      \includegraphics[trim=0.0in 0.0in 0.0in 0.0in,clip,width=1.10\textwidth]{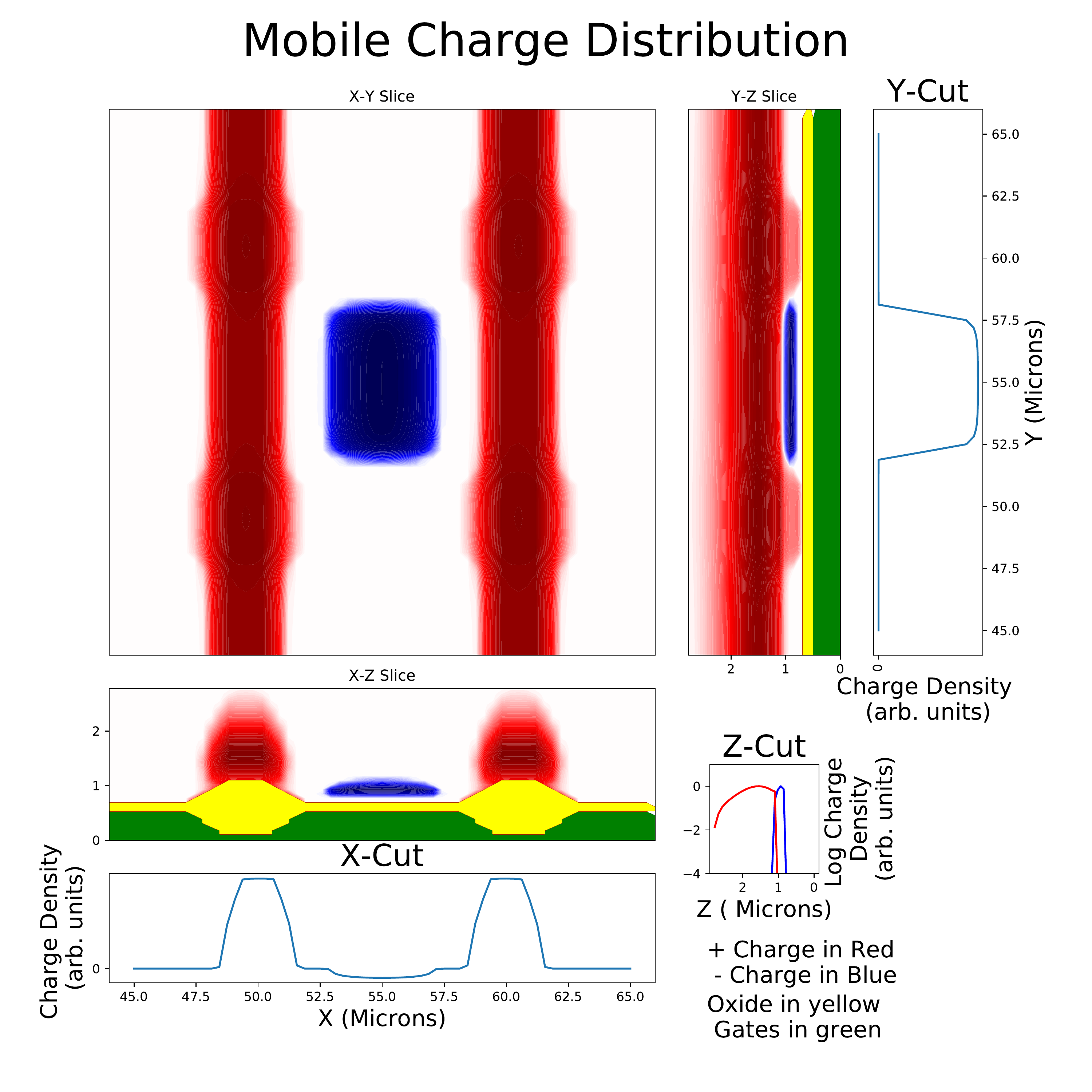}
    \end{center}
    \end{minipage}
  \caption{A set of 2D projections of the distribution of fixed and mobile charges near the bottom of the ITL STA3800C CCD.  Again, the center pixel has 100,000 electrons and the surrounding pixels are empty.  The Si/SiO2 interface, where the device silicon begins, is at a z-coordinate of about 0.69 in the channel region and 1.09 in the channel stop region.}

  \label{Charge_plot}
\end{figure}

Once the simulation has run, the potential, electric fields, and charge carrier densities are available throughout the simulation volume.  What one chooses to visualize depends on the problem being studied.  Here we have chosen three examples (Figures \ref{Summary_plot}, \ref{1D_plot}, and \ref{Charge_plot}) of the type of data which is available.

\section{Validation with measured data}
\label{Validation}
To successfully model a CCD, an accurate physical characterization of the CCD is needed.  Details such as dopant densities, layer thicknesses, and physical dimensions need to be known, at least approximately.  For modeling the CCDs from the two vendors which will be used in the LSST camera, we obtained detailed physical measurements, which have been described in \cite{CCD-Physical-Analysis}.  This has helped to make the simulations described in the next few sections as physically realistic as possible.  Most of the plots in this section can be reproduced using the examples at the {\carlito Poisson\_CCD} code site at \cite{Poisson-CCD-code}.

Regarding the parameters used in the simulations, there has been optimization of the numeric parameters, such as grid size, number of iterations, successive over-relation factor, etc. to ensure convergence and accuracy.  However, the physical parameters, such as physical dimensions, voltages, doping levels, oxide thicknesses, etc.  have all been directly measured.  In particular, these parameters have not been tuned to improve the fit with measured data shown in the following examples.

\subsection{Pixel distortions and pixel-pixel covariances}
\label{Covariances}
As has been extensively discussed in the literature (\cite{antilogus2014}, \cite{gruen2015}, \cite{guyonnet2015}, \cite{Lage_2017}), as charge builds up in the central region of bright objects, the stored charge repels additional incoming charge and broadens the profile of these objects.  The impact of the stored charge on the pixel shapes can be determined by measuring the pixel-pixel covariances on a large number of flat images (\cite{antilogus2014}, \cite{Coulton_2018}).  These covariances are calculated from a large number of flat pairs of varying intensity (see \cite{Coulton_2018} for example) as:

\begin{equation}
C_{i,j} = \frac{\sum_{I,J} (f_{I,J} - \bar{f}) (f_{I+i,J+j} - \bar{f})}{\bar{f}^2(N_{pix} - 1)}
\end{equation}

where $\rm f_{i,j}$ is the difference in flux between the two flats at pixel i,j, and $\rm N_{pix}$ is the number of pixels summed over.  

\begin {figure}[H]
    \centering
	\subfloat[b][\normalsize{Charge packet with 100,000 electrons}]{\includegraphics[trim=0.0in 0.0in 0.0in 0.0in,clip,width=0.40\textwidth]{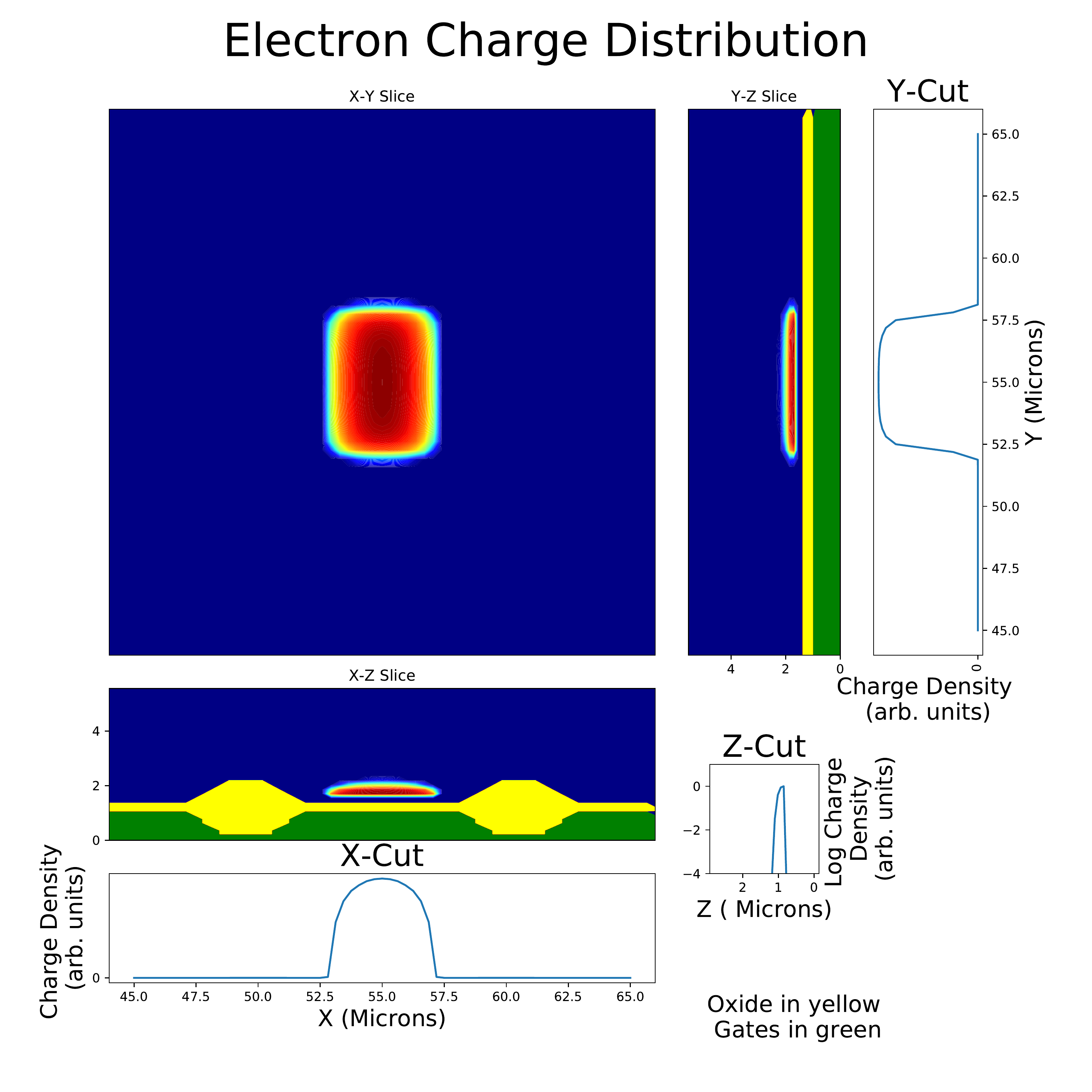}}    
	\subfloat[b][\normalsize{Pixel distortion from the central charge packet}]{\includegraphics[trim=1.0in 0.0in 1.0in 0.0in,clip,width=0.50\textwidth]{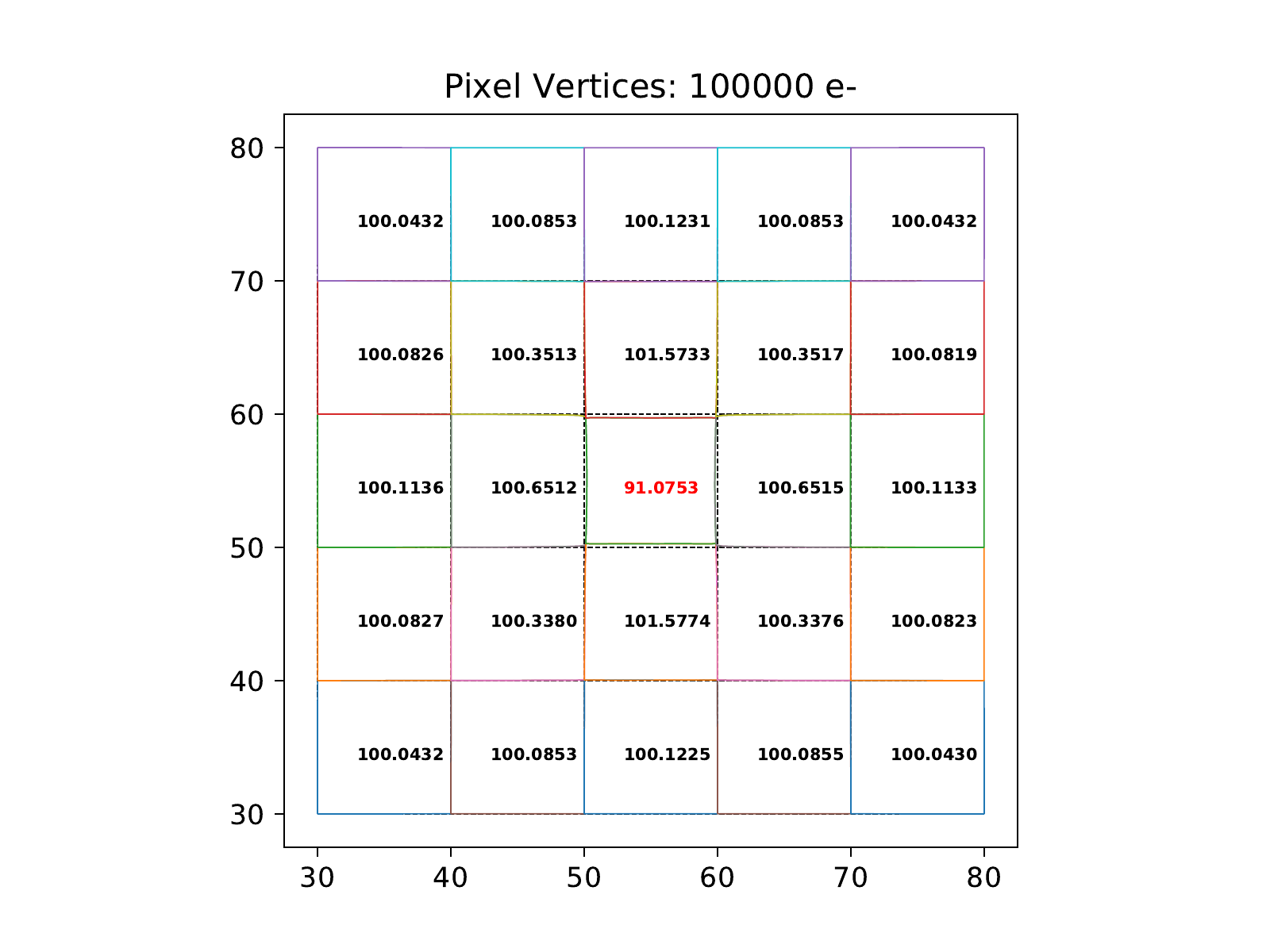}}
  \caption{Simulation of pixel distortions in an ITL chip when the central pixel contains 100,000 electrons and the surrounding pixels are empty.  X and Y are the lateral dimensions of the CCD, and Z is the thickness dimension.  The CCDs are 100 microns thick.  These distortions are obtained by solving Poisson's equation for the potentials in the CCD, then tracking electrons down and using a binary search to determine the pixel boundaries as described in the text. As expected, the central pixel loses area and the surrounding pixels all gain area. Note that the loss in area of the central pixel is greater than the sum of the area gains of the surrounding pixels because there are more distant pixels which are not plotted here and which also gain area.}
  \label{Pixel_Distortions}
  \end{figure}

\begin {figure}[H]
	\centering
	\subfloat[b][ITL Detector]{\includegraphics[trim=0.5in 0.0in 1.0in 0.0in,clip,width=0.88\textwidth]{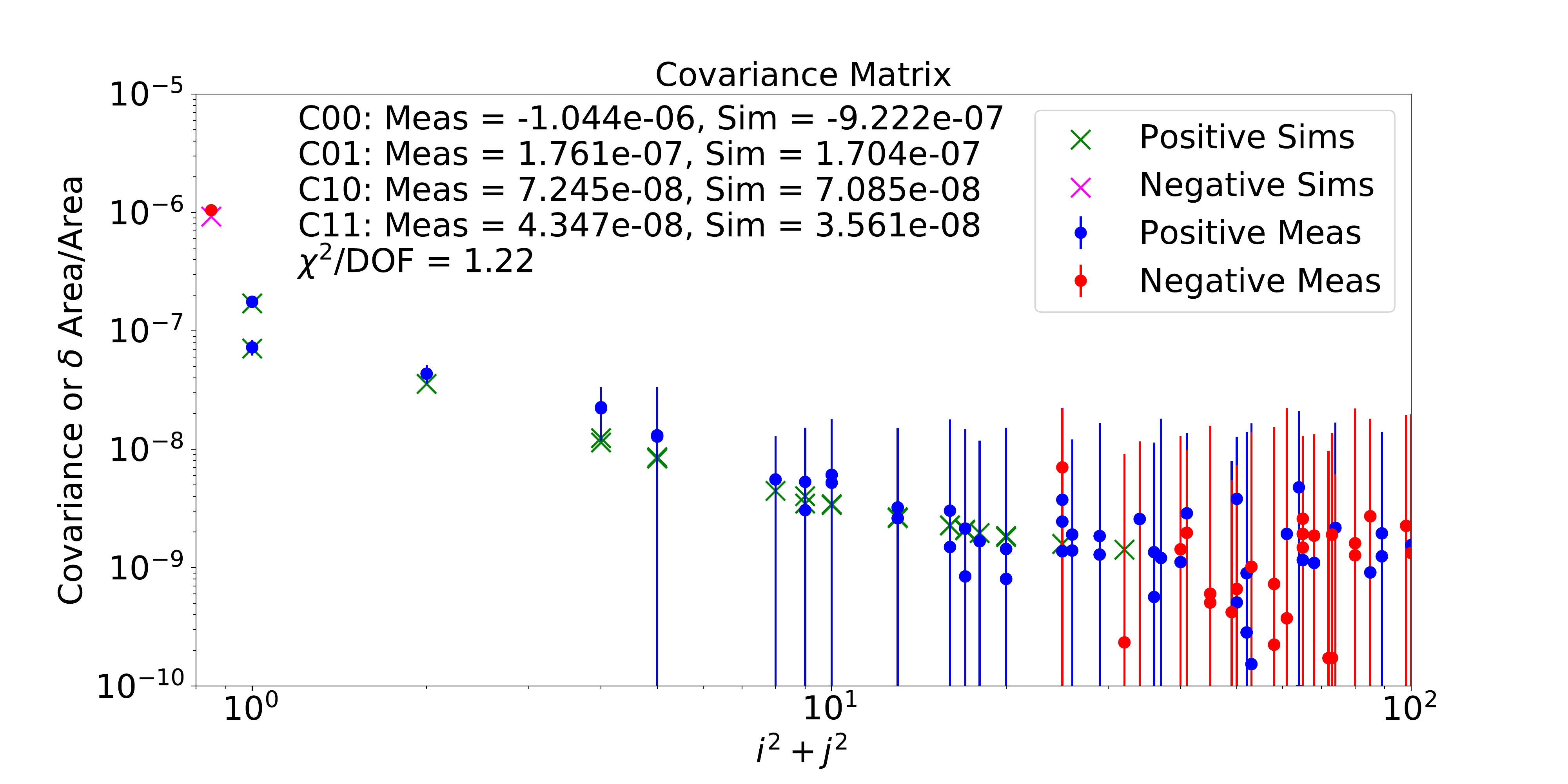}}\\
	\subfloat[b][E2V Detector]{\includegraphics[trim = 0.5in 0.0in 1.0in 0.0in, clip, width=0.88\textwidth]{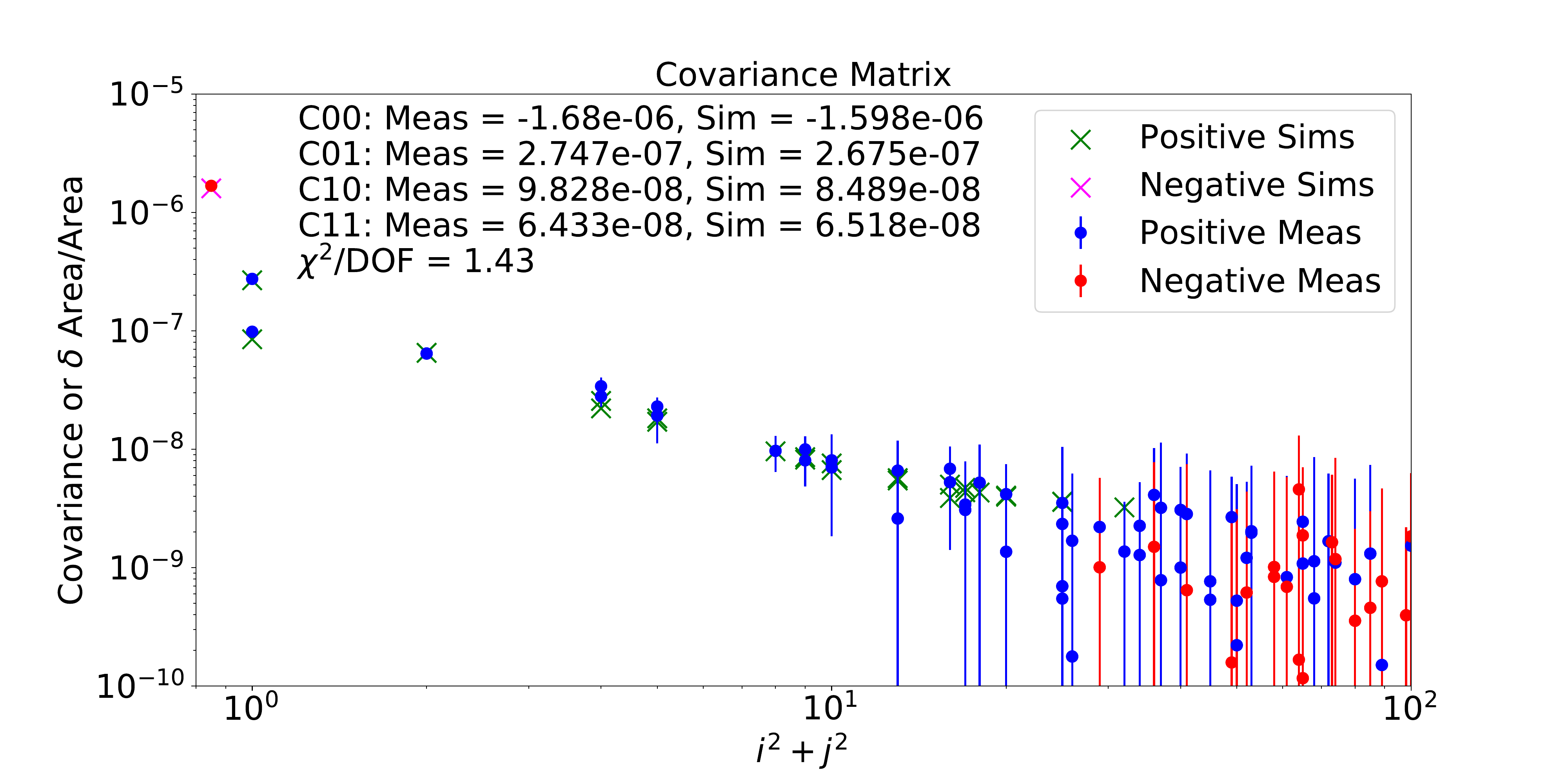}}
  \caption{Covariance measurements and simulations.  The simulated pixel area distortions (see Figure \ref{Pixel_Distortions}) accurately determine the measured pixel-pixel covariances as measured on flat pairs.  The circles are the measured covariances, as extracted by the code in the LSST image reduction pipeline as described in the text.  The crosses are the fractional area distortions as simulated by the {\carlito Poisson\_CCD} code and shown in  Figure \ref{Pixel_Distortions}.  The leftmost point (the central pixel) has been shifted to an X-axis value of 0.8 to allow plotting it on this log-log plot.  Both the E2V and ITL simulations have been informed by physical analysis of both chips, including SIMS dopant profiling and measurements of physical dimensions \cite{CCD-Physical-Analysis}.  Both the covariance measurements and the simulations have been normalized to the distortion caused by one electron.  The asymmetry of the nearest neighbor pixels is correctly modeled, and the simulated values agree with the measurements within the statistical errors.}
  \label{Correlations_Sims}
  \end{figure}

We have generated this data on a large number of flat pairs on LSST CCDs, measured on the UC Davis LSST beam simulator (\cite{tyson2014}, \cite{Lage_2017}), and would like to compare these results to simulations.  For a dataset of 100 flat pairs, each with 16 million pixels with an average signal level of 50,000 electrons, this is approximately $\rm 10^{14}$ electrons.  Directly simulating this data is out of the question.  However, we have found a simple way to run a single simulation which reproduces the measured pixel covariances.  In order to do this, we first simulate a situation where one pixel has a fixed amount of charge (typically 100,000 electrons), and all surrounding pixels are empty.  After solving for the potential and resulting electric field, we can track electrons down through the silicon.  As the electrons travel down through the silicon under the influence of the electric field, they eventually end up in one of the collecting wells.  A binary search is used to find the bifurcation points where electrons on one side of the bifurcation point end up in one pixel, and electrons on the other side of the bifurcation point end up in an adjacent pixel.   This binary search is performed with diffusion turned off in order not to introduce a stochastic element into the electron paths.  These bifurcation points are identifed as the pixel boundaries.  This allows us to characterize the distortion in the pixel boundaries which results from the central pixel charge.  A typical result of this process is shown in Figure \ref{Pixel_Distortions}.  The number of vertices used to define the distorted pixel shapes is an adjustable parameter.  A minimum of 12 vertices is needed to adequately define the distortion (4 corners plus 2 points per edge).  The distorted pixels shown in Figure \ref{Pixel_Distortions} and used in Figure \ref{Correlations_Sims} were simulated with a total of 132 vertices (4 corners plus 32 points per edge).  The distorted pixel shapes which result are what gives rise to the BF effect, with the central pixel losing area, which is gained by the surrounding pixels.  

We find that the area distortions which result accurately capture the measured pixel-pixel covariances.  Figure \ref{Correlations_Sims} shows the agreement between the measured pixel-pixel covariances on flat field images and the simulated area distortions, as measured and as simulated on LSST CCDs from both CCD vendors.   The agreement is quite good.  The asymmetry of the nearest neighbor pixels is correctly modeled, and the simulated values agree with the measurements within the statistical errors.  These simulations are run with the ``pixel-itl.cfg'' and ``pixel-e2v.cfg'' examples at \cite{Poisson-CCD-code}.  More details on these measurements and simulations are given in \cite{Lage_2017} and \cite{BF-Linearity}.

\subsection{Diffusion modeling and Fe55 tests}
\label{Fe55}
\begin {figure}[H]
\begin{center}
\includegraphics[trim=0.0in 0.0in 0.0in 0.4in,clip,width=0.75\textwidth]{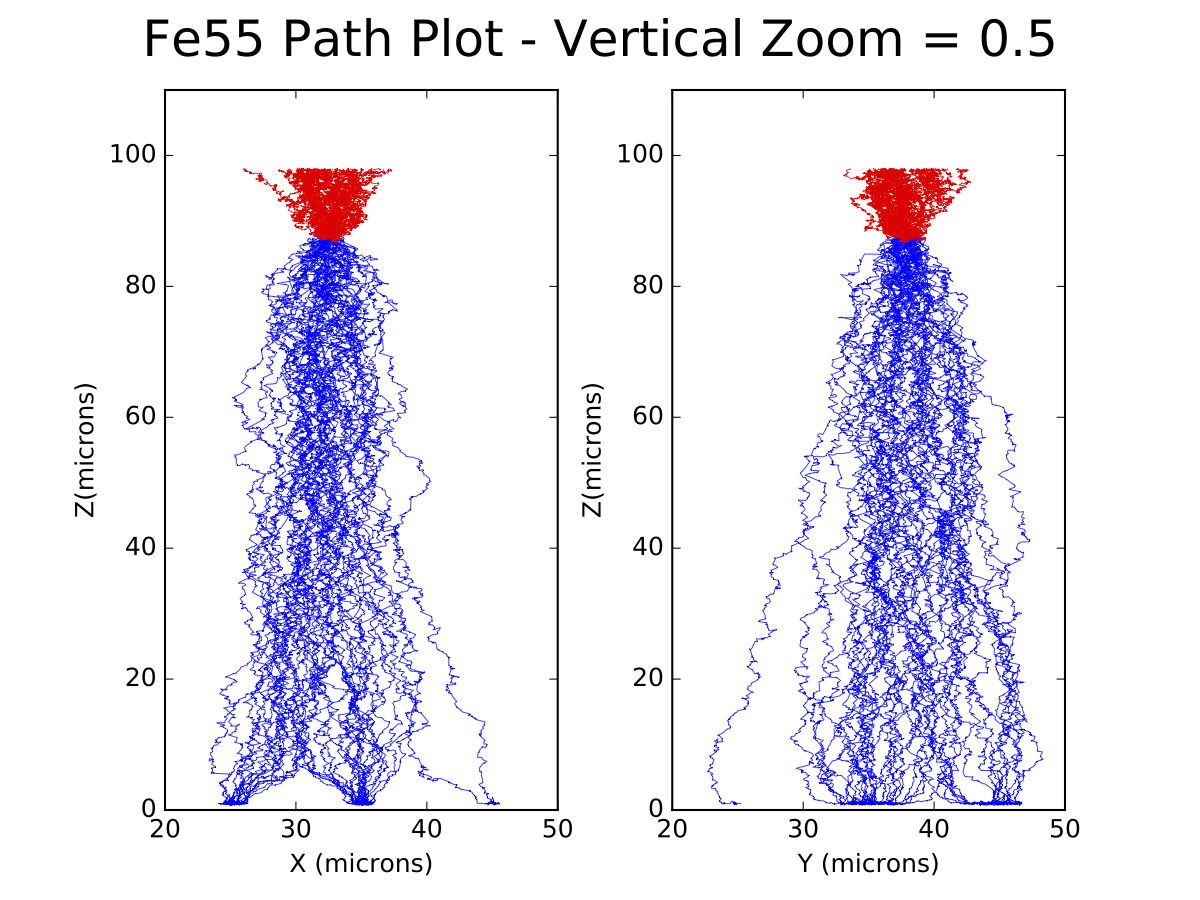}
   \caption{This shows a simulation of a typical $\rm Fe^{55}$ event in an ITL device.  Electrons are shown in blue (moving downward) and holes in red(moving upward).  The left panel is a slice in the serial direction, and the right panel is a slice in the parallel direction. Only a small fraction of the 1620 incident hole-electron pairs are plotted here. }
  \label{Fe55_event}
\end{center}
\end{figure}

\begin {figure}[H]
    \centering
    	\subfloat[b][ITL device - $\rm \chi^2/DOF = 1.4$]{\includegraphics[trim=1.0in 0.0in 1.0in 0.0in,clip,width=0.49\textwidth]{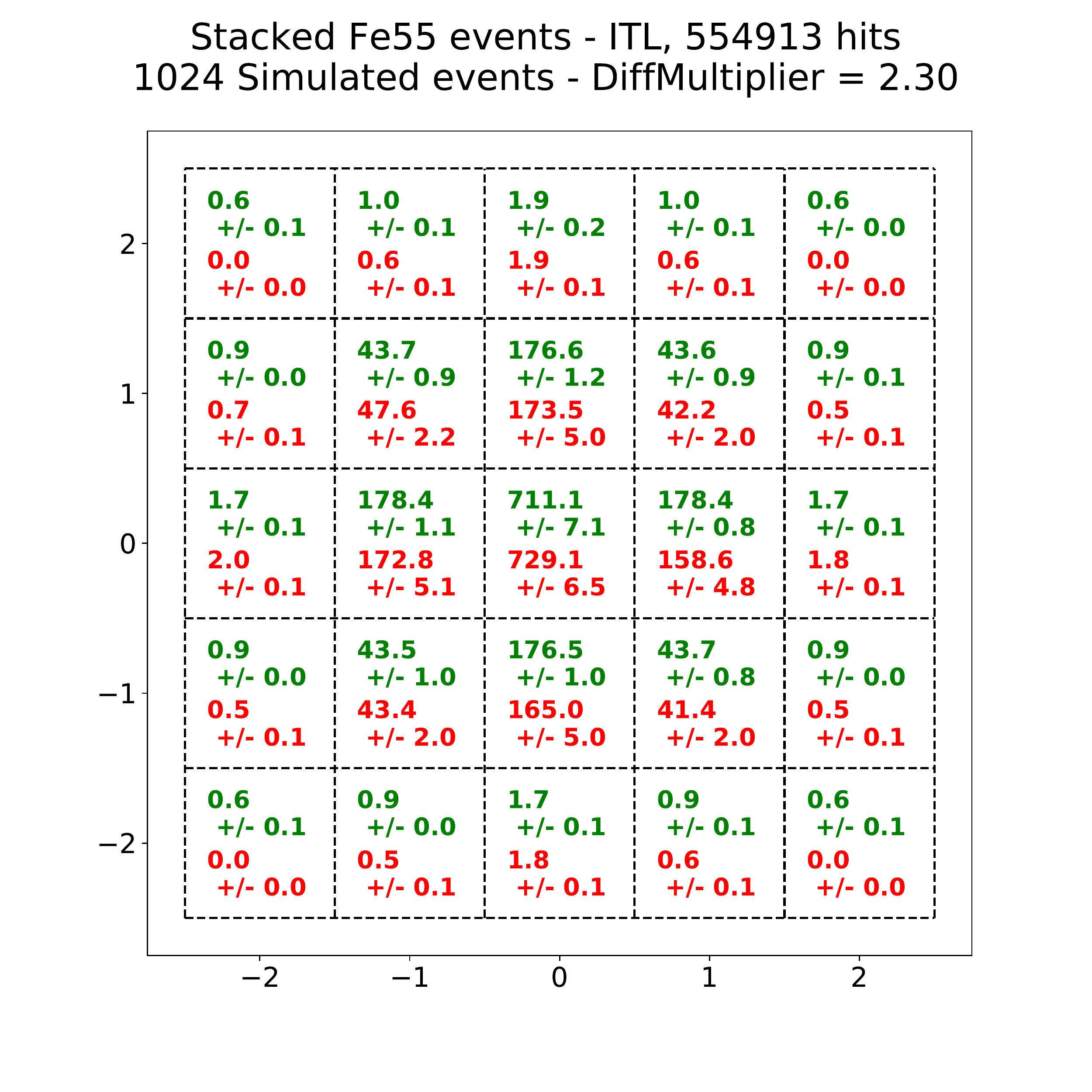}}
        \subfloat[b][E2V device - $\rm \chi^2/DOF = 2.1$]{\includegraphics[trim=1.0in 0.0in 1.0in 0.0in,clip,width=0.49\textwidth]{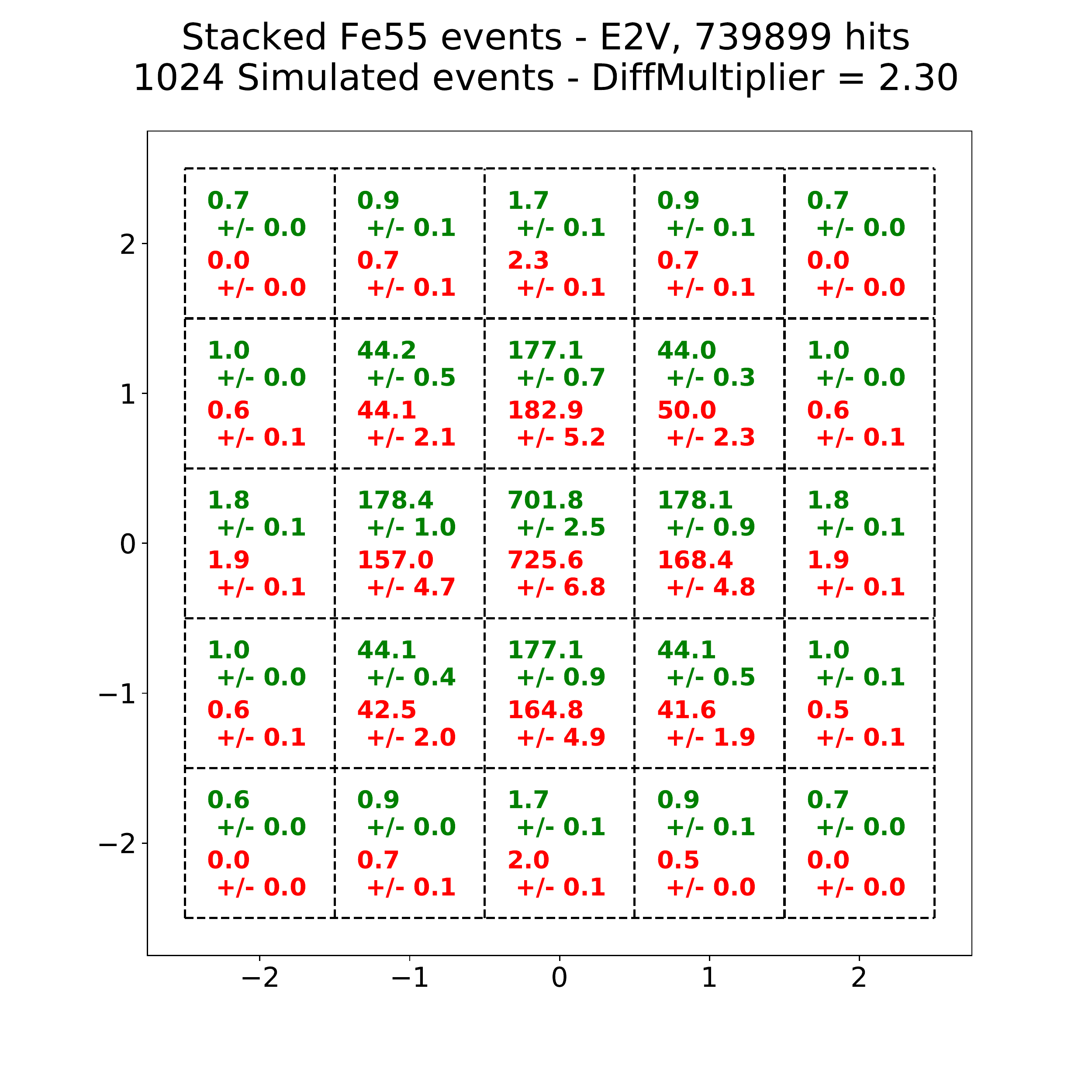}}    
  \caption{Comparison of $\rm Fe^{55}$ event stacked pixel maps between measurement and simulation for both ITL and E2V devices.  The numbers in each pixel are the average number of electrons in each pixel, when the event is centered on the center pixel(2,2).  The measurements (top, in green) are a stack of several hundred thousand events, and the errors of the measurements are one sigma values of the 16 amplifiers on one CCD.  The simulations (bottom, in red) are a stack of 1024 simulated events, and the errors are statistical.  The spread of the charge cloud due to diffusion is well modeled.}
  \label{Fe55_stacked}
\end{figure}

CCDs are routinely characterized by exposing the CCD to an $\rm Fe^{55}$ source. (see \cite{janesick2001scientific}, for example).  The radioactive decay produces X-rays with a known energy, with the $\rm K^\alpha$ peak being the strongest peak, typically producing 1620 hole-electron pairs when photoelectrically absorbed in the silicon.  These carriers then propagate down to the collecting wells, where they are collected and counted.  Because of diffusion, the carriers, which are initially produced in a small volume, spread out and occupy several pixels.  To simulate these events, a special module was written.  Normally photoelectrons propagate one at a time, without influence from neighboring carriers.  But the carriers produced in the $\rm Fe^{55}$ event are produced in a short time, so interactions between the carriers might be important.  The code takes the like carrier repulsion and opposite carrier attraction into account, and the parameters ``Fe55ElectronMult'' and ``Fe55HoleMult'' can be used to turn off or modify this interaction if desired.  Figure \ref{Fe55_event} shows a typical event, and Figure \ref{Fe55_stacked} shows stacked pixel maps compared between measurements and simulations.  The spread of the charge cloud due to diffusion is well modeled. This simulation is run with the ``fe55.cfg'' example at \cite{Poisson-CCD-code}.

\subsection{Saturation and blooming}
\begin {figure}[H]
\centering
  \subfloat[b][\normalsize{Surface full well vs bloomed full well. Reproduced from \cite{janesick2001scientific}.}]{\includegraphics[trim=0.0in 0.0in 0.0in 0.0in,clip,width=0.35\textwidth]{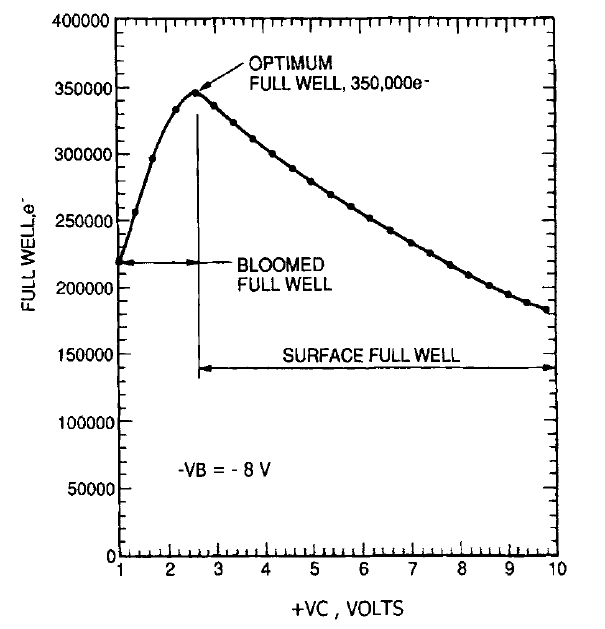}}
  \hspace{5mm}
  \subfloat[b][\normalsize{Similar measurements on the ITL STA3800C}]{\includegraphics[trim=0.0in 0.0in 0.0in 0.5in,clip,width=0.45\textwidth]{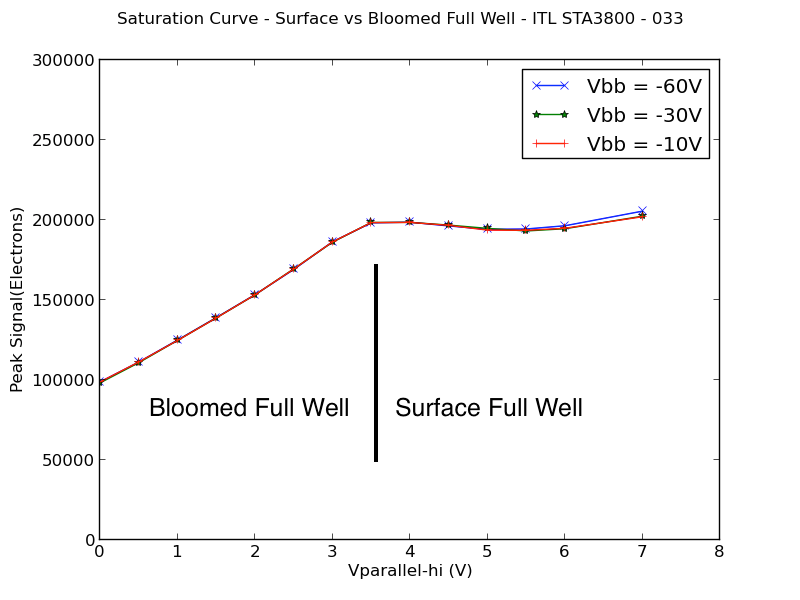}}
   \caption{As discussed in Janesick \cite{janesick2001scientific}, depending on the parallel gate high voltage, saturation can occur either at the silicon surface or above the collecting wells.  This is illustrated in Figures \ref{Sat2} and \ref{Sat3}.  The difference in shape in the surface full well condition in the two cases is not well understood at present. }
  \label{Sat1}
\end{figure}

\begin {figure}[H]
\begin{minipage}{0.69\textwidth}
      \begin{center}
  \begin{tikzpicture}        
    \node[anchor=south west,inner sep=0] (image) at (0,0) {\includegraphics[trim=0.0in 0.5in 0.0in 0.0in,clip,width=0.95\textwidth]{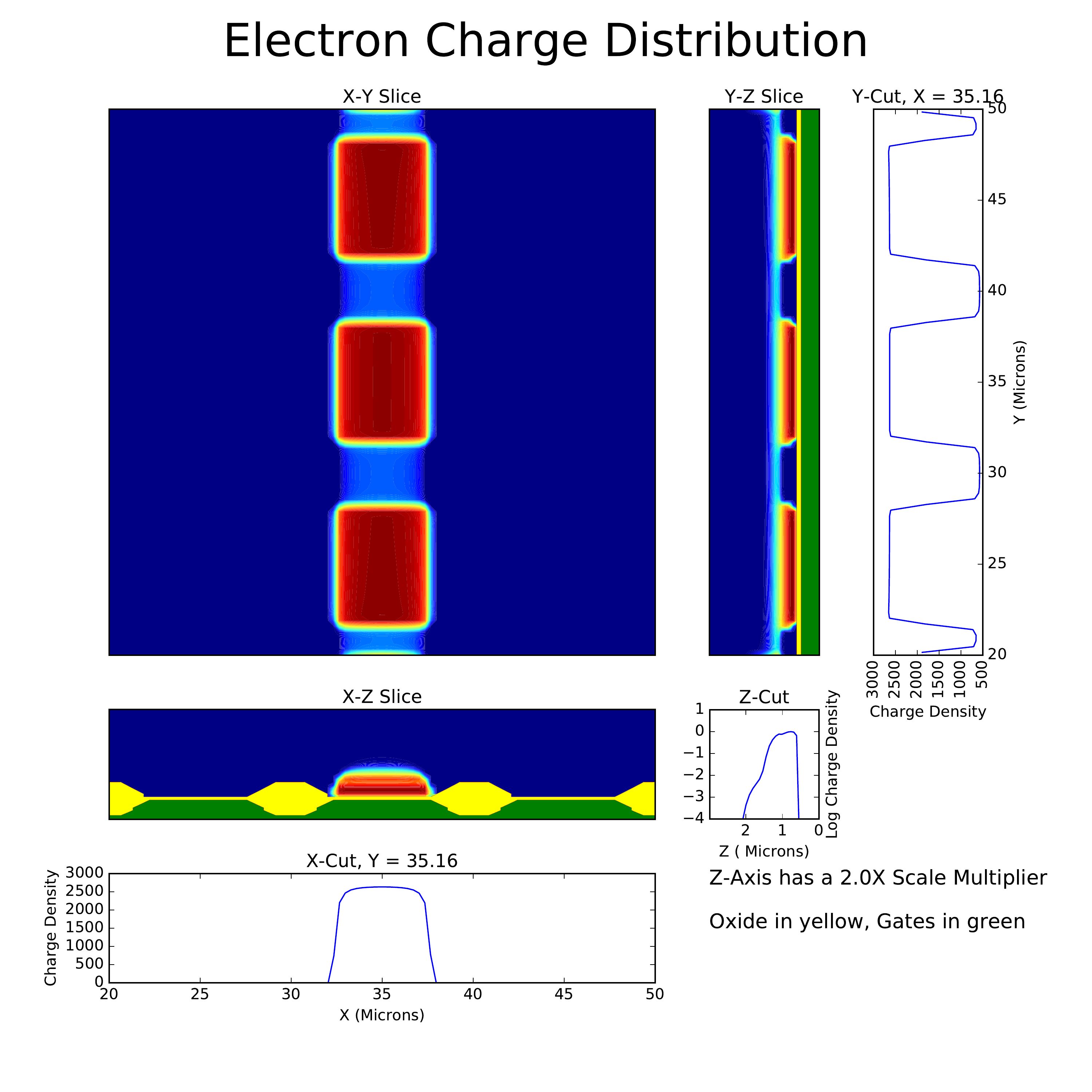}};
    \draw [line width=0.50mm, yellow] (8.1,5.9) circle (2mm);
  \end{tikzpicture}
    \end{center}
    \end{minipage}
    \begin{minipage}{0.29\textwidth}
    \begin{center}
  \begin{tikzpicture}        
    \node[anchor=south west,inner sep=0] (image) at (0,0) {\includegraphics[trim=0.2in 0.0in 5.0in 0.4in,clip,width=0.95\textwidth]{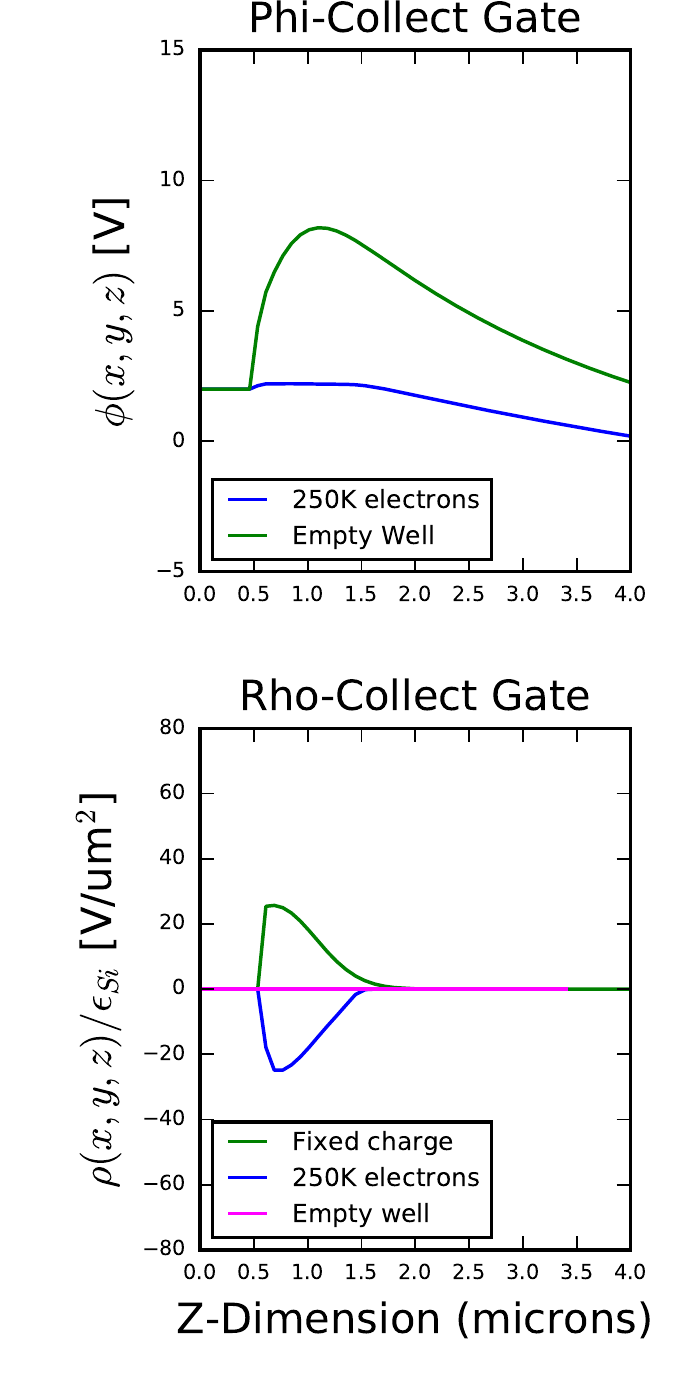}};
    \draw [line width=0.50mm, red] (1.7,6.9) circle (2mm);
  \end{tikzpicture}
    \end{center}
    \end{minipage}
   \caption{Simulation of the bloomed full well condition.  Each of the three pixels contains 250,000 electrons, and the parallel high voltage is 2.0V.  The yellow circle shows where charge is blooming above storage wells.  The red circle shows that the potential in the storage well is still above that at the gate interface, keeping charge away from the surface.}
  \label{Sat2}
\end{figure}

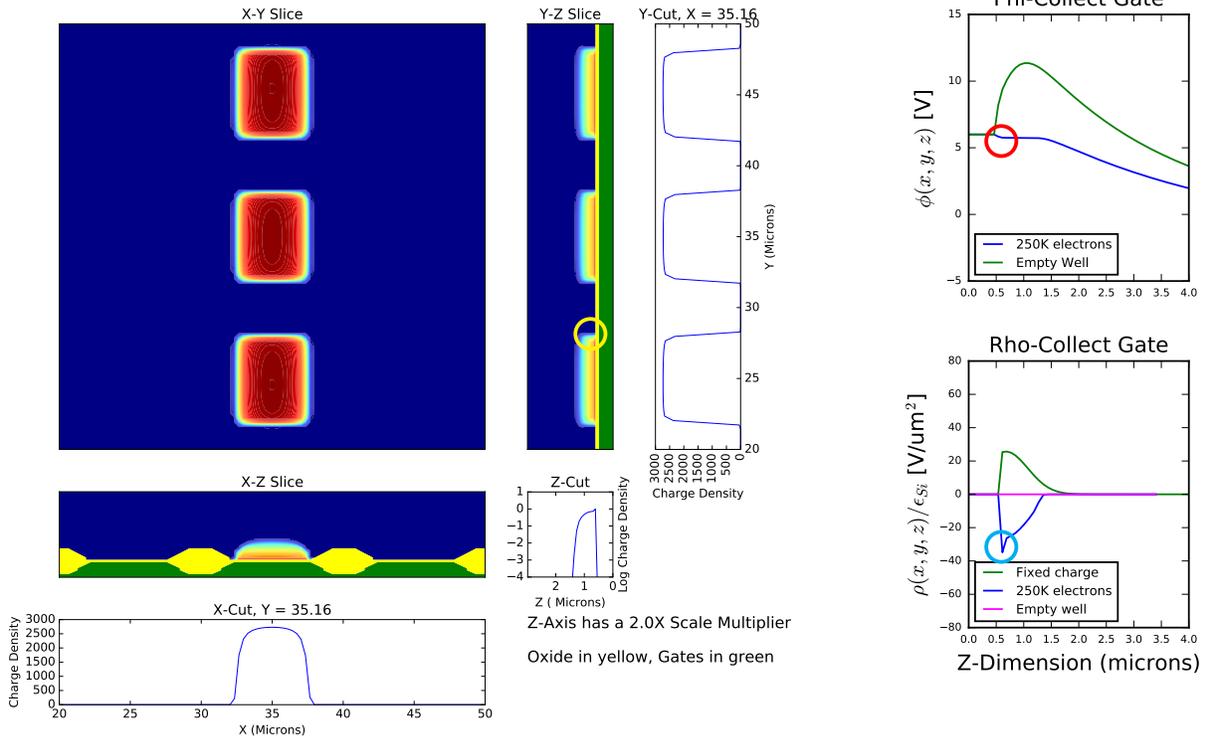
\begin {figure}[H]
\begin{minipage}{0.69\textwidth}
      \begin{center}
  \begin{tikzpicture}        
    \node[anchor=south west,inner sep=0] (image) at (0,0) {\includegraphics[trim=0.0in 0.5in 0.0in 0.0in,clip,width=0.95\textwidth]{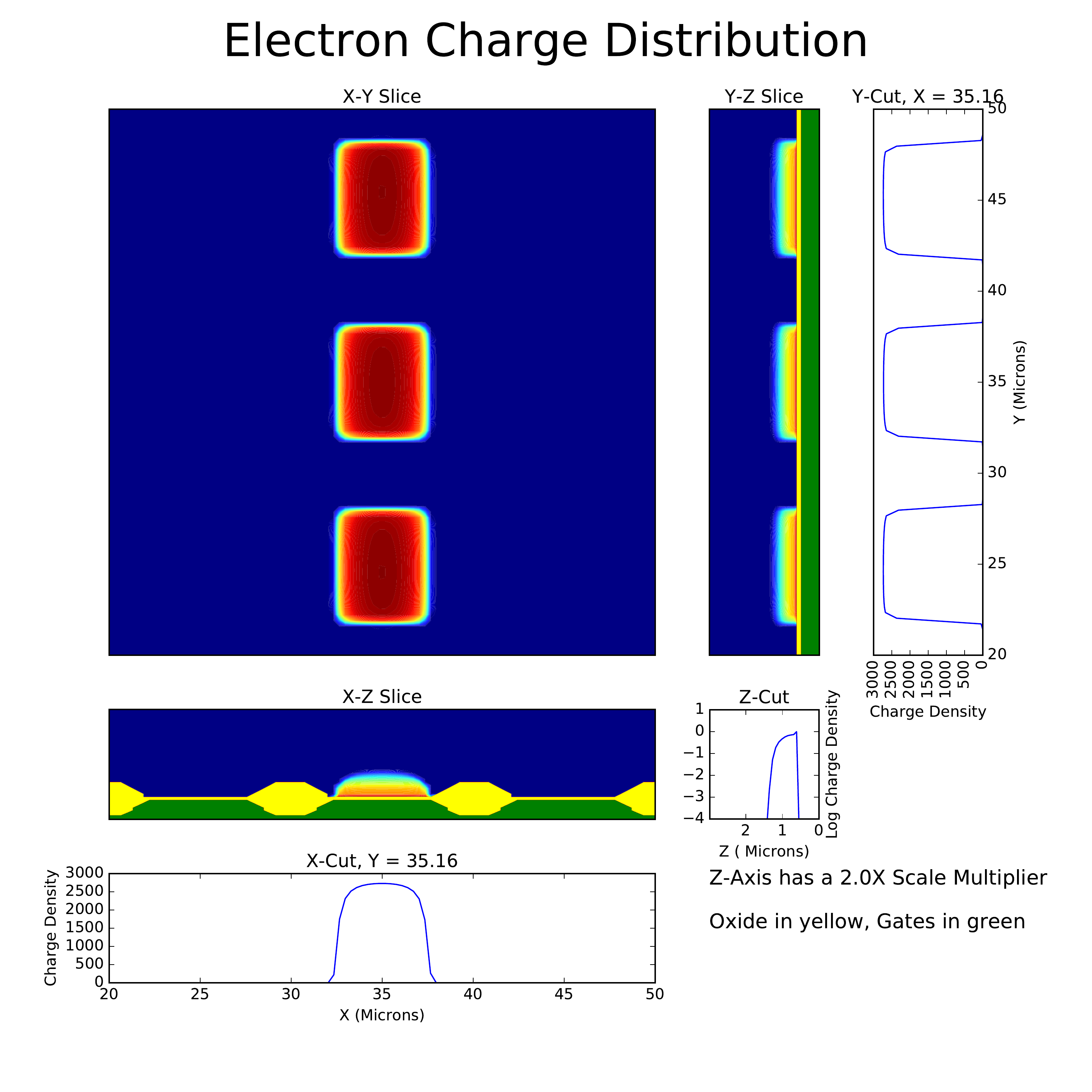}};
    \draw [line width=0.50mm, yellow] (8.2,5.6) circle (2mm);
  \end{tikzpicture}
    \end{center}
    \end{minipage}
    \begin{minipage}{0.29\textwidth}
    \begin{center}
  \begin{tikzpicture}        
    \node[anchor=south west,inner sep=0] (image) at (0,0) {\includegraphics[trim=0.2in 0.0in 5.0in 0.4in,clip,width=0.95\textwidth]{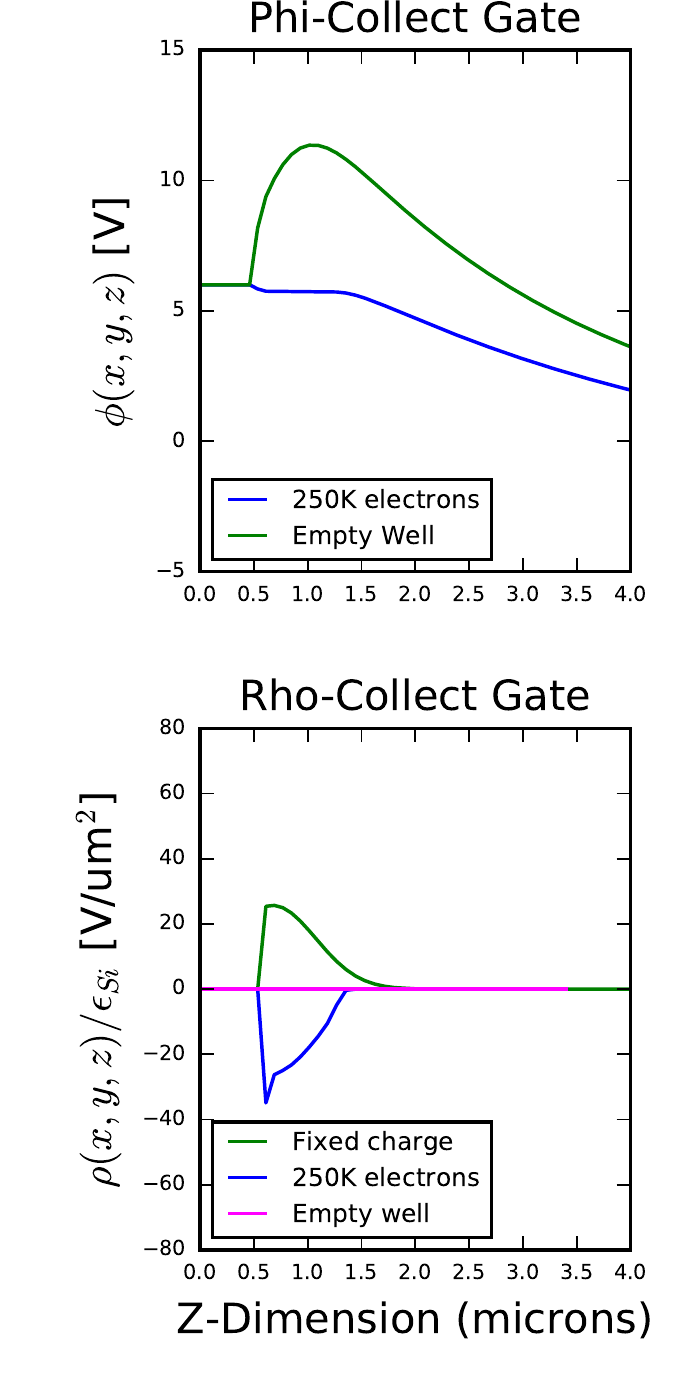}};
    \draw [line width=0.50mm, red] (1.8,7.5) circle (2mm);
    \draw [line width=0.50mm, cyan] (1.8,2.1) circle (2mm);    
  \end{tikzpicture}
    \end{center}
    \end{minipage}
   \caption{Simulation of the surface full well condition.  Each of the three pixels contains 250,000 electrons, and the parallel high voltage is 6.0V.  The yellow circle shows where charge is blooming along the silicon surface.  The red circle shows that the potential in the storage well is now below that at the gate interface, causing added charge to be added at the surface.  The cyan circle shows the surface spike of added charge.}
  \label{Sat3}
\end{figure}

\begin {figure}[H]
\centering
  \subfloat[b][Spot images at different parallel high voltages.]{\includegraphics[trim=0.0in 0.0in 0.0in 0.0in,clip,width=0.85\textwidth]{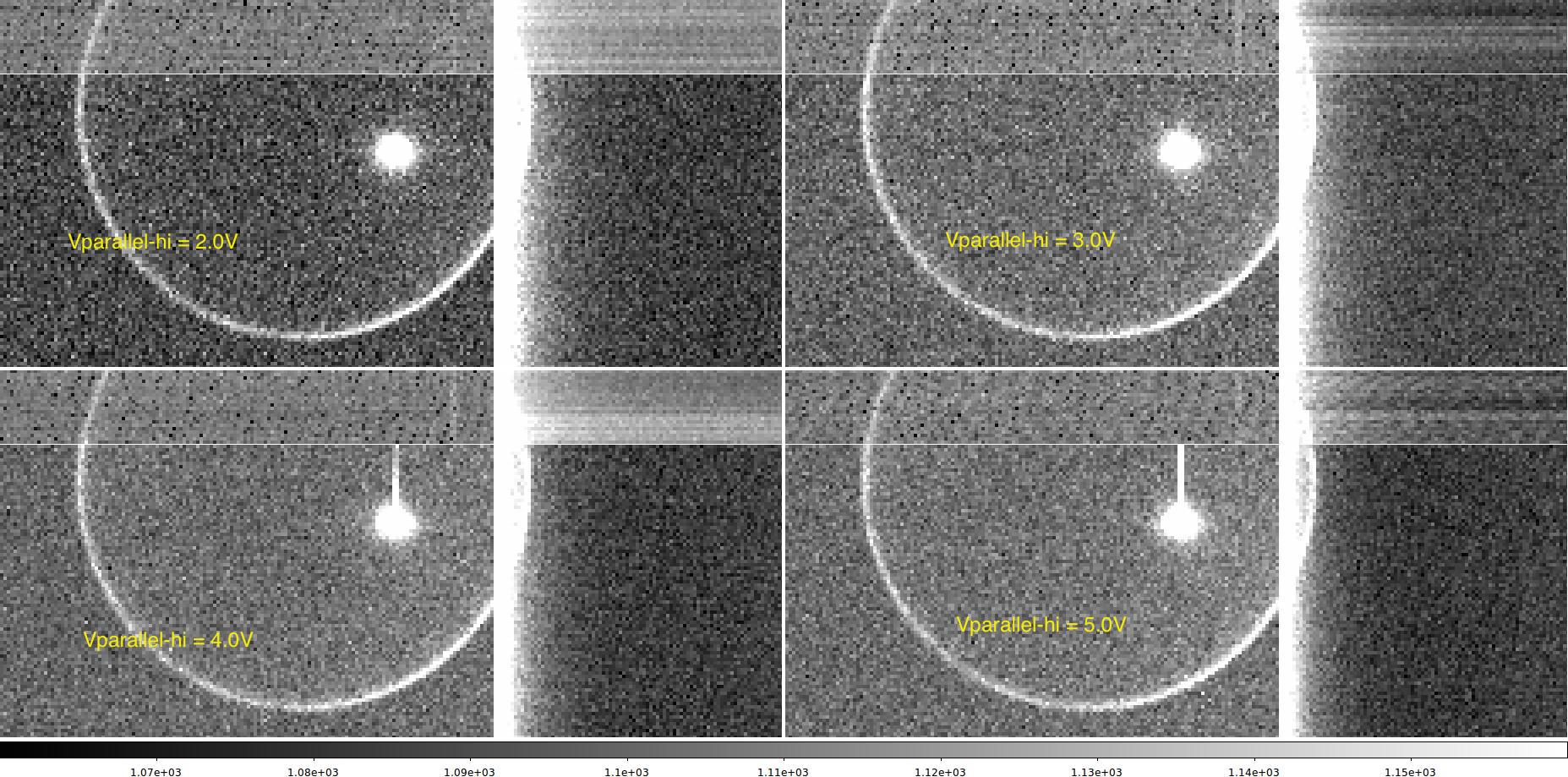}}\\
  \subfloat[b][\normalsize{Parallel high voltage of 2.5V - bloomed full well.}]{\includegraphics[trim=0.0in 0.0in 0.0in 0.0in,clip,width=0.49\textwidth]{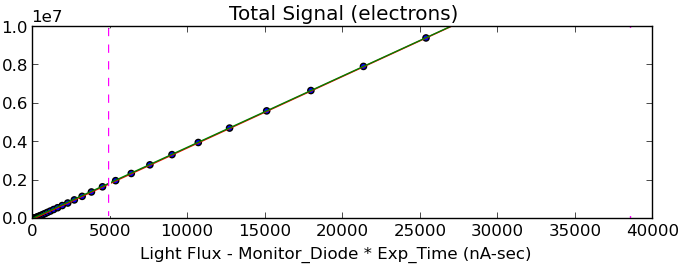}}
  \subfloat[b][\normalsize{Parallel high voltage of 6.0V - surface full well.}]{\includegraphics[trim=0.0in 0.0in 0.0in 0.0in,clip,width=0.49\textwidth]{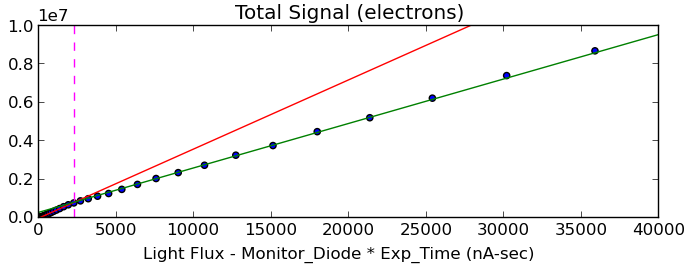}}
   \caption{Spot images in the bloomed full well and surface well conditions.  In the top panel, we see that we begin to see ``trailing'' when we enter the surface full well condition, because charges are trapped at the silicon surface.  In the bottom two panels, the vertical dotted line indicates the onset of saturation.  In the bottom left panel, we see that in the bloomed full well condition the total charge in the saturated spot increases linearly with flux and no charge is lost.  In the bottom right panel, we see that in the surface full well condition, charge begins to be lost as we enter saturation.  We believe this charge recombines at surface traps at the silicon-silicon dioxide interface.}
  \label{Sat4}
\end{figure}

A large number of simulations have been run to better understand saturation and blooming in the ITL STA3800C device.  These simulations have been very helpful to understand the physics of the device, because in the simulator one can get information which is simply not accessible to measurement.  Figure \ref{Sat1} shows the distinction between the ``bloomed full well'' condition, where charge begins to bloom above the charge storage regions, and the ``surface full well'' condition, where charge blooms along the silicon surface.  This effect is discussed in detail in Janesick \cite{janesick2001scientific}.  The simulations in Figures \ref{Sat2} and \ref{Sat3} reproduce these conditions, and these simulations illustrate the difference between these conditions.  Measurements of saturated spots in the surface full well condition, shown in Figure \ref{Sat4} also show that in the surface full well condition, charge is lost to traps at the silicon-silicon dioxide interface.  Thus it is apparent that the surface full well condition is to be avoided.

\begin {figure}[H]
  \centering
  \subfloat[b][Simulation with increasing charge, showing onset of blooming.]{\includegraphics[trim=0.0in 0.0in 0.0in 0.0in,clip,width=0.45\textwidth]{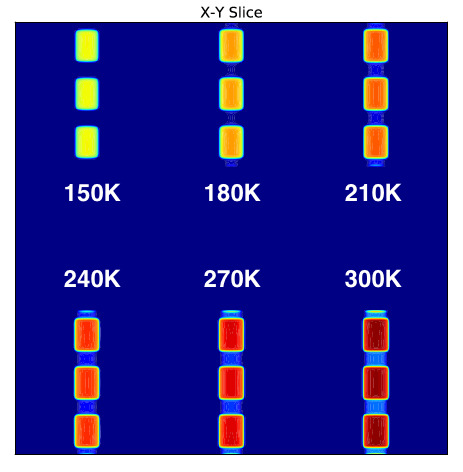}}
  \subfloat[b][Quantification of barrier height.]{\includegraphics[trim=3.2in 2.4in 3.3in 2.5in,clip,width=0.54\textwidth]{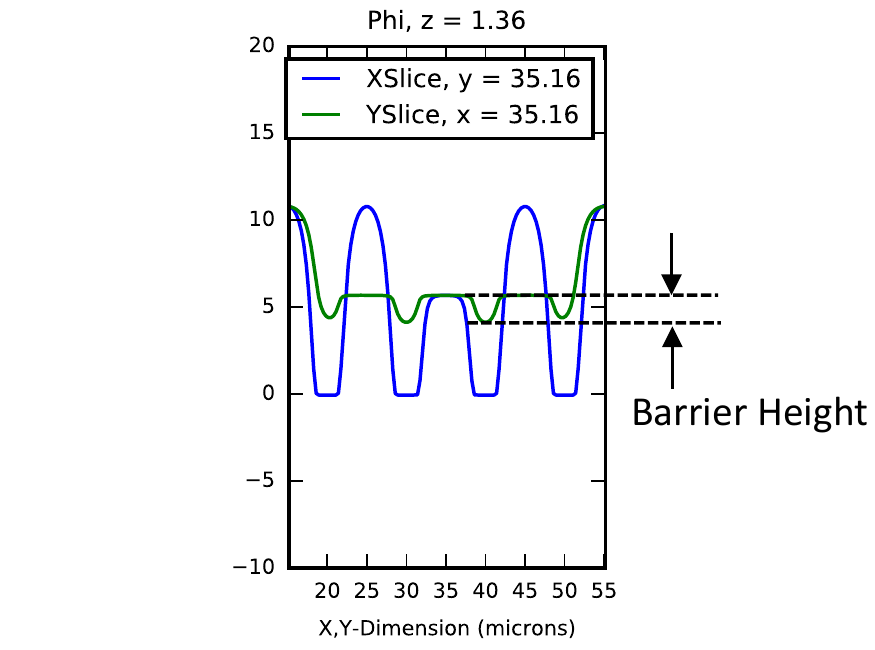}}\\
  \subfloat[b][Barrier height as a function of pixel charge.]{\includegraphics[trim=0.0in 0.0in 0.0in 0.6in,clip,width=0.49\textwidth]{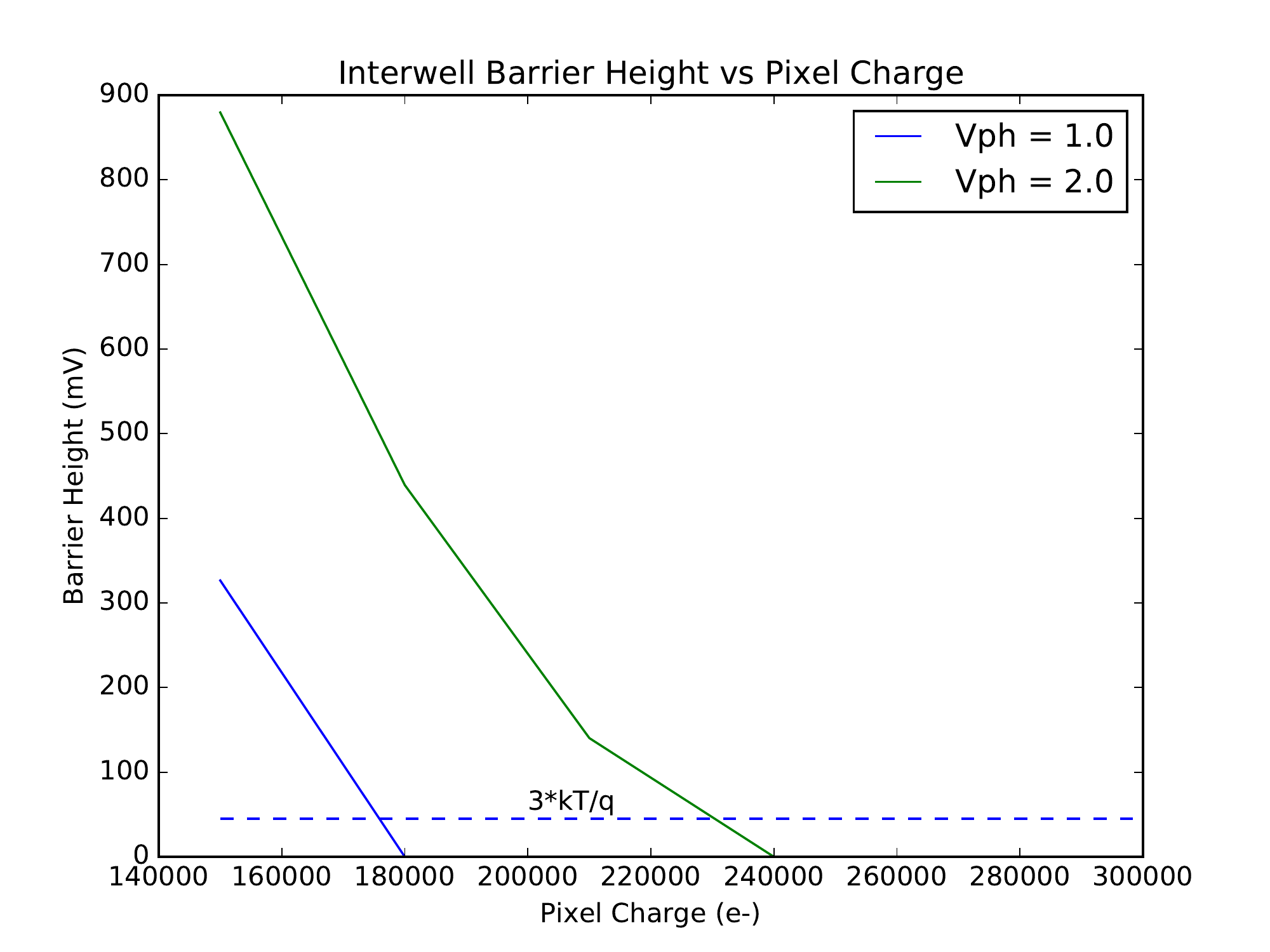}}
  \hspace{2mm}
  \subfloat[b][Measurements and simulations of onset of saturation.]{\includegraphics[trim=0.0in 0.0in 0.0in 0.5in,clip,width=0.49\textwidth]{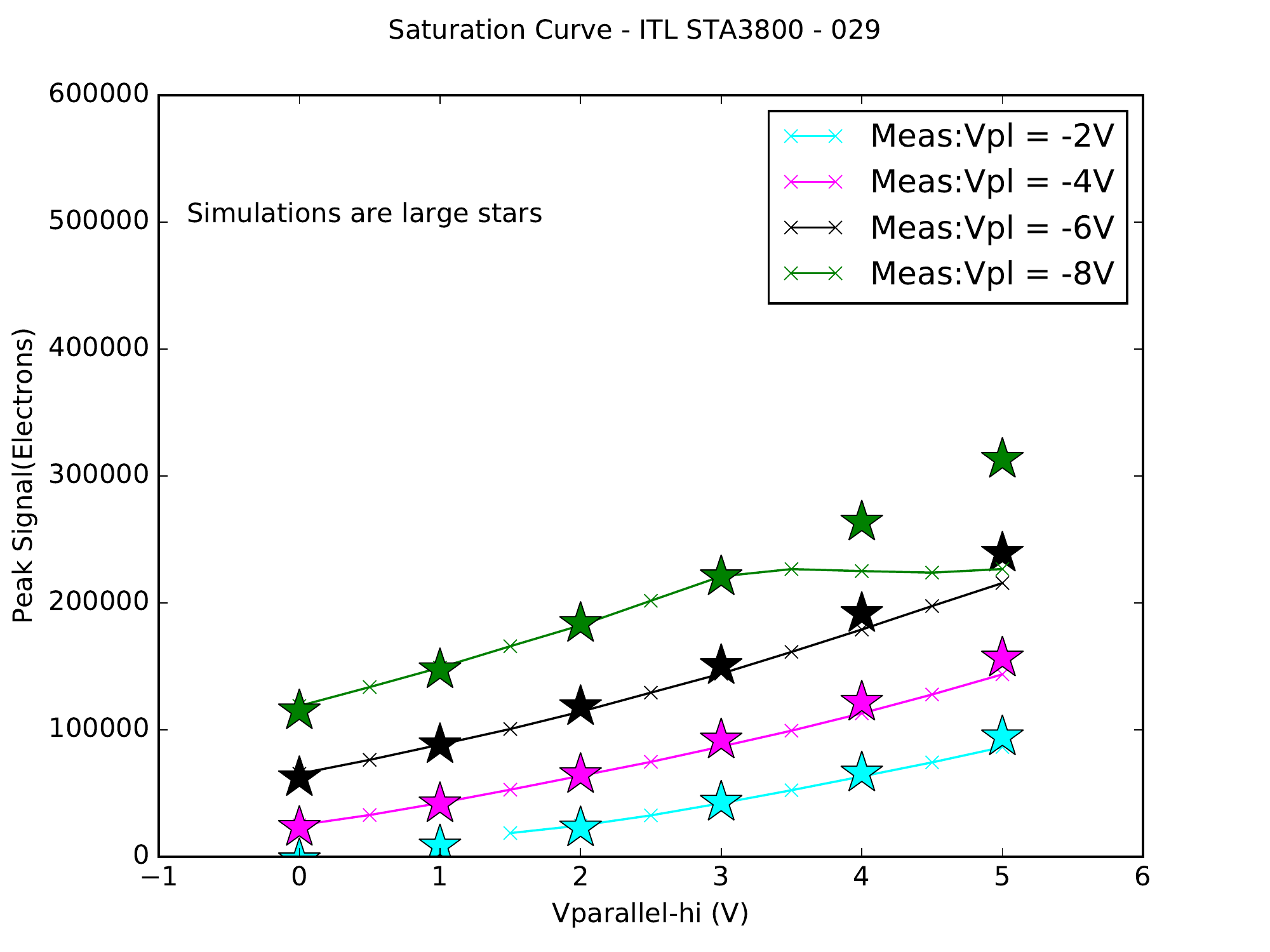}}
   \caption{Measurements and simulations of the onset of saturation as a function of parallel low and high voltages.  Panel (a) shows the simulations which are run, where the pixel charges are increased until saturation is seen.  From these simulations, the barrier height between pixels is quantified as a function of pixel charge, as shown in panels (b) and (c).  Panel (d) shows that this methodology accurately reproduces measurements of the onset of saturation over most of the range.  The simulation fails when we enter the strong surface full well condition, where there is significant charge loss.}
  \label{Sat5}
\end{figure}

We can go further and quantitatively reproduce measurements of the onset of saturation as a function of parallel low and high voltages, as shown in Figure \ref{Sat5}.  By quantifying the barrier height between the storage wells, we show that saturation occurs when the barrier height drops below a certain value.  The fit is good except in the strong surface full well condition, because the charge loss that occurs is not included in the simulations.  It would be possible to modify the simulations to include this effect, but since the surface full well condition is to be avoided, this was deemed to be not worth the effort.

\subsection{Astrometric shifts at array edges}
\begin {figure}[H]
  \begin{tikzpicture}
    \node[anchor=south west,inner sep=0] (image) at (0,0) {\includegraphics[trim=0.0in 0.2in 0.0in 0.0in,clip,width=0.99\textwidth]{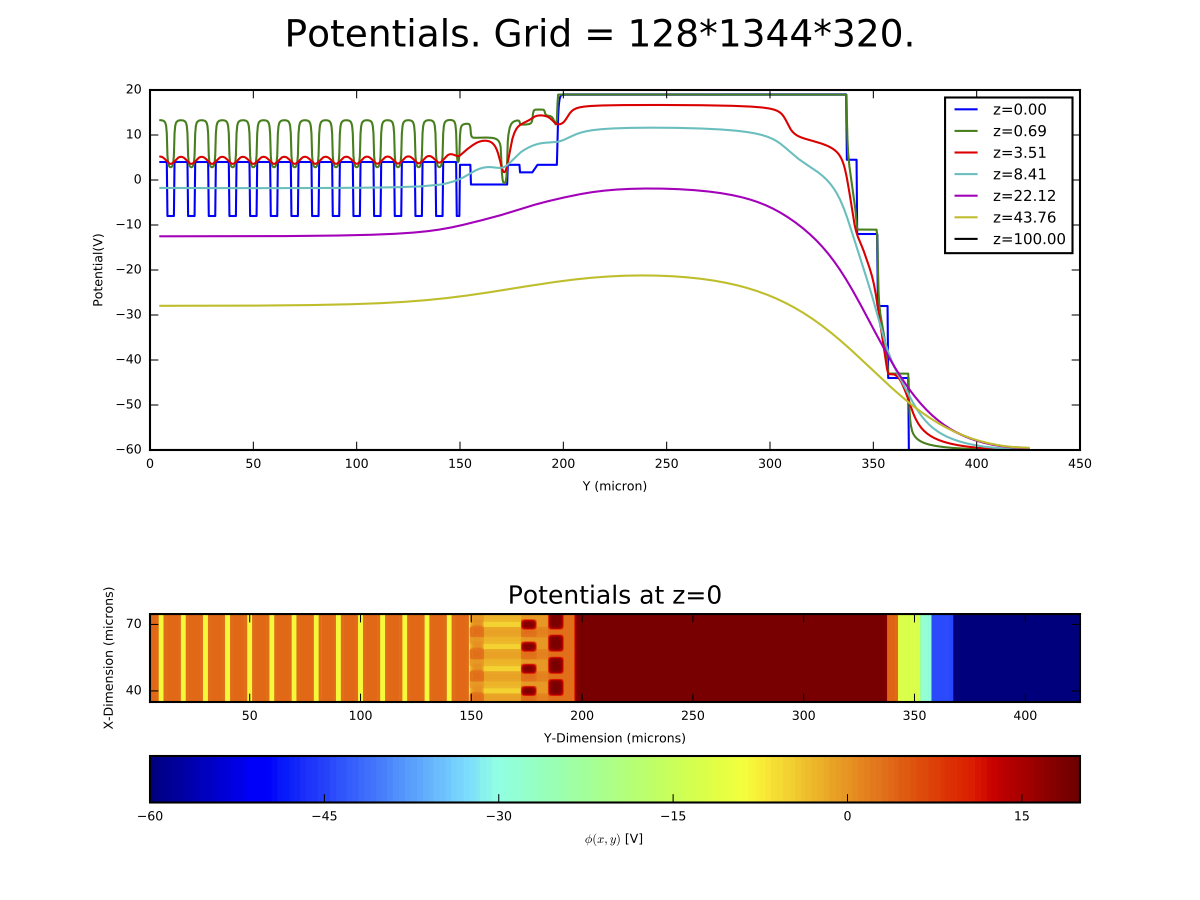}};
    \draw [line width=0.25mm, black] (2.0,4.2) -- (2.0,5);
    \node[align=center,text=black,font={\small\bfseries}] at (4.5,4.6) {Pixel Array};
    \draw [line width=0.25mm, black] (6.7,4.2) -- (6.7,5);
    \node[align=center,text=black,font={\small\bfseries}] at (7.5,4.6) {Serials};
    \draw [line width=0.25mm, black] (8.3,4.2) -- (8.3,5);
    \node[align=center,text=black,font={\small\bfseries}] at (10.7,4.6) {``Scupper''};
    \draw [line width=0.25mm, black] (12.4,4.2) -- (12.4,5);
    \node[align=center,text=black,font={\small\bfseries}] at (13.2,4.6) {Guard \\ Rings};
    \draw [line width=0.25mm, black] (14.0,4.2) -- (14.0,5);
    \node[align=center,text=black,font={\small\bfseries}] at (16.0,4.6) {Chip Edge};
    \node[align=center,text=black,fill=white,font={\small\bfseries}] at (8.4,0.4) {Potential(V)};    
    \end{tikzpicture}
   \caption{This shows the basic simulation which is run to determine the astrometric shifts at the array edge.  The simulation includes a narrow strip of pixels and continues until the edge of the chip.}
  \label{Edge_setup}
  \end{figure}
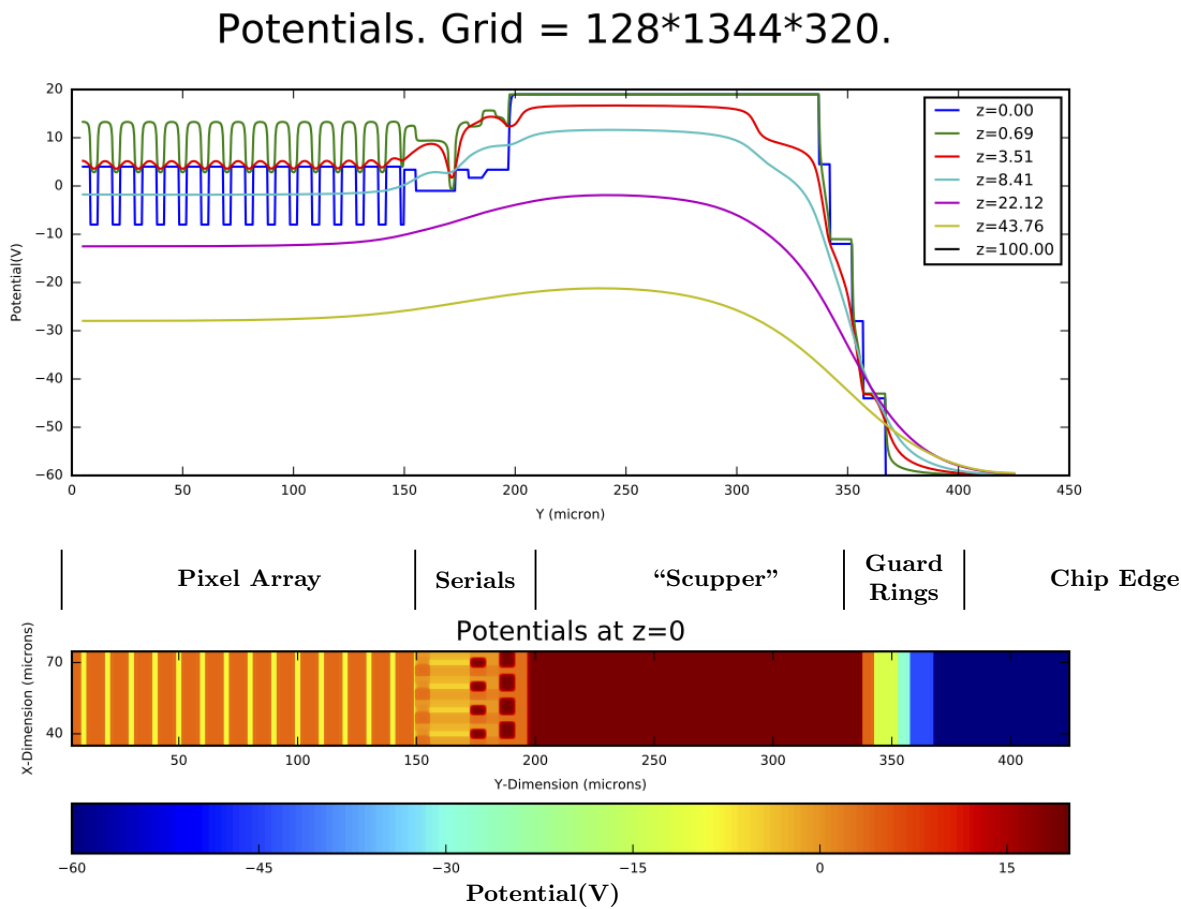

\begin {figure}[H]
\begin{center}
\includegraphics[trim=0.0in 0.0in 0.0in 0.0in,clip,width=0.55\textwidth]{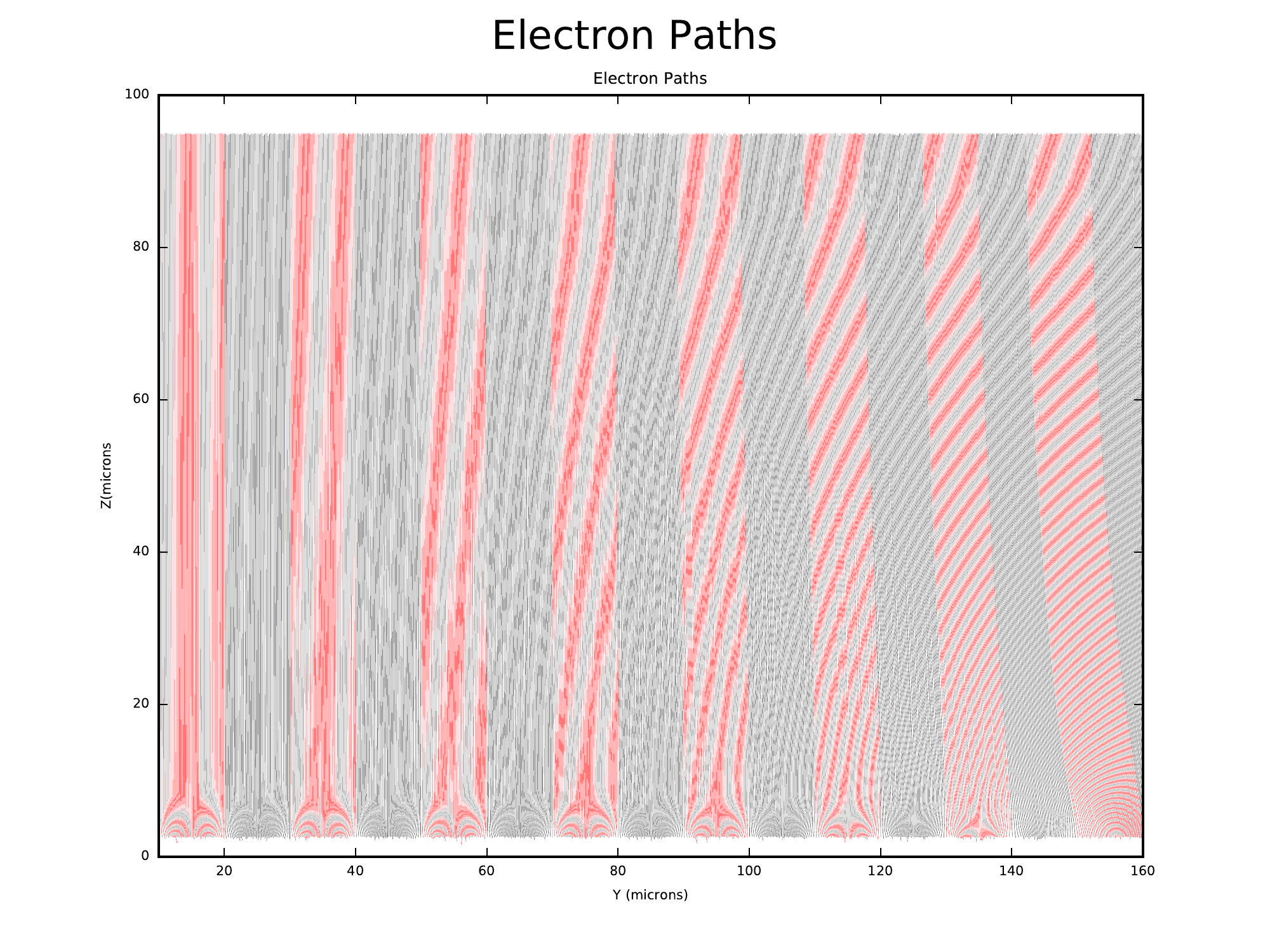}
   \caption{This shows the deviation of electron paths near the edge of the chip due to the lateral electric fields.  Diffusion is off for this plot}
  \label{Edge_paths}
\end{center}
\end{figure}

\begin {figure}[H]
\begin{center}
\includegraphics[trim=0.0in 0.0in 0.0in 0.0in,clip,width=0.95\textwidth]{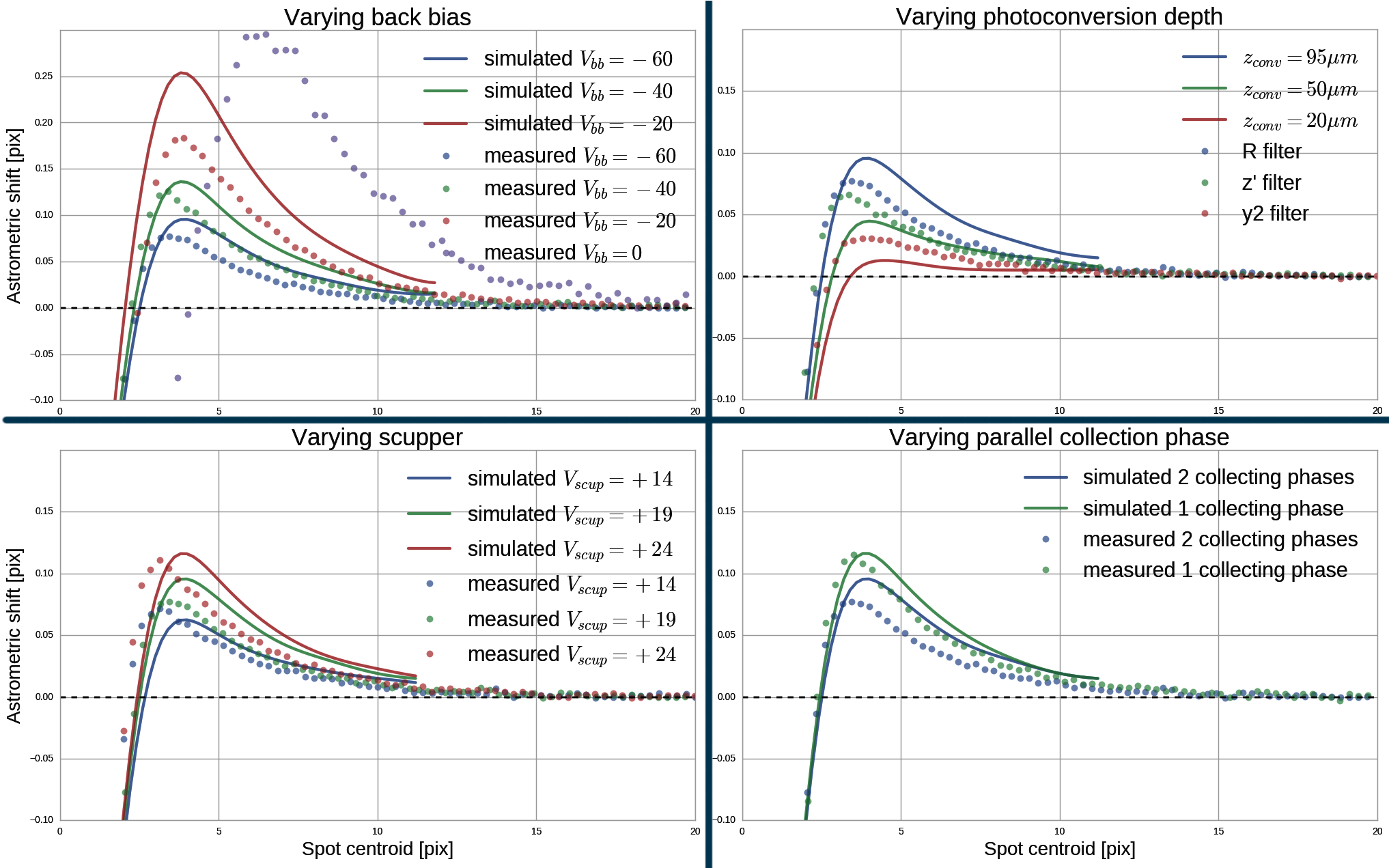}
   \caption{This shows the measured and simulated astrometric shifts at the array edge, with several parameters varied.  The simulation captures the major trends of all of these variables.  The simulation was not run for Vbb=0, where the chip is not fully depleted.}
  \label{Edge_shift}
\end{center}
\end{figure}

In addition to simulations of the pixel arrays, simulations can be done of the peripheral circuitry, as shown in the next two sections.  At the edges of the pixel array, lateral electric fields from the surrounding circuitry can introduce pixel boundary shifts, which lead to measurable astrometric shifts.  This effect has been characterized on the UC Davis LSST Optical Simulator (\cite{tyson2014}, \cite{bradshaw2015}).  We have then built a simulation of a portion of the pixel array which extends to the chip edge to compare the measurements to the simulations.  In addition to the pixel array, it is necessary to build into the simulation the appropriate ``fixed voltage regions'' at the edge of the chip.  Figure \ref{Edge_setup} shows the setup of this simulation.  The bending of the equipotential lines near the edge of the pixel array betray the presence of a lateral electric field which deflects incoming electrons.  Figure \ref{Edge_paths} shows these simulated paths (with diffusion turned off), and the edge deflection is apparent.  In addition to this deflection, there is a second effect which influences the astrometric shift, which is that as the measured spot begins to ``fall off' the edge of the array, there is a resulting shift in the opposite direction.  These two competing effects lead to the astrometric shifts seen in Figure \ref{Edge_shift}.  The shift has been characterized for a number of different measurement conditions.  The simulation, while not perfect, captures all of the trends correctly.  These simulations are run with the ``edge.cfg'' example at \cite{Poisson-CCD-code}.

\subsection{Output transistor characteristics}
\begin {figure}[H]
\centering
      \subfloat[\normalsize{Photograph of the output circuitry}]{\includegraphics[trim=0.0in -2.0in 0.0in 0.0in,clip,width=0.20\textwidth,valign=b]{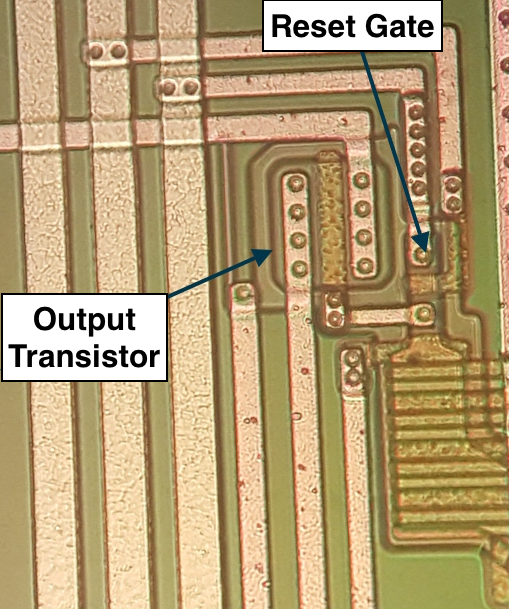}}
      \hspace{2mm}
      \subfloat[\normalsize{Simulation of the same region}]{\includegraphics[trim=0.0in 2.5in 0.5in 0.0in,clip,width=0.30\textwidth,valign=b]{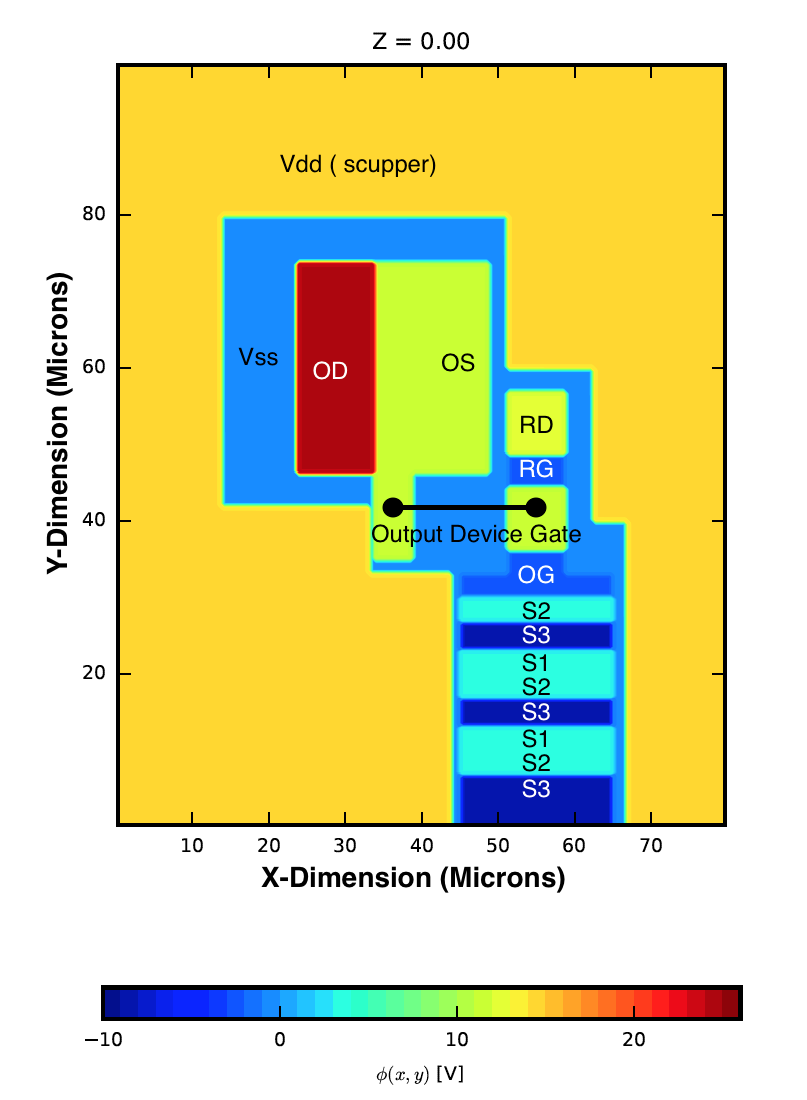}}
    \subfloat[\normalsize{Simulated and measured I-V}]{\includegraphics[trim=0.5in 0.0in 1.5in 1.8in,clip,width=0.45\textwidth,valign=b]{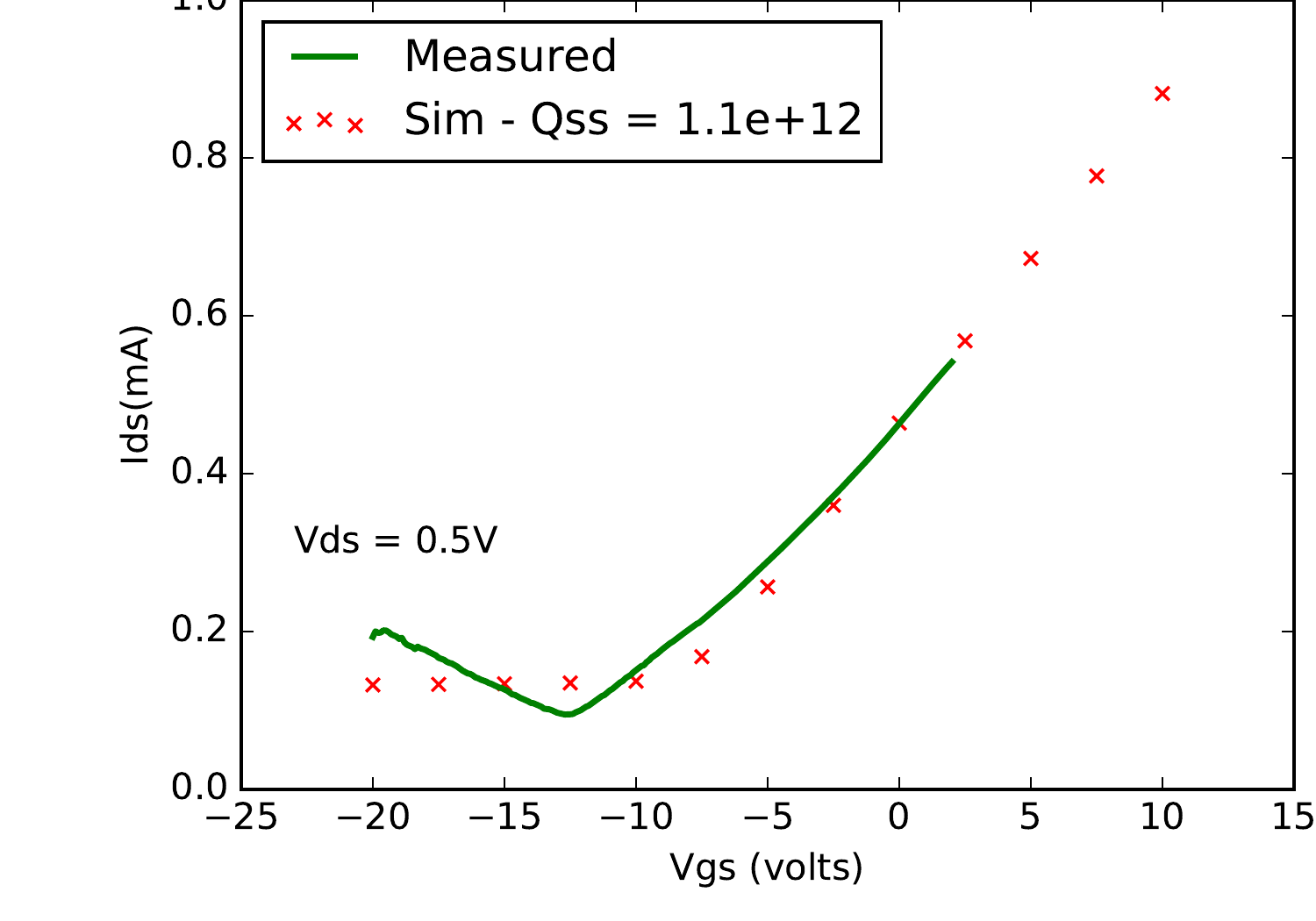}}
  \caption{Simulation of the ITL STA3800C output transistor I-V characteristics, compared to measurements.}
  \label{Output_transistor}
\end{figure}

This section shows simulations that were done to model the output transistor of the ITL STA3800C.  Figure \ref{Output_transistor} shows the simulated and measured $\rm I_d - V_g$ characteristics.  We obtained a relatively good fit of the transistor turn-on.  This simulation is run with the ``trans.cfg'' example at \cite{Poisson-CCD-code}.

\subsection{Other qualitative tests}
\label{Other_tests}
\begin {figure}[H]
    \centering
    	\subfloat[b][Simulated tree rings with 10\% dopant variation]{\includegraphics[trim=0.0in 0.5in 0.0in 1.0in,clip,width=0.85\textwidth]{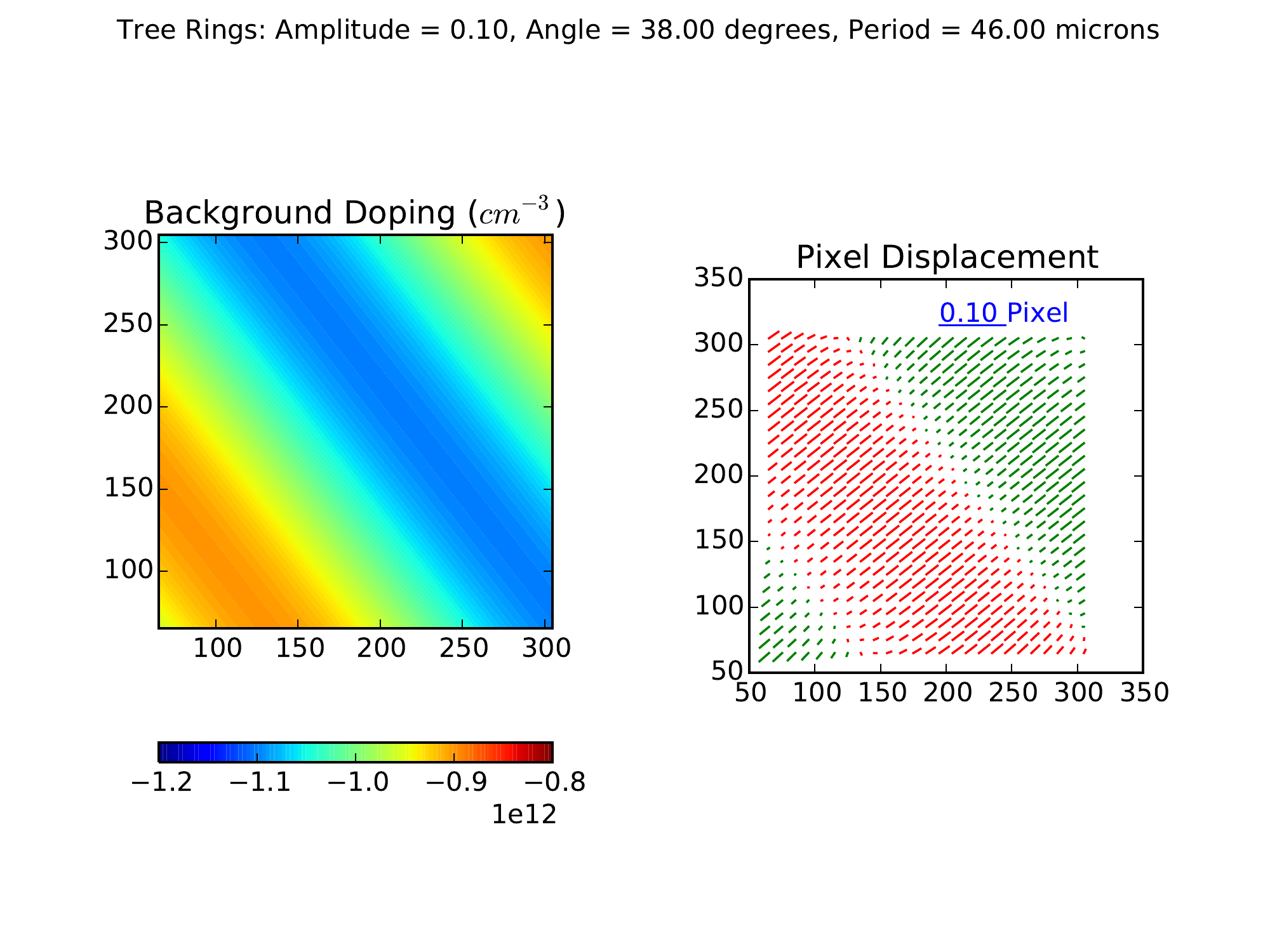}}\\
    	\subfloat[b][Simulated tree rings with 3\% dopant variation]{\includegraphics[trim=0.0in 0.5in 0.0in 1.0in,clip,width=0.85\textwidth]{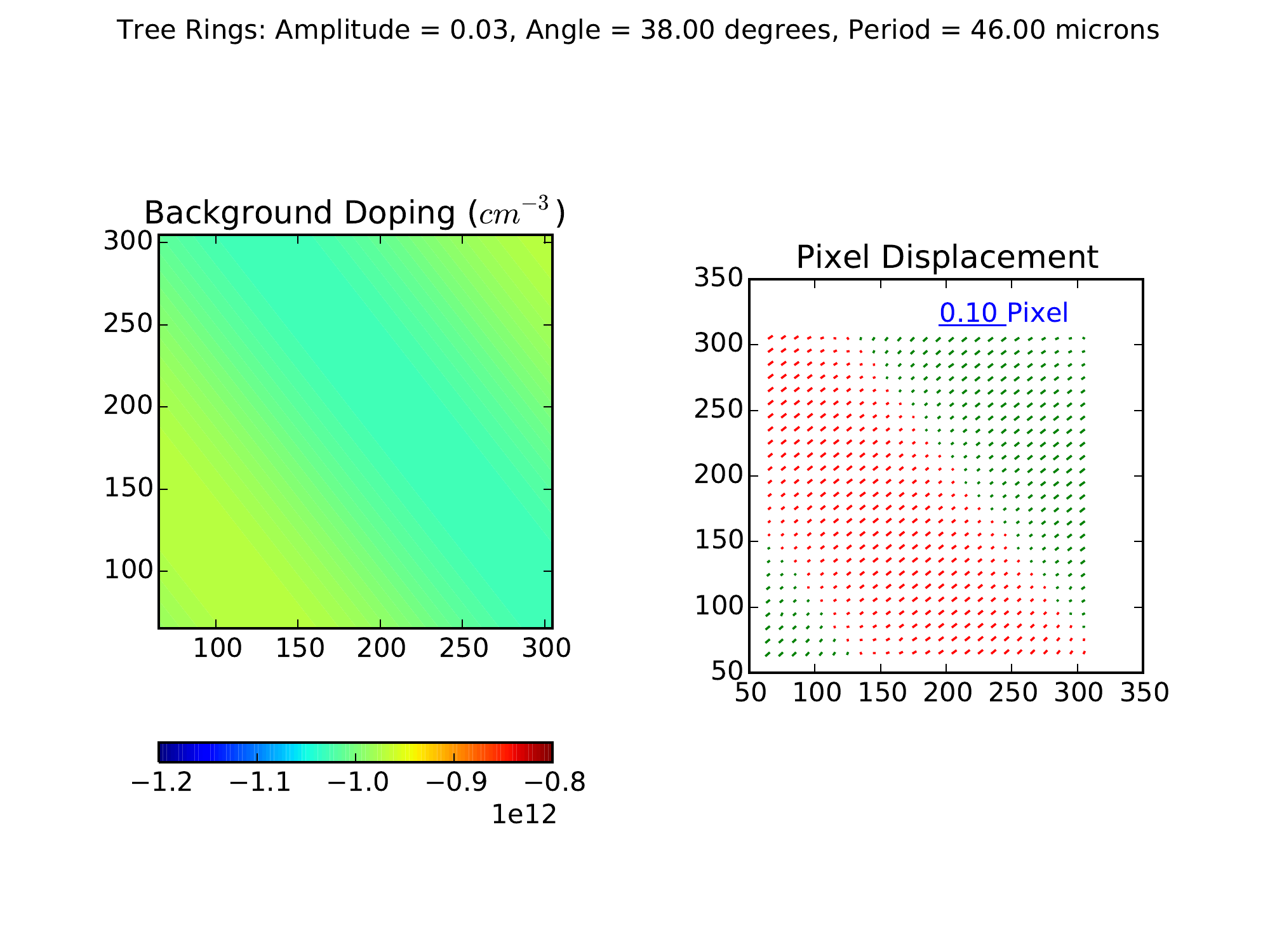}}
  \caption{Simulations of ``tree rings''.  A sinusoidal dopant variation in the silicon bulk is introduced, varying in X and Y, and constant in Z.  The resulting astrometric pixel shifts are characterized.  The X and Y axes are in pixels.  The lower value is more consistent with actual observations.}
  \label{Tree_rings}
  \end{figure}

Some other tests have been run which have given results which are qualitatively reasonable, but which have not been compared with quantitative measurements, and three of these are reviewed here. The first of these are known as ``tree rings''.  As is well known, periodic dopant variations introduced during the growth of the silicon boule can lead to measurable variations in flat fields, as well as introducing astrometric pixel shifts (see, for example, \cite{plazas2014}).  This effect has been successfully simulated, as shown in Figure \ref{Tree_rings}. This simulation is run with the ``treering.cfg'' example at \cite{Poisson-CCD-code}.

\begin {figure}[H]
    \centering
    	\subfloat[b][$\rm T_{Si} = 100 \mu m$; $\rm V_{BB} = -60V$]{\includegraphics[trim=4.0in 0.5in 0.5in 2.2in,clip,width=0.49\textwidth]{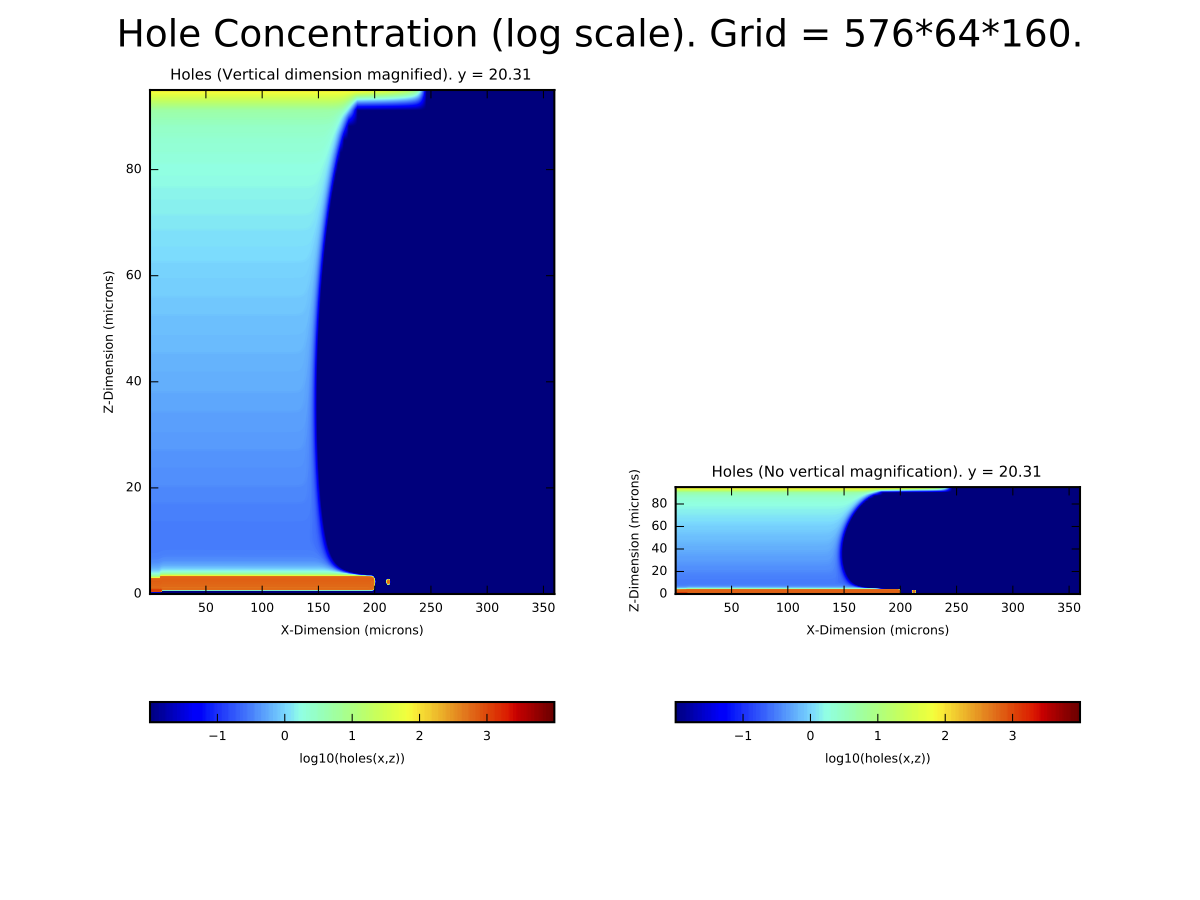}}
    	\subfloat[b][$\rm T_{Si} = 200 \mu m$; $\rm V_{BB} = -9V$]{\includegraphics[trim=4.0in 0.5in 0.5in 2.2in,clip,width=0.49\textwidth]{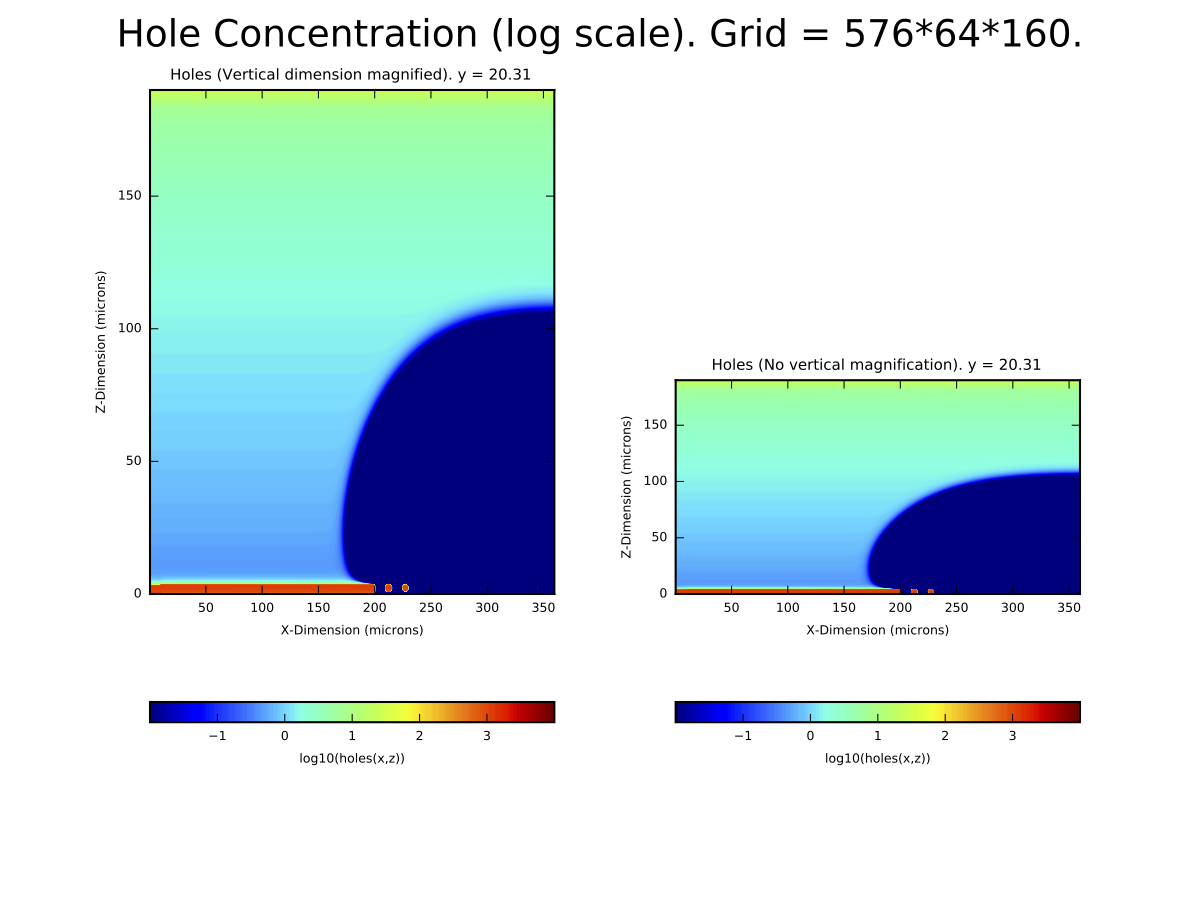}}        

  \caption{Simulations of the $\rm V_{BB}$ guard rings for two different conditions.  In both cases, the top of the simulation region is where photons are incident, and the left-hand edge is the edge of the CCD.  The imaging region begins at the right-hand edge of the simulation region and continues to the right.  The colors are the log of the hole concentration (in code units), as shown in the colorbar.  The imaging region in the left-hand simulation is fully depleted, while in the right-hand one, which is thicker and has a lower bias voltage, it is not.}
  \label{Vbb_connection}
  \end{figure}

A second interesting simulation is of the backside substrate connection $\rm V_{BB}$.  It is often questioned how this bias voltage, which is only connected to the CCD frontside, is conducted to the backside when the CCD is fully depleted.  The answer is that there is an undepleted region near the chip edge which serves this purpose.  It is interesting to simulate this connection, and the result is shown in Figure \ref{Vbb_connection}.

\begin {figure}[H]
    \centering
\includegraphics[trim=2.0in 0.0in 0.0in 0.9in,clip,width=0.75\textwidth]{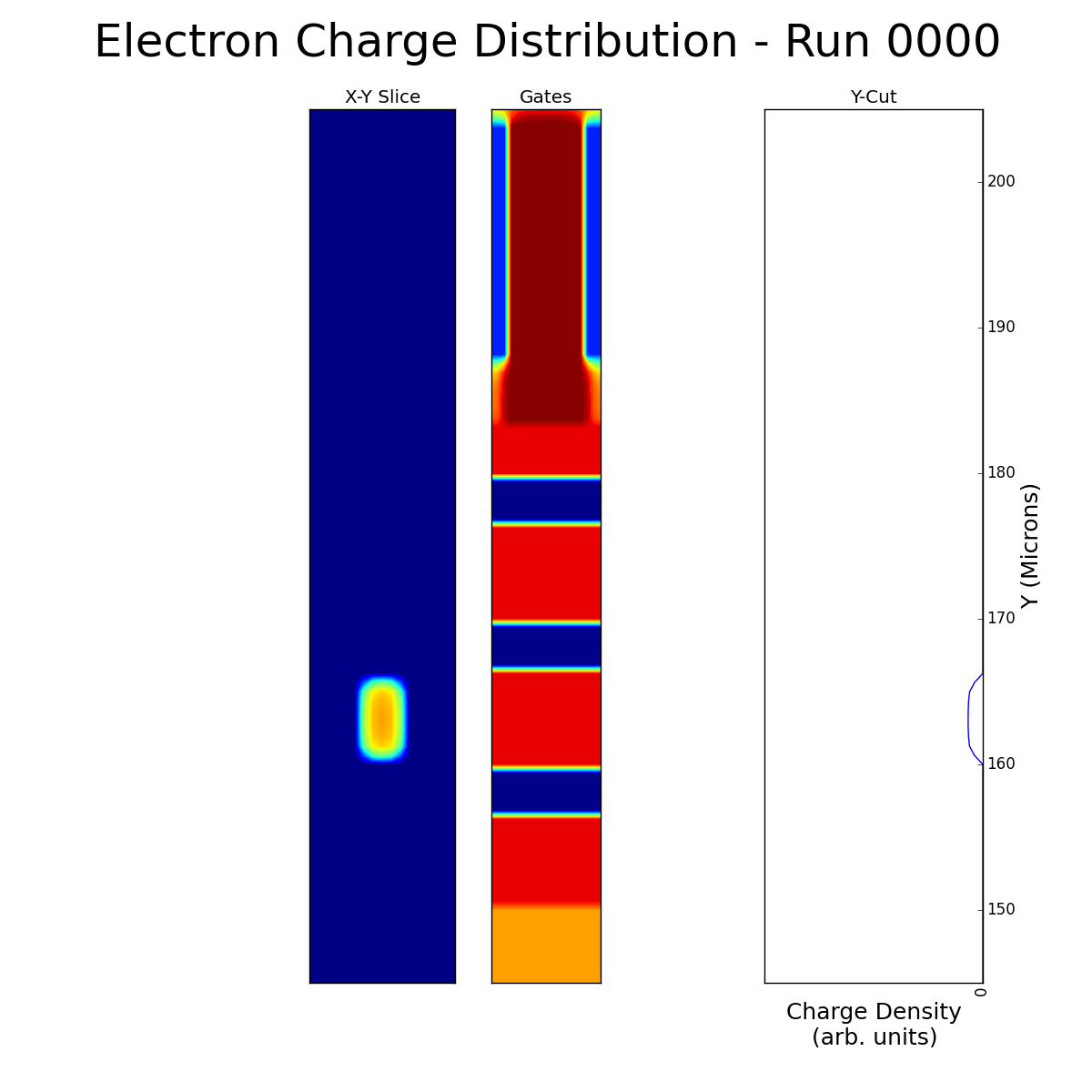}
  \caption{A single frame of a movie showing the transport of a charge packet.  The left panel shows the charge distribution, the center panel shows the surface potential, and the right panel shows a 1D slice through the charge distribution.  In the actual movie (available at \cite{Poisson-CCD-code}), the charge packet can be seen propagating through the parallel chain and into the serial register.}
  \label{Movie_frame}
  \end{figure}

While the simulator solves for the condition of the CCD in equilibrium, and does not do transient simulations, it is possible to simulate transient effects by repeatedly solving for the state of the CCD with small incremental changes to the boundary conditions.  These can then be stitched together to form a movie of the results.  An example of the parallel charge transport is shown in Figure \ref{Movie_frame}.  Several movies constructed in this way are available at \cite{Poisson-CCD-code}. 

\section{Conclusion}
We have presented a software package optimized for simulating astronomical CCDs.  The code has been optimized to be as physically realistic as possible, and accurately simulates many aspects of CCD behavior.  The package is open source and freely available \cite{Poisson-CCD-code}, and has been validated against a number of different types of CCD measurements.  It has proven useful for analyzing sensor effects in the CCDs being used to construct the LSST focal plane, and insights gained from running these simulations are being incorporated into the software stack to be used for instrument signature removal in the LSST images.  It is hoped that it will find additional uses in the future.

\section{Acknowledgements}
Kirk Gilmore's help in setting up the hardware and software for the CCDs from both vendors has also been invaluable.  Perry Gee has provided key support in software and networking.  Merlin Fisher-Levine has patiently helped with my understanding and implementation of the LSST data management software stack.  David Kirkby contributed code for including different filters.   For author contributions, C.L. developed the {\carlito Poisson\_CCD} code, ran most of the studies described here, and drafted the paper.  AB carried out the study of astrometric shifts at array edges, and provided test and debug support.  JAT supervised the project.  Financial support from DOE grant DE-SC0009999 and Heising-Simons Foundation grant 2015-106 are gratefully acknowledged.  This paper has undergone internal review in the LSST Dark Energy Science Collaboration.  The internal reviewers were Andrew Rasmussen, Douglas Tucker, and Dan Weatherill, and their inputs are much appreciated. 

\bibliographystyle{unsrt}
\bibliography{ccd}

\begin{thebibliography}{10}

\bibitem{Smith_Nobel}
G.E. {Smith}.
\newblock Nobel lecture: The invention and early history of the ccd.
\newblock {\em Rev. Mod. Phys.}, 82:2307--2312, Aug 2010.

\bibitem{Poisson-CCD-code}
C.~{Lage}.
\newblock {Poisson solver for LSST CCDs}, November 2019.
\newblock \url{https://github.com/craiglagegit/Poisson_CCD}.

\bibitem{LSST_2019}
{\v Z}.~{Ivezi{\'c}}, S.~M. {Kahn}, J.~A. {Tyson}, B.~{Abel}, E.~{Acosta},
  R.~{Allsman}, D.~{Alonso}, Y.~{AlSayyad}, S.~F. {Anderson}, J.~{Andrew}, and
  et~al.
\newblock {LSST: From Science Drivers to Reference Design and Anticipated Data
  Products}.
\newblock {\em \apj}, 873:111, March 2019.

\bibitem{oconnor2019uniformity}
P.~O'Connor.
\newblock {Uniformity and Stability of the LSST Focal Plane}, 2019.
\newblock arXiv:1907.00995.

\bibitem{oconnor_2016}
P.~O'Connor, P.~Antilogus, P.~Doherty, J.~Haupt, S.~Herrmann, M.~Huffer,
  C.~Juramy-Giles, J.~Kuczewski, S.~Russo, C.~Stubbs, and R.~Van Berg.
\newblock {Integrated system tests of the LSST raft tower modules}.
\newblock In Andrew~D. Holland and James Beletic, editors, {\em High Energy,
  Optical, and Infrared Detectors for Astronomy VII}, volume 9915, pages 327 --
  338. International Society for Optics and Photonics, SPIE, 2016.

\bibitem{ITL_website}
University of~Arizona.
\newblock {Imaging Technology Laboratory}, 2016.
\newblock \url{http://www.itl.arizona.edu}.

\bibitem{E2V_website}
{Teledyne E2V}, 2019.
\newblock \url{https://www.teledyne-e2v.com/products/space/}.

\bibitem{Villasenor_2017}
J.~Villasenor, G.~Prigozhin, J.P. Doty, C.~Lage, S.~Yazdi, G.~Ricker, and
  R.~Vanderspek.
\newblock Reach-through effect in deep depletion {TESS} {CCDs}.
\newblock {\em Journal of Instrumentation}, 12(04):C04025--C04025, apr 2017.

\bibitem{HADJIDIMOS2000177}
A.~Hadjidimos.
\newblock Successive overrelaxation (sor) and related methods.
\newblock {\em Journal of Computational and Applied Mathematics},
  123(1):177--199, 2000.
\newblock Numerical Analysis 2000. Vol. III: Linear Algebra.

\bibitem{plazas2014}
A.~A. {Plazas}, G.~M. {Bernstein}, and E.~S. {Sheldon}.
\newblock {On-Sky Measurements of the Transverse Electric Fields' Effects in
  the Dark Energy Camera CCDs}.
\newblock {\em \pasp}, 126(942):750, Aug 2014.

\bibitem{antilogus2014}
P.~{Antilogus}, P.~{Astier}, P.~{Doherty}, A.~{Guyonnet}, and N.~{Regnault}.
\newblock {The brighter-fatter effect and pixel correlations in CCD sensors}.
\newblock {\em Journal of Instrumentation}, 9:C03048, March 2014.

\bibitem{gruen2015}
D.~{Gruen}, G.~M. {Bernstein}, M.~{Jarvis}, B.~{Rowe}, V.~{Vikram}, A.~A.
  {Plazas}, and S.~{Seitz}.
\newblock {Characterization and correction of charge-induced pixel shifts in
  DECam}.
\newblock {\em Journal of Instrumentation}, 10:C05032, May 2015.

\bibitem{guyonnet2015}
A.~{Guyonnet}, P.~{Astier}, P.~{Antilogus}, N.~{Regnault}, and P.~{Doherty}.
\newblock {Evidence for self-interaction of charge distribution in
  charge-coupled devices}.
\newblock {\em \aap}, 575:A41, March 2015.

\bibitem{Lage_2017}
C.~{Lage}, A.~{Bradshaw}, and J.~A. {Tyson}.
\newblock {Measurements and simulations of the brighter-fatter effect in CCD
  sensors}.
\newblock {\em Journal of Instrumentation}, 12:C03091, Mar 2017.

\bibitem{Coulton_2018}
W.R. {Coulton}, R.~{Armstrong}, K.M. {Smith}, R.H. {Lupton}, and D.N.
  {Spergel}.
\newblock {Exploring the Brighter-fatter Effect with the Hyper Suprime-Cam}.
\newblock {\em The Astronomical Journal}, 155(6):258, May 2018.

\bibitem{CCD-Physical-Analysis}
C.~{Lage}.
\newblock {Physical and electrical analysis of LSST sensors}, 2019.
\newblock \url{https://arxiv.org/abs/1911.09577}.

\bibitem{sze1981physics}
S.M. Sze.
\newblock {\em Physics of Semiconductor Devices}.
\newblock Wiley-Interscience publication. John Wiley \& Sons, 1981.

\bibitem{rafferty1985iterative}
C.S. {Rafferty}, M.R. {Pinto}, and R.W. {Dutton}.
\newblock Iterative methods in semiconductor device simulation.
\newblock {\em IEEE Transactions on Electron Devices}, 32(10):2018--2027, 1985.

\bibitem{Gummel}
H.~K. {Gummel}.
\newblock A self-consistent iterative scheme for one-dimensional steady state
  transistor calculations.
\newblock {\em IEEE Transactions on Electron Devices}, 11(10):455--465, 1964.

\bibitem{briggs2000multigrid}
W.L. {Briggs}, S.F. {McCormick}, et~al.
\newblock {\em A multigrid tutorial}, volume~72.
\newblock Siam, 2000.

\bibitem{NumericalRecipes}
W.~H. {Press}, S.~A. {Teukolsky}, W.~T. {Vetterling}, and B.~P. {Flannery}.
\newblock {\em Numerical Recipes}.
\newblock Cambridge University Press, third edition, 2007.

\bibitem{jacobini1977}
C.~{Jacobini}, C.~{Canali}, G.. {Ottaviani}, and A.~{Quaranta}.
\newblock {A review of some transport properties of silicon}.
\newblock {\em Solid-State Electronics}, 20:77--89, 1977.

\bibitem{green_1990}
M.A. {Green}.
\newblock Intrinsic concentration, effective densities of states, and effective
  mass in silicon.
\newblock {\em Journal of Applied Physics}, 67(6):2944--2954, 1990.

\bibitem{tyson2014}
J.~A. {Tyson}, J.~{Sasian}, K.~{Gilmore}, A.~{Bradshaw}, C.~{Claver},
  M.~{Klint}, G.~{Muller}, G.~{Poczulp}, and E.~{Resseguie}.
\newblock {LSST optical beam simulator}.
\newblock In {\em High Energy, Optical, and Infrared Detectors for Astronomy
  VI}, volume 9154 of {\em Society of Photo-Optical Instrumentation Engineers
  (SPIE) Conference Series}, page 915415, July 2014.

\bibitem{BF-Linearity}
C.~{Lage}.
\newblock {Linearity and correction of the BF effect in LSST sensors}, 2019.
\newblock \url{https://arxiv.org/abs/1911.09567}.

\bibitem{janesick2001scientific}
J.R. Janesick.
\newblock {\em Scientific Charge-coupled Devices}.
\newblock Press Monograph Series. SPIE Press, 2001.

\bibitem{bradshaw2015}
A.~{Bradshaw}, C.~{Lage}, E.~{Resseguie}, and J.~A. {Tyson}.
\newblock {Mapping charge transport effects in thick CCDs with a dithered array
  of 40,000 stars}.
\newblock {\em Journal of Instrumentation}, 10:C04034, April 2015.

\end{thebibliography}

\begin{appendices}
\section{List of configuration parameters}
\renewcommand{\thefigure}{A.\arabic{figure}}
\setcounter{figure}{0}
\label{Parameter_Appendix}
\begin{tabular}{|l|c|c|l|} 
\hline
Name                            &       type    &       Default &        Description \\
\hline
AddTreeRings                  	&	int	&	0	&	 0-No tree rings, 1-Add tree rings\\ 
\hline 
BackgroundDoping              	&	float	&	-1e+12	&	 Background doping in $\rm cm^{-3}$ \\ 
\hline 
BottomSteps                   	&	int	&	1000	&	 Number of diffusion steps each electron takes\\
                                &               &               &        while logging final charge location\\ 
\hline 
BuildQFeLookup                	&	int	&	0	&	 0-Don't build look-up table, 1-build look-up table\\ 
\hline 
CCDTemperature                	&	float	&	173.00	&	 Temperature in K\\ 
\hline 
CalculateZ0                   	&	int	&	0	&	 0 - don't calculate - Use ElectronZ0, \\
                                &               &               &        1 - calculate from filter and SED.\\
\hline 
ChannelDepth                  	&	float	&	1.00	&	 Square profile depth in microns\\ 
\hline 
ChannelDoping                 	&	float	&	5e+11	&	 Square profile doping in $\rm cm^{-3}$ \\
\hline 
ChannelDose                   	&	float	&	5e+11	&	 Gaussian profile dose in $\rm cm^{-2}$ \\
\hline 
ChannelPeak                   	&	float	&	0.00	&	 Gaussian peak depth in microns \\ 
\hline 
ChannelProfile                	&	int	&	0	&        0 = Square profile, N = N Gaussian profiles       \\
\hline 
ChannelSigma                  	&	float	&	0.50	&	 Gaussian sigma in microns \\ 
\hline 
ChannelStopDepth              	&	float	&	1.00	&	 Square profile depth in microns\\ 
\hline 
ChannelStopDoping             	&	float	&	5e+11	&	 Square profile doping in $\rm cm^{-3}$ \\
\hline 
ChannelStopDose               	&	float	&	5e+11	&	 Gaussian profile dose in $\rm cm^{-2}$ \\
\hline 
ChannelStopDotCenter          	&	float	&	5.00	&	 Center position in microns from pixel bottom\\ 
\hline 
ChannelStopDotDepth           	&	float	&	1.00	&	 Square profile depth in microns\\ 
\hline 
ChannelStopDotDoping          	&	float	&	5e+11	&	 Square profile doping in $\rm cm^{-3}$ \\
\hline 
ChannelStopDotDose            	&	float	&	5e+11	&	 Gaussian profile dose in $\rm cm^{-2}$ \\
\hline 
ChannelStopDotHeight          	&	float	&	0.00	&	 Height in microns \\ 
\hline 
ChannelStopDotPeak            	&	float	&	0.00	&	 Gaussian peak depth in microns \\ 
\hline 
ChannelStopDotProfile         	&	int	&	0	&        0 = Square profile, N = N Gaussian profiles       \\
\hline 
ChannelStopDotSigma           	&	float	&	0.50	&	 Gaussian sigma in microns \\ 
\hline 
ChannelStopDotSurfaceCharge   	&	float	&	0.00	&	 Surface charge in $\rm cm^{-2}$ \\
\hline 
ChannelStopPeak               	&	float	&	0.00	&	 Gaussian peak depth in microns \\ 
\hline 
ChannelStopProfile            	&	int	&	0	&        0 = Square profile, N = N Gaussian profiles       \\
\hline 
ChannelStopSideDiff           	&	float	&	FieldOxideTaper	&  Side diffusion in microns	 \\ 
\hline 
ChannelStopSigma              	&	float	&	0.50	&	 Gaussian sigma in microns \\ 
\hline 
ChannelStopSurfaceCharge      	&	float	&	0.00	&	 Surface charge in $\rm cm^{-2}$ \\
\hline 
ChannelStopWidth              	&	float	&	1.00	&	 Width in microns \\ 
\hline 
ChannelSurfaceCharge          	&	float	&	0.00	&	 Surface charge in $\rm cm^{-2}$ \\
\hline 
CollectedCharge[i][j]         	&	int	&	0	&	 Number of electron in well i,j \\ 
\hline 
CollectingPhases              	&	int	&	1	&	 Number of collecting phases\\ 
\hline 
Continuation                  	&	int	&	0	&	 0-No continuation, 1-continue at LastCont..Step\\ 
\hline 
DiffMultiplier                	&	float	&	2.30	&	 Used to adjust amount of diffusion\\ 
\hline 
ElectronMethod                	&	int	&	0	&	 Controls electron calculation\\
	                        &               &               &        0 - Leave electrons where they land from tracking (old)\\
			        &               &               &        1 - Set QFe (QFe is always used in Fixed Regions)\\
                                &               &               &        If 1 is specified, you must provide QFe lookup params\\
				&               &               &        2 - Electron conservation and constant QFe (best)\\
\hline 
ElectronZ0Area                	&	float	&	100.00	&	 Initial Z value for electrons calculating pixel areas\\ 
\hline 
ElectronZ0Fill                	&	float	&	100.00	&	 Initial Z value for electrons filling pixel wells\\ 
\hline 
EquilibrateSteps              	&	int	&	100	&	  Number of diffusion steps each electron takes after\\
                                &               &               &         reaching bottom and before beginning to log charge.\\ 
\hline 
\end{tabular}
\newpage
\begin{tabular}{|l|c|c|l|} 
\hline
Name                            &       type    &       Default &        Description \\
\hline
Fe55CloudRadius               	&	float	&	0.20	&	 Cloud radius in microns\\ 
\hline 
Fe55ElectronMult              	&	float	&	1.00	&	 Used to adjust cloud electron attraction/repulsion\\ 
\hline 
Fe55HoleMult                  	&	float	&	1.00	&	 Used to adjust cloud hole attraction/repulsion\\ 
\hline 
FieldOxide                    	&	float	&	0.40	&	 Thickness in microns\\ 
\hline 
FieldOxideTaper               	&	float	&	0.50	&	 Taper width in microns\\ 
\hline 
FilledPixelCoords[i][j]       	&	float	&	2.00	&        Center x,y (in microns) of well i,j\\ 
\hline 
FilterBand                    	&	str	&	none	&	 One of u,g,r,i,z,y\\ 
\hline 
FilterFile                    	&	str	&	notebooks/depth\_pdf.dat	& SED file	 \\ 
\hline 
FixedRegionBCType          	&	int	&	0	&	Fixed region description\\ 
\hline 
FixedRegionDoping          	&	int	&	0	&	Fixed region description\\
\hline 
FixedRegionOxide           	&	int	&	0	&	Fixed region description\\
\hline 
FixedRegionQFe            	&	float	&	100.00	&	Fixed region description\\
\hline 
FixedRegionQFh             	&	float	&	-100.00	&	Fixed region description\\
\hline 
FixedRegionVoltage         	&	float	&	0.00	&	Fixed region description\\
\hline 
FringeAngle                   	&	float	&	0.00	&	 Fringe parameter for PixelBoundaryTestType=3\\ 
\hline 
FringePeriod                  	&	float	&	0.00	&	 Fringe parameter for PixelBoundaryTestType=3\\ 
\hline 
GateGap                       	&	float	&	0.00	&        Gap between paralel gates in microns (experimental)\\
\hline 
GateOxide                     	&	float	&	0.15	&	 Thickness in microns\\ 
\hline 
GridsPerPixelX                	&	int	&	16	&	 Grids per pixel at ScaleFactor = 1 \\ 
\hline 
GridsPerPixelY                	&	int	&	16	&	 Grids per pixel at ScaleFactor = 1 \\ 
\hline
LastContinuationStep          	&	int	&	0	&	 Used when continuing a stopped simulation\\ 
\hline 
LogEField                     	&	int	&	0	&	 0 - don't calculate E-Field, 1 - Calc. and store E-Field\\
\hline 
LogPixelPaths                 	&	int	&	0	&	 0 - only the final (z~0) point is logged, \\
                                &               &               &        1 - Entire path is logged\\
\hline 
NQFe                          	&	int	&	81	&	 Number of steps in QFe look-up table\\ 
\hline 
NZExp                         	&	float	&	10.00	&	 Non-linear z axis exponent \\ 
\hline 
NumDiffSteps                  	&	int	&	1	&	  A speed/accuracy trade-off. A value of 1 uses the\\
                                &               &               &         theoretical diffusion step.  A higher value takes \\
                                &               &               &         larger steps. Experimental\\ 
\hline 
NumElec                       	&	int	&	1000	&	 Number of electrons to be traced between field recalc.\\ 
\hline 
NumPhases                     	&	int	&	3	&	 Number of parallel phases\\ 
\hline 
NumSteps                      	&	int	&	100	&	 Number of steps, each one adding NumElec electrons\\ 
\hline 
NumVertices                   	&	int	&	2	&	 Number of vertices per side for the pixel area calc.\\
	      		        &               &               &        Since there are also 4 corners, there will be:\\
			        &               &               &        (4 * NumVertices + 4) vertices in each pixel\\
\hline 
NumberofFilledWells           	&	int	&	0	&	 Self-explanatory\\ 
\hline 
NumberofFixedRegions          	&	int	&	0	&	 Self-explanatory\\ 
\hline 
NumberofPixelRegions          	&	int	&	0	&	 Self-explanatory\\ 
\hline 
Nx                            	&	int	&	160	&	 Number of grids in x at ScaleFactor = 1 \\ 
\hline 
Ny                            	&	int	&	160	&	 Number of grids in y at ScaleFactor = 1 \\ 
\hline 
Nz                            	&	int	&	160	&	 Number of grids in z at ScaleFactor = 1 \\ 
\hline 
Nzelec                        	&	int	&	32	&	 No. grids in z in elec \& hole grids at ScaleFactor = 1 \\ 
\hline 
PhotonList                    	&	str	&	PhotonList&	 Photon list filename \\ 
\hline 
PixelAreas                    	&	int	&	0	&	 -1 - Don't calc areas, N - calc areas every Nth step\\ 
\hline 
PixelBoundaryLowerLeft        	&	float	&	2.00	&	 x,y coordinates of PixelBoundary lower left corner\\ 
\hline 
PixelBoundaryNx               	&	int	&	9	&	 Number of pixels in PixelBoundary\\ 
\hline 
PixelBoundaryNy               	&	int	&	9	&	 Number of pixels in PixelBoundary\\ 
\hline 
PixelBoundaryStepSize         	&	float	&	2.00	&	 Used with PixelBoundaryTestType=0\\ 
\hline 
PixelBoundaryTestType         	&	int	&	0	&	 0-Trace uniform grid, 1-TraceGaussian spot,\\
                                &               &               &        2,4-TraceRegion, 5-TraceList, 6-Fe55 cloud.\\ 
\hline 
\end{tabular}
\newpage
\begin{tabular}{|l|c|c|l|} 
\hline
Name                            &       type    &       Default &        Description \\
\hline 
PixelBoundaryUpperRight       	&	float	&	2.00	&	 x,y coordinates of PixelBoundary lower left corner\\
\hline 
PixelRegionLowerLeft       	&	float	&	2.00	&	 PixelRegion is used for PBTestType 0,2,4\\ 
\hline 
PixelRegionUpperRight      	&	float	&	2.00	&	 PixelRegion is used for PBTestType 0,2,4\\ 
\hline 
PixelSizeX                    	&	float	&	-1.00	&	 Pixel size in microns \\ 
\hline 
PixelSizeY                    	&	float	&	-1.00	&	 Pixel size in microns     \\ 
\hline 
QFemax                        	&	float	&	10.00	&	 Max QFe in look-up table\\ 
\hline 
QFemin                        	&	float	&	5.00	&	 Min QFe in look-up table\\ 
\hline 
SaturationModel               	&	int	&	0	&	 Experimental\\ 
\hline 
SaveData                      	&	int	&	1	&	 0 - Save only Pts, N save phi,rho,E every Nth step \\ 
\hline 
SaveElec                      	&	int	&	1	&	 0 - Save only Pts, N save Elec every Nth step \\ 
\hline 
SaveMultiGrids                	&	int	&	0	&	 0 - Don't save subgrids, 1 - Save all of the grids at all scales   \\ 
\hline 
ScaleFactor                   	&	int	&	1	&	 Power of 2 that sets the grid size \\ 
\hline 
Seed                          	&	int	&	77	&	 Pseudo random number seed \\ 
\hline 
SensorThickness               	&	float	&	100.00	&	 Sensor Thickness in microns   \\ 
\hline 
Sigmax                        	&	float	&	1.00	&	 Gaussian spot sigma in microns\\ 
\hline 
Sigmay                        	&	float	&	1.00	&	 Gaussian spot sigma in microns\\ 
\hline 
SimulationRegionLowerLeft     	&	float	&	2.00	&	 x,y coordinates of lower left corner of entire simulation\\ 
\hline 
TopAbsorptionProb             	&	float	&	0.00	&	 Probability an electron is absorbed if it reaches the top surface\\ 
\hline 
TreeRingAmplitude             	&	float	&	0.00	&	 Fractional amplitude (i.e. 0.10 is a 10\% variation)\\ 
\hline 
TreeRingAngle                 	&	float	&	0.00	&	 Direction of variation in degrees (0:constant in x)\\ 
\hline 
TreeRingPeriod                	&	float	&	0.00	&	 Period of sinusoidal variation in microns\\ 
\hline
Vbb                           	&	float	&	-50.00	&	 Back bias voltage\\ 
\hline 
VerboseLevel                  	&	int	&	1	&	 0 - minimal output, 1 - normal, 2 - more verbose 3-dump everything \\ 
\hline 
Vparallel\_lo                  	&	float	&	-8.00	&	 Parallel Low Voltage \\ 
\hline 
Vparallel\_hi                 	&	float	&	4.00	&	 Parallel High Voltage \\ 
\hline 
XBCType                       	&	int	&	1	&	 0 - Free BC, 1 - Periodic BC \\ 
\hline 
Xoffset                       	&	float	&	0.00	&	 Shift of Gaussian spot center\\ 
\hline 
YBCType                       	&	int	&	1	&	 0 - Free BC, 1 - Periodic BC \\ 
\hline 
Yoffset                       	&	float	&	0.00	&	 Shift of Gaussian spot center \\ 
\hline 
iterations                    	&	int	&	1	&	 Number of VCycles \\ 
\hline 
ncycle                        	&	int	&	100	&	 Number of SOR cycles at each resolution \\ 
\hline 
outputfilebase                  &	str	&       Test	&	Output filename base \\ 
\hline 
outputfiledir                   &	str	&	data	&	Output filename directory \\ 
\hline 
qfh                           	&	float	&	-100.00	&	 Hole quasi-Fermi level - applies everywhere except Fixed regions\\ 
\hline 
w                             	&	float	&	1.90	&	 Successive Over-Relaxation factor \\ 
\hline 
\end{tabular}

\section{Description of example configuration files included with the code.}
\renewcommand{\thefigure}{A.\arabic{figure}}
\setcounter{figure}{0}
\label{Example_Appendix}

The figures in this paper were generated with v1.0 of the  {\carlito Poisson\_CCD} code, at \cite{Poisson-CCD-code}. There are a total of 14 examples included with the code.  Each example is in a separate directory in the data directory, and has a configuration file of the form *.cfg. The parameters in the *.cfg files are commented to explain (hopefully) the purpose of each parameter, and a detailed listing of all configuration parameters is in Appendix \ref{Parameter_Appendix}. Python plotting routines are included with instructions below on how to run the plotting routines and the expected output.  The plot outputs are placed in the data/*/plots files, so you can see the expected plots without having to run the code.  If you edit the .cfg files, it is likely that you will need to customize the Python plotting routines as well.  The estimated run times given here are what was seen on a dual-core 1.4 GHz Intel Core i5 laptop computer.

\begin{itemize}
  \item Example 1: data/smallpixel/smallpixel.cfg
    \begin{enumerate}
      \item Purpose: A single pixel and surroundings.  The central pixel contains 100,000 electrons. No electron tracking or pixel boundary plotting is done.  The subgrids are saved so one can look at convergence.  This is useful for getting things set up rapidly 
      \item Syntax: src/Poisson data/smallpixel/smallpixel.cfg
      \item Expected run time: $\rm  < 1 minute$.
      \item Plot Syntax: python pysrc/Poisson\_Small.py data/smallpixel/smallpixel.cfg 0
      \item Plot Syntax: python pysrc/Poisson\_Convergence.py data/smallpixel/smallpixel.cfg 0        
    \end{enumerate}

  \item Example 2: data/pixel0/pixel.cfg
    \begin{enumerate}
      \item Purpose: A 9x9 grid of pixels at low resolution (ScaleFactor=1).  The central pixel contains 100,000 electrons. No electron tracking or pixel boundary plotting is done.
      \item Syntax: src/Poisson data/pixel0/pixel.cfg
      \item Expected run time: $\rm  \approx 1 minute$.
      \item Plot Syntax: python pysrc/Poisson\_Plots.py data/pixel0/pixel.cfg 0
      \item Plot Syntax: python pysrc/ChargePlots.py data/pixel0/pixel.cfg 0 2
    \end{enumerate}

  \item Example 3: data/pixel-itl/pixel.cfg
    \begin{enumerate}
      \item Purpose: A 9x9 grid of pixels at higher resolution (ScaleFactor=2).  The central pixel contains 100,000 electrons. The parameters are set up for the ITL STA3800C CCD.  This should give physically meaningful results, and is what was used in the published papers.  After solving Poisson's equation, electron tracking is done to determine the pixel distortions.
      \item Syntax: src/Poisson data/pixel-itl/pixel.cfg
      \item Expected run time: $\rm  \approx 30 minutes$.
      \item Plot Syntax: python pysrc/Poisson\_Plots.py data/pixel-itl/pixel.cfg 0
      \item Plot Syntax: python pysrc/ChargePlots.py data/pixel-itl/pixel.cfg 0 2
      \item Plot Syntax: python pysrc/VertexPlot.py data/pixel-itl/pixel.cfg 0 2
      \item Plot Syntax: python pysrc/Area\_Covariance\_Plot.py data/pixel-itl/pixel.cfg 0                
    \end{enumerate}

  \item Example 4: data/pixel-e2v/pixel.cfg
    \begin{enumerate}
      \item Purpose: A 9x9 grid of pixels at higher resolution (ScaleFactor=2).  The central pixel contains 100,000 electrons. The parameters are set up for the E2V CCD250 CCD.  This should give physically meaningful results, and is what was used in the published papers.  After solving Poisson's equation, electron tracking is done to determine the pixel distortions.
      \item Syntax: src/Poisson data/pixel-e2v/pixel.cfg
      \item Expected run time: $\rm  \approx 30 minutes$.
      \item Plot Syntax: python pysrc/Poisson\_Plots.py data/pixel-e2v/pixel.cfg 0
      \item Plot Syntax: python pysrc/ChargePlots.py data/pixel-e2v/pixel.cfg 0 2
      \item Plot Syntax: python pysrc/VertexPlot.py data/pixel-e2v/pixel.cfg 0 2
      \item Plot Syntax: python pysrc/Area\_Covariance\_Plot.py data/pixel-e2v/pixel.cfg 0                
    \end{enumerate}

  \item Example 5: data/pixel1/pixel.cfg
    \begin{enumerate}
      \item Purpose: A 9x9 grid of pixels at higher resolution (ScaleFactor=2).  The central pixel contains 100,000 electrons. The parameters are set up for the ITL STA3800C CCD.  This is included as an example of ElectronMethod=1, which iterates to find the value of the electron Quasi-Fermi level.  After solving Poisson's equation, electron tracking is done to determine the pixel distortions.
      \item Syntax: src/Poisson data/pixel1/pixel.cfg
      \item Expected run time: $\rm  \approx 45 minutes$.
      \item Plot Syntax: python pysrc/Poisson\_Plots.py data/pixel1/pixel.cfg 0
      \item Plot Syntax: python pysrc/ChargePlots.py data/pixel1/pixel.cfg 0 2
      \item Plot Syntax: python pysrc/VertexPlot.py data/pixel1/pixel.cfg 0 2
      \item Plot Syntax: python pysrc/Area\_Covariance\_Plot.py data/pixel1/pixel.cfg 0                
    \end{enumerate}

  \item Example 6: data/smallcapf/smallcap.cfg
    \begin{enumerate}
      \item Purpose: A simple field oxide capacitor for testing convergence.
      \item Syntax: src/Poisson data/smallcapf/smallcap.cfg
      \item Expected run time: $\rm  \approx 1 minute$.
      \item Plot Syntax: python pysrc/Poisson\_CapConvergence.py data/smallcapf/smallcap.cfg 0        
    \end{enumerate}

  \item Example 7: data/smallcapg/smallcap.cfg
    \begin{enumerate}
      \item Purpose: A simple gate oxide capacitor for testing convergence.
      \item Syntax: src/Poisson data/smallcapg/smallcap.cfg
      \item Expected run time: $\rm  \approx 1 minute$.
      \item Plot Syntax: python pysrc/Poisson\_CapConvergence.py data/smallcapg/smallcap.cfg 0        
    \end{enumerate}

  \item Example 8: data/edgerun/edge.cfg
    \begin{enumerate}
      \item Purpose: A simulation of the pixel distortion at the top and bottom edges of the ITL STA3800C CCD.
      \item Syntax: src/Poisson data/edgerun/edge.cfg
      \item Expected run time: $\rm  \approx 15 minutes$.
      \item Plot Syntax: python pysrc/Poisson\_Edge.py data/edgerun/edge.cfg 0
      \item Plot Syntax: python pysrc/Pixel\_Shift.py data/edgerun/edge.cfg 0                
    \end{enumerate}

  \item Example 9: data/iorun/io.cfg
    \begin{enumerate}
      \item Purpose: A simulation of the output transistor region of the ITL STA3800C CCD, including the last few serial stages.
      \item Syntax: src/Poisson data/iorun/io.cfg
      \item Expected run time: $\rm  \approx 5 minutes$.
      \item Plot Syntax: python pysrc/Poisson\_IO.py data/iorun/io.cfg 0
      \item Plot Syntax: python pysrc/ChargePlots\_IO.py data/iorun/io.cfg 0                
    \end{enumerate}

  \item Example 10: data/treering/treering.cfg
    \begin{enumerate}
      \item Purpose: A simulation of a 9x9 pixel region with a sinusoidal treering dopant variation introduced, so one can see the resulting pixel distortions.
      \item Syntax: src/Poisson data/treering/treering.cfg
      \item Expected run time: $\rm  \approx 20 minutes$.
      \item Plot Syntax: python pysrc/VertexPlot.py data/treering/treering.cfg 0 4
    \end{enumerate}

  \item Example 11: data/fe55/pixel.cfg
    \begin{enumerate}
      \item Purpose: A simulation of a 5x5 pixel region with a single Fe55 event above the center pixel.  To determine the distribution of pixel counts, one needs to run many of these simulations with a random center point.
      \item Syntax: src/Poisson data/fe55/pixel.cfg
      \item Expected run time: $\rm  \approx 5 minutes$.
      \item Plot Syntax: python pysrc/Poisson\_Fe55.py data/fe55/pixel.cfg 0 40
    \end{enumerate}

  \item Example 12: data/transrun/trans.cfg
    \begin{enumerate}
      \item Purpose: A simulation of the output transistor of the ITL STA3800C CCD.  In order to plot the I-V characteristic, this needs to be re-run with different gate voltage values.
      \item Syntax: src/Poisson data/transrun/trans.cfg
      \item Expected run time: $\rm  \approx 10 minutes$.
      \item Plot Syntax: python pysrc/Poisson\_Trans.py data/transrun/trans.cfg 0
      \item Plot Syntax: python pysrc/ChargePlots\_Trans.py data/transrun/trans.cfg 0
      \item Plot Syntax: python pysrc/MOSFET\_Calculated\_IV.py data/transrun/trans.cfg 0        
    \end{enumerate}

  \item Example 13: data/satrun/sat.cfg
    \begin{enumerate}
      \item Purpose: A simulation of the saturation level, as determined by the barrier lowering.  Again, this needs to be run multiple times with varying parallel voltages.
      \item Syntax: src/Poisson data/satrun/sat.cfg
      \item Expected run time: $\rm  \approx 20 minutes$.
      \item Plot Syntax: python pysrc/Poisson\_Sat.py data/satrun/sat.cfg 0
      \item Plot Syntax: python pysrc/ChargePlots.py data/satrun/sat.cfg 0 9
      \item Plot Syntax: python pysrc/Barrier.py data/satrun/sat.cfg 0
      \item Plot Syntax: python pysrc/Plot\_SatLevel.py data/satrun/sat.cfg 0                
    \end{enumerate}

  \item Example 14: data/bfrun/bf.cfg
    \begin{enumerate}
      \item Purpose: This builds up a Gaussian spot in the center of a 9x9 pixel region by sequentially solving Poisson's equation, adding electrons, and repeating.  As written, this is done 80 times, building up the central pixel charge to about 80,000 electrons.
      \item Syntax: src/Poisson data/bfrun/bf.cfg
      \item Expected run time: $\rm \approx 8 hours$.
      \item Plot Syntax: python Poisson\_Plots.py data/bfrun/bf.cfg 80
      \item Plot Syntax: python ChargePlots.py data/bfrun/bf.cfg 80 9    
      \item Plot Syntax: python Plot\_BF\_Spots.py data/bfrun/bf.cfg 0 NumSpots 80 (Assumes this was run NumSpots times with random center locations)
    \end{enumerate}

\end{itemize}

\end{appendices}

\end{document}